\documentclass[a4paper,12pt]{article}

\usepackage{amsmath,amssymb,epsfig}

\newcommand{\beq}{\begin{equation}}
\newcommand{\eeq}{\end{equation}}
\newcommand{\ba}{\begin{array}}
\newcommand{\ea}{\end{array}}
\newcommand{\bea}{\begin{eqnarray}}
\newcommand{\eea}{\end{eqnarray}}
\newcommand{\bean}{\begin{eqnarray*}}
\newcommand{\eean}{\end{eqnarray*}}
\newcommand{\eref}[1]{(\ref{#1})}
\newcommand{\sref}[1]{\S\ref{#1}}
\newcommand{\fref}[1]{Figure~\ref{#1}}

\newcommand{\tr}{\mathop{\rm Tr}}
\newcommand{\comment}[1]{}

\newcommand{\CM}{{\cal M}}
\newcommand{\CN}{{\cal N}}
\newcommand{\CZ}{{\cal Z}}
\newcommand{\cO}{{\cal O}}
\newcommand{\cC}{{\cal C}}
\newcommand{\cX}{{\cal X}}
\newcommand{\IP}{\mathbb{P}}
\newcommand{\IC}{\mathbb{C}}
\newcommand{\IV}{\mathbb{V}}
\newcommand{\IZ}{\mathbb{Z}}
\newcommand{\f}{{\cal F}^{\flat}}
\newcommand{\firr}[1]{{}^{{\rm Irr}}\!{\cal F}^{\flat}_{#1}}

\newcommand{\vev}[1]{\langle #1 \rangle}
\newtheorem{theorem}{\bf THEOREM}

\newtheorem{conjecture}{\bf CONJECTURE}

\newtheorem{observation}{\bf OBSERVATION}

\newcommand{\setall}{\setcounter{equation}{0}
        \setcounter{theorem}{0}}

\newcommand{\tmat}[1]{{\tiny \left(\begin{matrix} #1 \end{matrix}\right)}}

\begin{document}
\begin{flushright}
Bicocca-FT-07-17\\
CERN-PH-TH/2007-266;
SISSA 98/2007/EP\\
Imperial/TP/08/AH/01\\
NI07096\\ 
\end{flushright}
\vskip 0.25in

\renewcommand{\thefootnote}{\fnsymbol{footnote}}
\centerline{{\Huge The Master Space of $\CN=1$ Gauge Theories}}
~\\
{\bf
Davide Forcella${}^{1}$\footnote{\tt forcella@sissa.it}, 
Amihay Hanany${}^{2}$\footnote{\tt a.hanany@imperial.ac.uk, hanany@physics.technion.ac.il}, 
Yang-Hui He${}^{3}$\footnote{\tt hey@maths.ox.ac.uk}, 
Alberto Zaffaroni${}^{4}$\footnote{\tt alberto.zaffaroni@mib.infn.it}
}
~\\
~\\
{\hspace{-1in}
\scriptsize
\begin{tabular}{ll}
  ${}^1$ &{\it International School for Advanced Studies (SISSA/ISAS) \& INFN-Sezione di Trieste, via Beirut 2, I-34014, Trieste, Italy} \\
  &{\it PH-TH Division, CERN CH-1211 Geneva 23, Switzerland}\\
  ${}^2$ 
  &{\it Department of Physics 
Technion, Israel Institute of Technology 
Haifa 32000, Israel 
}\\
 &{\it Theoretical Physics Group, Blackett Laboratory, 
Imperial College, London SW7 2AZ, U.K. 
}\\
  ${}^3$
  & {\it Collegium Mertonense in Academia Oxoniensis, Oxford, OX1 4JD, U.K.}\\
  & {\it Mathematical Institute, Oxford University, 24-29 St.\ Giles', Oxford, OX1 3LB, U.K.}\\
  & {\it Rudolf Peierls Centre for Theoretical Physics, Oxford University, 1 Keble Road, OX1 3NP, U.K.}\\
  ${}^4$
  & {\it Universit\`{a} di Milano-Bicocca and INFN, sezione di Milano-Bicocca, Piazza della Scienza, 3; I-20126 Milano, Italy}
\end{tabular}
}

\begin{abstract}
The full moduli space $\CM$ of a class of $\CN=1$ supersymmetric gauge theories is studied. For gauge theories living on a stack of D3-branes at Calabi-Yau singularities $\cX$, $\CM$ is a combination of the mesonic and baryonic branches.
In consonance with the mathematical literature, the single brane moduli space is called the master space $\f$. Illustrating with a host of explicit examples, we exhibit many algebro-geometric properties of the master space such as when $\f$ is toric Calabi-Yau, behaviour of its Hilbert series, its irreducible components and its symmetries. 
In conjunction with the plethystic programme, we investigate the counting of BPS gauge invariants, baryonic and mesonic, using the geometry of $\f$ and show how its refined Hilbert series not only engenders the generating functions for the counting but also beautifully encode ``hidden'' global symmetries of the gauge theory which manifest themselves as symmetries of the complete moduli space $\CM$ for $N$ number of branes.
\end{abstract}

\setcounter{footnote}{0}
\renewcommand{\thefootnote}{\arabic{footnote}}

\newpage
\tableofcontents

\section{Introduction}\setall
The vacuum moduli space, $\CM$, of supersymmetric gauge theories is one
of the most fundamental quantities in modern physics. It is given by the
vanishing of the scalar potential as a function of the scalar components
of the superfields of the field theory. This, in turn, is the set of
zeros, dubbed flatness, of so-called {\bf D-terms} and {\bf F-terms},
constituting a parameter, or moduli, space of solutions describing the
vacuum. The structure of this space is usually complicated, and should be best cast in the language of algebraic varieties.  Typically, $\CM$ consists of a union of various branches, such as the mesonic branch or the baryonic branch, the Coulomb branch or the Higgs branch; the names are chosen according to some characteristic property of the specific branch.

It is a standard fact now that $\CM$ can be phrased in a succinct
mathematical language: it is the symplectic quotient of the space of
F-flatness, by the gauge symmetries provided by the D-flatness. We will
denote the space of F-flatness by $\f$ and symmetries prescribed by
D-flatness as $G_{D^{\flat}}$, then we have
\begin{equation}\label{M}
 \CM \simeq \f // G_{D^{\flat}} \ .
\end{equation}
Using this language, we see that $\f$ is a parent space whose quotient is a moduli space. In
the mathematical literature, this parent is referred to as the {\bf
master space} \cite{master} and to this cognomen we shall
adhere\footnote{We thank Alastair King to pointing this out to us.}.

In the context of certain string theory backgrounds, $\CM$ has an elegant geometrical realisation. When D-branes are transverse to an affine (non-compact)
Calabi-Yau space $\cX$, a supersymmetric gauge theory exists on the world-volume of the branes. 
Of particular interest is, of course, when the  gauge theory, prescribed by D3-branes, is in four-dimensions. Our main interest is the IR
physics of this system, where all Abelian symmetry decouples and the gauge
symmetry is fully non-Abelian, 
typically given by products of $SU(N)$ groups. The Abelian factors
are not gauged but rather appear as global baryonic symmetries of the
gauge theory. 

Under these circumstances, the moduli space $\CM$ is a
combined mesonic branch and baryonic branch. These branches are not necessarily separate (irreducible) components of $\CM$ but are instead in most cases
intrinsically merged into one or more components in $\CM$. Even when mesonic
and baryonic directions are mixed, it still makes sense to talk about
the more familiar mesonic moduli space ${}^{{\rm mes}}\!{\cal M}$, as the sub-variety of $\CM$ parameterized by
mesonic operators only. Since mesonic operators have zero
baryonic charge, and thus invariant under the $U(1)$ Abelian factors,
the mesonic moduli space can be obtained as a further quotient of 
$\CM$ by the Abelian symmetries:

\begin{equation}
{}^{{\rm mes}}\!{\cal M} \simeq \CM // U(1)_{D^{\flat}} \ .
\label{mesmod}
\end{equation}

We are interested in the theory of physical $N$ 
branes probing the singularity; 
the gauge theory on the worldvolume is then superconformal.

It is of particular interest to consider the case of a single D3-brane
transverse to the Calabi-Yau three-fold $\cX$, which will enlighten the geometrical
interpretation of the moduli space. 
Since the motion of the D-brane is parameterized by this transverse space, part of the vacuum moduli space $\CM$ is, {\it per constructionem}, precisely the space which the brane probes. The question of which part of the moduli space is going to be clarified in detail in this paper. For now it is enough to specify that for a single D-brane it is the mesonic branch: 
\beq
\CM \supset {}^{{\rm mes}}\!{\cal M} \simeq \cX \simeq \mbox{non-compact Calabi-Yau threefold transverse to 
D3-brane.}
\eeq

In this paper we are interested in studying the full moduli space $\CM$. 
In general, $\CM$ will be an algebraic variety of dimension greater than three. In the case of a single D3-brane, $N=1$, the
IR theory has no gauge group and the full moduli space $\CM$ is given by
the space of F-flatness $\f$. Geometrically, $\f$ is a $\IC^{dim \f -3}$ fibration over the mesonic moduli space $\cX$ given by relaxing the
$U(1)$ D-term constraints in \eref{mesmod}. Physically, $\f$ is obtained
by adding {\it baryonic} directions to the mesonic moduli space. Of course,  
we can not talk about baryons for $N=1$ but we can alternatively interpret
these directions as Fayet-Iliopoulos (FI) parameters in the stringy realization of the $N=1$ gauge theory. Indeed on the world-volume of a single D-brane there is a collection of $U(1)$ gauge groups, each giving rise to a FI parameter, which 
relax the D-term constraint in \eref{mesmod}. When these FI parameters acquire vacuum expectation values they induce non-zero values for the collection of fields in the problem and this is going to be taken to be the full moduli space  $\CM\equiv \f$. If one further restricts the moduli space to zero baryonic number we get the mesonic branch which is $\cX$, the Calabi-Yau itself.

For $N>1$ number of physical branes, the situation is more subtle. 
The mesonic moduli space,
probed by a collection of $N$ physical branes, 
is given by the symmetrized product
of $N$ copies of $\cX$ \footnote{Cf. ~\cite{Berenstein:2002ge} for a consistency analysis of this identification.}. The full moduli space $\CM$ is a bigger algebraic
variety of more difficult characterization. One of the purposes of this paper is to elucidate this situation and to show how the properties of $\CM$
for arbitrary number of branes are encoded in the master space $\f$ for a 
single brane.
In view of the importance of the master space $\f$ for one brane even for
larger $N$, we will adopt the important convention 
that, in the rest of the paper, the word {\bf master space}
and the symbol $\f$ will refer to the $N=1$ case, unless explicitly stated.

The symplectic quotient structure of \eref{M} should immediately suggest
toric varieties in the $N=1$ case. Indeed, the case of $\cX$ being a {\bf toric Calabi-Yau
space} has been studied extensively over the last decade.
The translation between
the physics data of the D-brane world volume gauge theory and the
mathematical data of the geometry of the toric $\cX$ was initiated in
\cite{Douglas:1997de,Beasley:1999uz,Feng:2000mi}. In the computational
language of \cite{Feng:2000mi}, the process of arriving at the toric
diagram from the quiver diagram plus superpotential was called the
forward algorithm, whereas the geometrical computation of the
gauge theory data given the toric diagram was called the inverse
algorithm. The computational intensity of these algorithms,
bottle-necked by finding dual integer cones, has been a technical
hurdle.

Only lately it is realized that the correct way to think
about toric geometry in the context of gauge theories is through the
language of {\bf dimer models} and {\bf brane tilings}
\cite{dimers,tilings}. Though the efficiency of this new perspective
has far superseded the traditional approach of the partial resolutions
of the inverse algorithm, the canonical toric language of the latter
is still conducive to us, particularly in studying objects beyond
$\cX$, and in particular, $\f$. We will thus make extensive use of this 
language as well as the modern one of dimers.

Recently, a so-called {\bf plethystic programme}
\cite{pleth,Butti:2006au,Forcella:2007wk,Butti:2007jv,Forcella:2007ps} has been
advocated in counting the gauge invariant operators of supersymmetric
gauge theories, especially in the above-mentioned D-brane quiver theories.
For mesonic BPS operators, the fundamental generating function turns out to
be the Hilbert series of $\cX$ \cite{pleth}. The beautiful fact is that the full
counting \cite{Forcella:2007wk}, including baryons as well, 
is achieved
with the Hilbert series of $\f$ for one brane! Indeed, mesons have gauge-index
contractions corresponding to closed paths in the quiver diagram and the
quotienting by $G_{D^{\flat}}$ achieves this. Baryons, on the other
hand, have more general index-contractions and correspond to all paths
in the quiver; whence their counting should not involve the quotienting
and the master space should determine their counting.

In light of the discussions thus far presented, it is clear that the
master space $\f$ of gauge theories, especially those arising from toric
Calabi-Yau threefolds, is of unquestionable importance. It is therefore
the purpose of this paper to investigate their properties in detail.
We exhibit many algebro-geometric properties of the master space $\f$
for one brane, including its decomposition into irreducible components,
its symmetry and
the remarkable property of the biggest component of being always
{\bf toric Calabi-Yau} if $\cX$ is.
We point out that even though we mainly concentrate on the master space $\f$
for one brane, we are able, using the operator counting technique, to extract important information about the complete moduli space $\mathcal{M}$, information such as its symmetries for arbitrary number of branes.

The organisation of the paper is as follows. In \sref{s:master} we introduce the concept of the master space in detail, starting with various computational approaches, emphasizing on the Hilbert series and toric presentation, and then launching into a wealth of illustrative examples. We recapitulate at the end of the section on the key abstract properties of the master space while reviewing the plethystic programme which counts gauge invariants given the Hilbert series. We then, in \sref{s:branch}, discuss how the master space, and indeed, the moduli space of supersymmetric theories, are generically reducible and have various branches which we will obtain by primary decomposition. We shall see how certain lower dimensional components of one theory causes it to flow to another.
Another remarkable feature of gauge theories arising from underlying geometry such as those living on world-volumes of D-brane probes at Calabi-Yau singularities is that the symmetries of the master space can manifest themselves as hidden global symmetries of the gauge theory. In \sref{s:hidden} we examine how such symmetries beautifully exhibit themselves in the plethystics of the Hilbert series of the master space by explicitly arranging themselves into representations of the associated Lie algebra. Finally, we part with concluding remarks and outlooks in \sref{s:conc}.

Due to the length of this paper, we find it expedient to supplement it with a companion essay \cite{letter}. The reader who wishes for a quick tour of the high-lights is referred thereto.

\section{The Master Space}\label{s:master}\setall
It was realized in \cite{Beasley:1999uz} that for a single D3-brane probing a toric Calabi-Yau threefold $\cX$, the space of solutions to the F-terms (or, in the notation of Section 1.2 in Cit.~Ibid., the commuting variety ${\CZ}$) is also a toric variety, albeit of higher dimension. In particular, for a quiver theory with $g$ nodes, it is of dimension $g+2$. Thus we have the first important property for the master space $\f$ for a toric $U(1)^g$ quiver theory:
\begin{equation}\label{dimF}
\dim(\f) = g+2 \ .
\end{equation}
This can be seen as follows. The F-term equations are invariant  under a $(\mathbb{C}^*)^{g+2}$ action,
given by the three mesonic symmetries of the toric quiver, one R and two flavor
symmetries, as well as the $g-1$ baryonic symmetries, 
including the anomalous ones.
This induces an action of $(\mathbb{C}^*)^{g+2}$ on the master space.
Moreover, the dimension of $\f$ is exactly $g+2$ as the following
field theory argument shows. We know that the mesonic moduli space has
dimension three, being isomorphic to the transverse Calabi-Yau manifold $\cX$.
As described in the introduction, the mesonic moduli space is obtained as the solution of both F-term and D-term constraints for the $U(1)^g$ quiver theory. The full master space is obtained by relaxing the $U(1)$ D-term constraints. Since an overall $U(1)$ factor is decoupled, we have $g-1$ independent baryonic parameters corresponding to the values of the $U(1)$ D-terms, which, by abuse of language, we can refer to as FI terms. As a result, the dimension of the master space is $g+2$, given by three mesonic parameters plus $g-1$ FI terms.

A second property of the Master Space is its reducibility. We will see in several examples below that it decomposes into several irreducible components, the largest of which turns out to be of the same dimension as the master space, and more importantly the largest component is a toric Calabi-Yau manifold in $g+2$ complex dimensions which is furthermore a cone over a Sasaki Einstein manifold of real dimension $2g+3$. Examples follow below, as well as a proof that it is Calabi-Yau in section \sref{MS}.

Having learned the dimension of the master space, let us now see a few warm-up examples of what the space looks like. Let us begin with an illustrative example of perhaps the most common affine toric variety, viz., the Abelian orbifold.

\subsection{Warm-up: An Etude in $\f$}
The orbifold $\IC^3 / \IZ_k \times \IZ_m$ is well-studied. It is an affine toric singularity whose toric diagrams are lattice points enclosed in a right-triangle of lengths $k \times m$ (cf.~e.g., \cite{fulton}). The matter content and
superpotential for these theories can be readily obtained from
brane-box constructions \cite{bb}. We summarize as follows:
\beq\label{zkzm}
\ba{cl}
\mbox{Gauge Group Factors:} & m n; \\
\mbox{Fields:} & \mbox{bi-fundamentals} 
  \{ X_{i,j}, \ Y_{i,j}, \ Z_{i,j} \} \mbox{ from node $i$ to node $j$}
  \\
  & (i,j) \mbox{ defined modulo } (k,m) \ , \qquad
  \mbox{ total } = 3 m n; \\
\mbox{Superpotential:} & 
W = \sum\limits_{i=0}^{k-1} \sum\limits_{j=0}^{m-1} 
  X_{i,j} Y_{i+1, j} Z_{i+1, j+1} -
  Y_{ij} X_{i, j+1} Z_{i+1, j+1} \ .
\\
\ea
\eeq
We point out here the important fact that in the notation above, when
either of the factors $(k,m)$ is equal to 1, the resulting theory is
really an $\CN=2$ gauge theory since the action of $\IZ_k \times
\IZ_m$ on the $\IC^3$ is chosen so that it degenerates to have a line
of singularities when either $k$ or $m$ equals 1. In other words, if
$m=1$ in \eref{zkzm}, we would have a $(\IC^2 / \IZ_k) \times \IC$
orbifold rather than a proper $\IC^3 / \IZ_k$ one
(in the language of \cite{su4}, this proper action would be called
``transitive''). We shall henceforth
be careful to distinguish these two types of 
orbifolds with this notation.

\subsubsection{Direct Computation}
Given the matter content and superpotential of an $\CN=1$ gauge theory, a convenient and algorithmic method of computation, emphasising \eref{M}, is that of \cite{Gray:2006jb}. We can immediately compute
$\f$ as an affine algebraic variety: it is an ideal in $\IC^{3 m n}$
given by the $3mn$ equations prescribed by $\partial W = 0$. The
defining equation is also readily obtained:
it is simply the image of the ring map $\partial W$ from 
$\IC^{3 m n} \rightarrow \IC^{3 m n}$, i.e., the syzygies of the $3mn$
polynomials given by $\partial W$.
To specify $\f$ explicitly as an affine algebraic variety, let us use the notation that
\beq
\label{DDE}
(d , \delta | p) := \mbox{affine variety of dimension $d$ and degree
  $\delta$ embedded in $\IC^p$ } \ .
\eeq

Subsequently, we present what $\f$ actually is as an algebraic variety
for some low values of $(m,k)$ in Table \ref{t:fzkzm}.
We remind the reader that, of course, quotienting these above spaces by $D^{\flat}$, which in the algorithm of \cite{Gray:2006jb} is also achieved by a ring map, should all give our starting point of $\cX = \IC^3 / \IZ_k \times \IZ_m$.

\begin{table}[t]
\begin{center}
$\ba{|c||c|c|c|c|c|}\hline
m \backslash k & 1 & 2 & 3 & 4 & 5 \\ \hline \hline
1 & (3,1|3) & (4,2|6) & (5,4|9) & (6,8|12) & (7,16|15) 
\\ \hline 
2 & (4,2|6) & (6,14|12) & (8,92|18) & (10,584|24) & (12, 3632|30)
\\ \hline
3 & (5,4|9) & (8,92|18) & (11,1620|27) & (14,26762|36) &
(17, 437038 | 45)
\\ \hline
\ea$
\end{center}
\caption{{\sf The master space $\f$ for $\IC^3 / \IZ_k \times \IZ_m$ as explicit algebraic varieties, for some low values of $k$ and $m$.}}
\label{t:fzkzm}
\end{table}

As pointed out above, the limit
of either $k$ or $m$ going to 1 in the theory prescribed in
\eref{zkzm} is really just $(\IC^2 / \IZ_k) \times \IC$ with the
$X$-fields, say, acting as adjoints. The first row and column
of \eref{t:fzkzm} should
thus be interpreted carefully since they are secretly $\CN=2$ theories
with adjoints. We shall study proper $\IC^3 / \IZ_k$ theories later.

\subsubsection{Hilbert Series}\label{s:hilb1}
One of the most important quantities which characterize an algebraic variety is the Hilbert series\footnote{Note, however, that the Hilbert series is not a topological invariant and does depend on embedding.}.
In \cite{pleth}, it was pointed out that it is also key to the problem of counting gauge invariant operators in the quiver gauge theory. Let us thus calculate this quantity for $\f$.

We recall that for a variety $M$ in $\IC[x_1,...,x_k]$, the Hilbert series is the generating function for the dimension of the graded pieces:
\beq
H(t; M) = \sum\limits_{i=-\infty}^{\infty} \dim_{\IC} M_i t^i \ ,
\eeq
where $M_i$, the $i$-th graded piece of $M$ can be thought of as the number of independent degree $i$ (Laurent) polynomials on the variety $M$. The most useful property of $H(t)$ is that it is a rational function in $t$ and can be written in 2 ways:
\beq\label{hs12}
H(t; M) = \left\{
\ba{ll}
\frac{Q(t)}{(1-t)^k} \ , & \mbox{ Hilbert series of First Kind} \ ;\\
\frac{P(t)}{(1-t)^{\dim(M)}} \ , & \mbox{ Hilbert series of Second Kind} \ . 
\ea
\right.
\eeq
Importantly, both $P(t)$ and $Q(t)$ are polynomials with {\em integer}
coefficients. The powers of the denominators are such that the leading
pole captures the dimension of the manifold and the embedding space,
respectively. In particular, $P(1)$ always equals the degree of the
variety. We can also relate the Hilbert series to the Reeb vector, which
elucidate in Appendix \ref{app:reeb}.

For now, let us present in Table \ref{hilbzkzm} the Hilbert series, in Second form, of some of the examples above in Table \ref{t:fzkzm}.
\begin{table}[t]
\begin{center}
$\ba{|c|c|c|c|}\hline
(k, m) & \f & \mbox{Hilbert Series } H(t; \f) \\ \hline
(2,2) & (6,14|12) & 
\frac{1 + 6\,t + 9\,t^2 - 5\,t^3 + 3\,t^4}{{\left( 1 - t \right) }^6}
\\ \hline
(2,3) & (8,92|18) & \frac{1 + 10\,t + 37\,t^2 + 47\,t^3 - 15\,t^4 +
  7\,t^5 + 5\,t^6}{{\left( 1 - t \right) }^8} \\ \hline
(2,4) & (10,584|24) &
\frac{1 + 14\,t + 81\,t^2 + 233\,t^3 + 263\,t^4 - 84\,t^5 + 4\,t^6 +
  71\,t^7 - 7\,t^8 + 8\,t^9}{{\left( 1 - t \right) }^{10}} \\ \hline
(3,3) &  (11,1620|27) &
\frac{1 + 16\,t + 109\,t^2 + 394\,t^3 + 715\,t^4 + 286\,t^5 - 104\,t^6
  + 253\,t^7 - 77\,t^8 + 27\,t^9}{{\left( 1 - t \right) }^{11}} \\
\hline
\ea$
\end{center}
\caption{{\sf The Hilbert series, in second form, of the master space $\f$ for $\IC^3 / \IZ_k \times \IZ_m$, for some low values of $k$ and $m$.}}
\label{hilbzkzm}
\end{table}
We see that indeed the leading pole is the dimension
of $\f$ and that the numerator evaluated at 1 (i.e., the coefficient of
the leading pole) is equal to the degree of $\f$. Furthermore, we point out that the Hilbert series thus far defined depends on a single variable $t$, we shall shortly discuss in \sref{s:hilb} and also from Appendix \ref{app:reeb} how to refine this to multi-variate and how these variables should be thought of as chemical potentials.

\subsubsection{Irreducible Components and Primary Decomposition}\label{s:irred}
The variety $\f$ may not be a single irreducible piece, but rather, be
composed of various components. This is a well recognized feature in supersymmetric gauge theories. The different components are typically called {\bf branches} of the moduli space, such as Coulomb or Higgs branches, mesonic or baryonic branches. Possibly the most famous case is the Seiberg-Witten solution to ${\cal N}=2$ supersymmetric gauge theories which deals mainly with the Coulomb branch but gives some attention to the other components on the moduli space which are generically called the Higgs branch. 

It is thus an interesting question to identify the different components since sometimes the massless spectrum on each component has its own unique features. We are naturally lead to a process to extract the various components which in the math literature is called {\bf primary decomposition} of the ideal corresponding to $\f$. This is an extensively studied algorithm in computational algebraic geometry (cf.~e.g.~\cite{m2}) and a convenient programme which calls these routines externally but based on the more familiar Mathematica interface is \cite{stringvacua}.

\paragraph{Example of $\IC^2 / \IZ_3$: }
Let us first exemplify with the case of $(\IC^2 /
\IZ_3) \times \IC$ (i.e., $(k,m) = (1,3)$). This case, having ${\cal N}=2$ supersymmetry, is known to have a Coulomb branch and a Higgs branch which is 
a combined mesonic and  baryonic branch. The superpotential is
\beq
W_{(\IC^2 / \IZ_3) \times \IC} = 
 X_{0,0}\,Y_{0,0}\,Z_{0,1} - 
 Y_{0,0}\,X_{0,1}\,Z_{0,1} + 
 X_{0,1}\,Y_{0,1}\,Z_{0,2} - 
 Y_{0,1}\,X_{0,2}\,Z_{0,2} +
 X_{0,2}\,Y_{0,2}\,Z_{0,0} - 
 Y_{0,2}\,X_{0,0}\,Z_{0,0}  
 \ ,
\eeq
composed of a total of 9 fields, where the $X$ fields are adjoint fields in the ${\cal N}=2$ vector multiplet of the corresponding gauge group. Here, since we are dealing with a single D-brane, these fields have charge 0. Hence, $\f$ is defined, as an ideal in
$\IC^9$, by 9 quadrics:
\beq\label{f-z3}\ba{rcl}
\f_{(\IC^2 / \IZ_3) \times \IC} &=& \{
- Y_{0,2}\,Z_{0,0} + Y_{0,0}\,Z_{0,1} \ , \;
- Y_{0,0}\,Z_{0,1} + Y_{0,1}\,Z_{0,2} \ , \;
  Y_{0,2}\,Z_{0,0} - Y_{0,1}\,Z_{0,2} \ , \\
&&  \left( X_{0,0} - X_{0,1} \right) \,Z_{0,1}\ ,  \;
  \left( X_{0,1} - X_{0,2} \right) \,Z_{0,2}\ ,  \;
  \left( -X_{0,0} + X_{0,2} \right) \,Z_{0,0}\ , \\
&&  \left( -X_{0,0} + X_{0,2} \right) \,Y_{0,2}\ ,  \;
  \left( X_{0,0} - X_{0,1} \right) \,Y_{0,0}\ ,  \; 
  \left( X_{0,1} - X_{0,2} \right) \,Y_{0,1}
\} \ .
\ea\eeq
Immediately one can see that on one branch, the so-called Higgs branch, which we shall denote as 
$\f_1$, the adjoint fields $X$ do not participate. Thus it is defined by the first 3 equations in \eref{f-z3}: 3 quadrics in 6 variables. 
Furthermore, one can
see that one of the quadrics is not independent. Therefore $\f_1$ is a complete intersection of 2
quadratics in $\IC^6$, of dimension 4. To this we must form a direct
product with the Coulomb branch which is parametrized by the $X$-directions, which turns out to be one dimensional
$X_{0,0} = X_{0,1} = X_{0,2}$ (in order to satisfy the remaining 6
equations from \eref{f-z3} such that the $Y$'s and $Z$'s are
non-zero). Hence, $\f$ is 5-dimensional (as we expect from \eref{dimF} since there are $g=3$ nodes), of degree 4, and composed of 2 quadrics in $\IC^6$ crossed with $\IC$.

Now, this example may essentially be observed with ease, more
involved examples requires an algorithmic approach, as we shall see in 
many cases which ensue. 
The decoupling of
the $X$'s is indicative of the fact that we have an non-transitive action
and this is indeed just an orbifold of $\IC^2$.

\begin{figure}[t]\begin{center}
$\ba{ccc}
\ba{l} \epsfxsize = 2.5cm\epsfbox{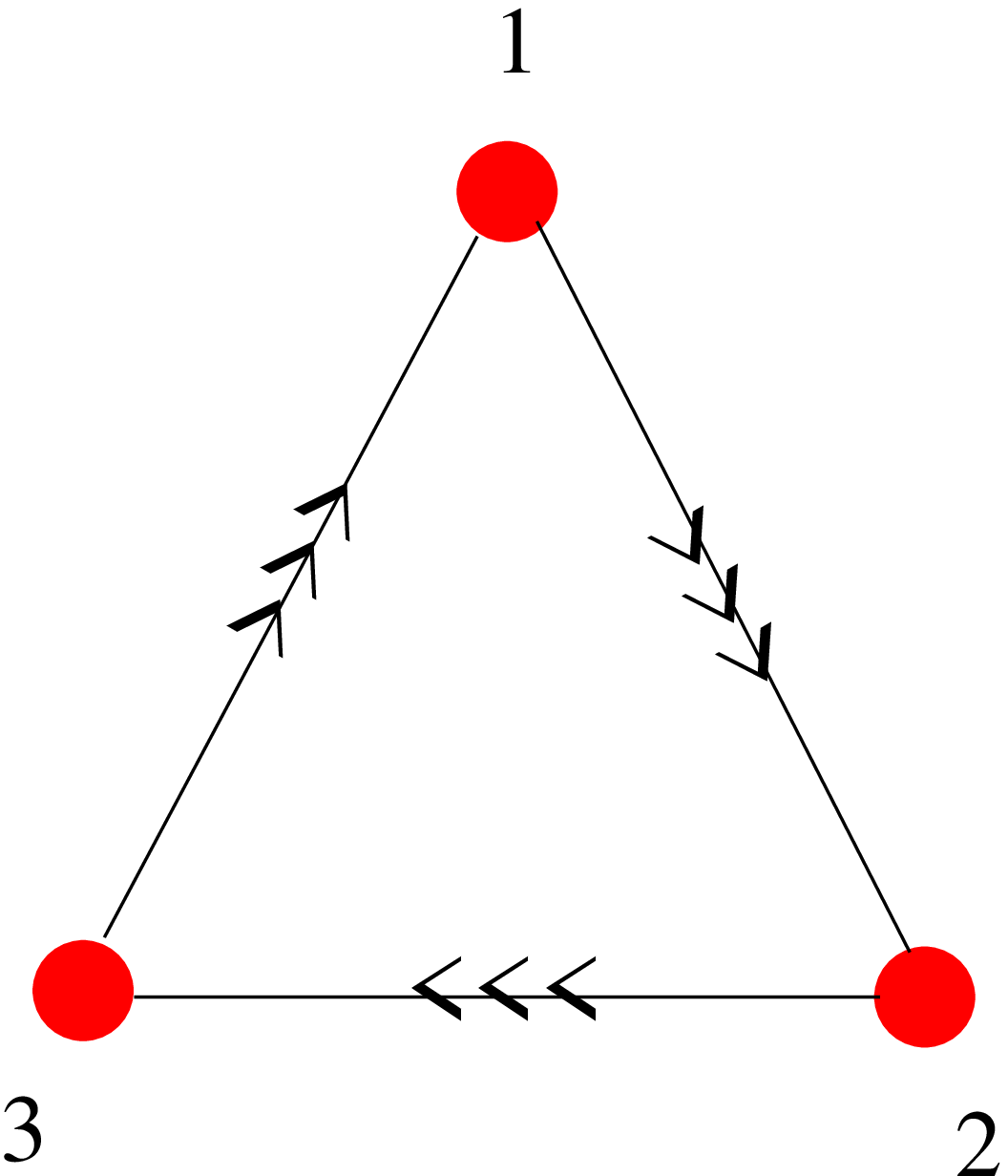} \ea
& \qquad &
\ba{l}
W_{\IC^3 / \IZ_3}=\epsilon_{\alpha\beta\gamma} X^{(\alpha)}_{12}
  X^{(\beta)}_{23} X^{(\gamma)}_{31} \\
\mbox{9 fields and 27 GIO's} \\
X^{(\alpha)}_{12}, X^{(\beta)}_{23},
X^{(\gamma)}_{31}, \alpha,\beta,\gamma=1,2,3
\ea
\ea$
\caption{{\sf The quiver diagram and superpotential for $dP_0$.}}
\label{f:dP0theory}
\end{center}\end{figure}

\paragraph{Example of $dP_0 = \IC^3 / \IZ_3$: }
Let us next study a proper orbifold $\IC^3 / \IZ_3$ with a
non-trivial action, say $(1,1,1)$, on the $\IC^3$. This is also referred to in the literature as $dP_0$, the cone over the zeroth del Pezzo surface. In other words, this is the total space of the line bundle $\cO_{\IP^2}(-3)$ over $\IP^2$. Here, there are 9 fields and the theory is summarized in \fref{f:dP0theory}.
Now, the F-terms are 
\beq\label{fc3z3}\ba{rcl}
\f_{\IC^3 / \IZ_3} & = & \{
-X_{2,3}^3\,X_{3,1}^2 + X_{2,3}^2\,X_{3,1}^3 \ , \;
X_{2,3}^3\,X_{3,1}^1 - X_{2,3}^1\,X_{3,1}^3\ , \;
-X_{2,3}^2\,X_{3,1}^1 + X_{2,3}^1\,X_{3,1}^2, \\
&&  X_{1,2}^3\,X_{3,1}^2 - X_{1,2}^2\,X_{3,1}^3\ , \;
-X_{1,2}^3\,X_{3,1}^1 + X_{1,2}^1\,X_{3,1}^3\ , \;
X_{1,2}^2\,X_{3,1}^1 - X_{1,2}^1\,X_{3,1}^2, \\
&&-X_{1,2}^3\,X_{2,3}^2 + X_{1,2}^2\,X_{2,3}^3\ , \;
  X_{1,2}^3\,X_{2,3}^1 - X_{1,2}^1\,X_{2,3}^3\ , \;
  -X_{1,2}^2\,X_{2,3}^1 +  X_{1,2}^1\,X_{2,3}^2
\}
\ea\eeq
and a direct primary decomposition shows that $\f$ is itself
irreducible and it is given, using the notation in (\ref{DDE}), as
\beq\label{FdP0}
\f_{\IC^3 / \IZ_3} \simeq (5,6 | 9), 
\eeq
a non-complete-intersection of 9 quadrics as given in \eref{fc3z3}, embedded in $\IC^9$. We see that the dimension is 5 since there are 3 nodes. The Hilbert
series (cf.~\cite{Butti:2007jv} and a re-derivation below) is
\begin{equation}
H(t; \f_{\IC^3 / \IZ_3}) = \frac {1+4t+t^2}{(1-t)^5} .
\label{HSdP0-1}
\end{equation}

\paragraph{Example of $\IC^3 / \IZ_2 \times \IZ_2$: }
Finally, take the case of $(k,m)=(2,2)$, or the Abelian orbifold $\IC^3 / \IZ_2 \times \IZ_2$, studied in detail by \cite{Beasley:1999uz,Feng:2000mi}. The reader is referred to \fref{f:pmC3Z2Z2} which we present in the next section.
Here, there are $2^2=4$ nodes and we expect the dimension of the master space to be 6.
Again, we can obtain $\f$ from \eref{zkzm} and perform primary decomposition
on it. We see, using \cite{m2}, that there are 4 irreducible components, three of which are merely coordinate planes and trivial. Along these directions the gauge theory admits an accidental supersymmetry enhancement to ${\cal N} = 2$ and each direction can be viewed as a Coulomb branch of the corresponding ${\cal N} = 2$ supersymmetric theory. 

The non-trivial piece of  $\f_{\IC^3 / \IZ_2 \times \IZ_2}$ is a Higgs branch and is an irreducible variety which we shall call $\firr{\IC^3 / \IZ_2 \times \IZ_2}$; it is also of dimension 6. Moreover, it is of degree 14, and is prescribed by the intersection of 15 quadrics in 12 variables. The Hilbert series for $\firr{\IC^3 / \IZ_2 \times \IZ_2}$ is given by
\begin{equation}\label{Hz2z2}
H(t;~\firr{\IC^3/\IZ_2\times \IZ_2})=\frac{1+6 t+6 t^2+t^3}{(1-t)^6} \ .
\end{equation}

\paragraph{Summary: }
We have therefore learned, from our few warm-up exercises, that one can compute $\f$ directly, its Hilbert series, dimension, degree, etc., using the methods of computational algebraic geometry, using, in particular, computer packages such as \cite{m2}. In general, the master space $\f$ need not be irreducible. We will see this in detail in the ensuing sections. The smaller components are typically referred to as {\bf Coulomb Branches} of the moduli space.

The largest irreducible component of the master space $\f$ will play a special 
r\^{o}le in our analysis and deserves a special notation. We will denote it
$\firr{~}$. In the toric case, it is also known as the
{\bf coherent component} of the master space.  In all our examples, this component actually has the same dimension as the full master space and, as we will see in detail in \sref{MS}, is in fact Calabi-Yau. Let us redo Table \ref{hilbzkzm}, now for the coherent component; we present the result in Table \ref{hilbzkzm-irr}.

\begin{table}[h]
\begin{center}
$\ba{|c|c|c|c|}\hline
(k, m) & \firr{~} & \mbox{Hilbert Series } H(t; \firr{~}) 
\\ \hline
(2,2) & (6,14|12) & 
\frac{1 + 6\,t + 6\,t^2 + t^3}{{\left( 1 - t \right) }^6}
\\ \hline
(2,3) & (8,92|18) & \frac{1 + 10\,t + 35\,t^2 + 35\,t^3 + 10\,t^4 +
  t^5}{{\left( 1 - t \right) }^8} \\ \hline
(2,4) & (10,584|24) &
\frac{1 + 14\,t + 78\,t^2 + 199\,t^3 + 199\,t^4 + 78\,t^5 + 14\,t^6 +
  t^7}{{\left( 1 - t \right) }^{10}} \\ \hline
(3,3) &  (11,1620|27) &
\frac{1 + 16\,t + 109\,t^2 + 382\,t^3 + 604\,t^4 + 382\,t^5 + 109\,t^6
  + 16\,t^7 + t^8}{{\left( 1 - t \right) }^{11}} \\
\hline
\ea$
\end{center}
\caption{{\sf 
The Hilbert series, in second form, of the coherent component of the master space $\f$ for $\IC^3 / \IZ_k \times \IZ_m$, for some low values of $k$ and $m$. The reader is referred to Table \ref{hilbzkzm} for contrast.}}
\label{hilbzkzm-irr}
\end{table}

We note that the degree and dimension of $\firr{~}$ is the same as that of $\f$, again suggesting that the smaller dimensional components are merely linear pieces. Nevertheless the linear pieces play a crucial r\^{o}le in the analysis of the physics of these theories since there is a rich structure of mixed Higgs and Coulomb branches; we will see this in \sref{s:branch}.
Moreover, we observe that the numerator now becomes symmetric (palindromic), a remarkable fact that will persist throughout our remaining examples; we will show why in \sref{MS}.

\subsubsection{Toric Presentation: Binomial Ideals and Toric Ideals}
\label{s:toric}
We have so far seen the application of computational algebraic geometry in studying the master space as an explicit algebraic variety. This analysis has not fully exploited the fact that $\f$ is in fact a toric variety; that is, we have been simplifying and primary decomposing the ideal corresponding to $\f$ without utilising its inherent combinatorial, toric nature. Now, given an ideal ${\cal I}$ of a polynomial ring, when each generator of ${\cal I}$, i.e., each polynomial equation, is written in the form ``monomial = monomial,'' then ${\cal I}$ is known as a {\bf toric ideal} and the variety thus defined will be toric \cite{sturmfels}. The F-terms arising from the partials of the superpotential in \eref{zkzm} clearly obey this condition and this is true for all toric Calabi-Yau spaces $\cX$.

A single matrix can in fact encode all the information about the ideal of $\f$, called the {\bf $K$-matrix} in \cite{Beasley:1999uz,Feng:2000mi}. For orbifolds of the form $\IC^3 / \IZ_n$ with action $(a,b,-1)$ the $K$-matrix were constructed in Eqs 4.1-3 of \cite{Douglas:1997de} and that of  $\IC^3 / \IZ_3 \times \IZ_3$, in \cite{Beasley:1999uz,Feng:2000mi} (see also \cite{Muto:2001gu,Sarkar:2000iz}). In general, the procedure is straight-forward: solve the F-terms explicitly so that each field can be written as a fraction of monomials in terms of a smaller independent set. Then, translate these monomials as a matrix of exponents; this is the $K$-matrix.

We have seen from above that the master space $\f$ and its coherent component $\firr{~}$ of a toric $U(1)^g$ quiver gauge theory is a variety of dimension $g+2$. The F-terms provide $E$ equations for the $E$ fields in the quiver. Not all of them are algebraically independent, since the F-terms are invariant
under the $(\mathbb{C}^*)^{g+2}$ toric action. It follows that the $E$
fields can be parameterized in terms of $g+2$ independent fields.   
$K$ is therefore a matrix of dimensions $g+2$ by $E$.

\paragraph{$\IC^3 / \IZ_3$ Revisited: }
For the $\IC^3 / \IZ_3$ example above let us illustrate the
procedure. Solving \eref{fc3z3}, we have that
\beq
{{X_{1,2}^1}= 
    {\frac{X_{1,2}^3\,X_{3,1}^1}{X_{3,1}^3}}}, \
  {{X_{1,2}^2}= 
    {\frac{X_{1,2}^3\,X_{3,1}^2}{X_{3,1}^3}}}, \ 
  {{X_{2,3}^1}= 
    {\frac{X_{2,3}^3\,X_{3,1}^1}{X_{3,1}^3}}}, \
  {{X_{2,3}^2}= 
    {\frac{X_{2,3}^3\,X_{3,1}^2}{X_{3,1}^3}}} \ .
\eeq
We see that there are 5 fields $\{X_{1, 2}^3, X_{2, 3}^3, X_{3, 1}^1,
  X_{3, 1}^2, X_{3, 1}^3\}$ which parameterize all 9 fields, signifying
  that $\f$ is 5-dimensional, as stated above.
Whence we can plot the 9 fields in terms of the 5 independent ones as:
\beq
\ba{c|ccccccccc}
& X_{1, 2}^1 & X_{1, 2}^2& X_{1, 2}^3& X_{2, 3}^1& X_{2, 3}^2& X_{2,
  3}^3&  X_{3, 1}^1& X_{3, 1}^2& X_{3, 1}^3 
\\ \hline
X_{1, 2}^3 & 1 & 1 & 1 & 0 & 0 & 0 & 0 & 0 & 0 \\
X_{2, 3}^3 & 0 & 0 & 0 & 1 & 1 & 1 & 0 & 0 & 0 \\
X_{3, 1}^1 & 1 & 0 & 0 & 1 & 0 & 0 & 1 & 0 & 0 \\
X_{3, 1}^2 & 0 & 1 & 0 & 0 & 1 & 0 & 0 & 1 & 0 \\
X_{3, 1}^3 & -1 & -1 & 0 & -1 & -1 & 0 & 0 & 0 & 1
\ea
\eeq
where we read each column as the exponent of the 5 solution
fields. This is the $K$-matrix, and captures the toric information
entirely. 
In particular, the number of rows $g+2$ of the $K$-matrix is the dimension of $\f$ and the columns of $K$ are the charge vectors of the toric action
of $(\mathbb{C}^*)^{g+2}$ on $\f$. 

The $K$-matrix gives a nice toric presentation
for the coherent component $\firr{~}$ of the master space. 
It defines an integer cone $\sigma_K^{\vee}$ in $\mathbb{Z}^{g+2}$
prescribed by the non-negative integer span of the columns of
$K$. Then, in the
language of \cite{fulton}, $\firr{~}$ as an algebraic (toric) variety of
dimension $g+2$, is given by
\beq
\firr{~} \simeq \mbox{Spec}_{\IC}[\sigma_K^{\vee} \cap \IZ^{g+2}] \ .
\eeq

Now, the toric diagram of the variety is not, customarily, given by 
$\sigma^{\vee}$, but, rather, by the dual cone $\sigma$. Let us denote
the generators of $\sigma$ as the matrix $T$, then, using the
algorithm in \cite{fulton}, we can readily find that
\beq\label{TdP0}
{\scriptsize
T = \left( \begin{matrix}
   0 & 0 & 1 & 1 & 0 & 0 \cr
   0 & 0 & 1 & 0 & 1 & 0 \cr
   1 & 0 & 0 & 0 & 0 & 1 \cr
   0 & 1 & 0 & 0 & 0 & 1 \cr
   0 & 0 & 1 & 0 & 0 & 1 \cr  \end{matrix} \right) \ .
}
\eeq
$T$ is a matrix of dimensions $g+2$ by $c$, where the number of its columns, $c$, is a special combinatorial number which is specific to the particular toric phase \cite{Feng:2000mi}. We recall that the dual cone consists of all lattice points which have non-negative inner product with all lattice points in the original cone. In terms of our dual matrices,
\beq\label{Pmat}
P := K^t \cdot T \ge 0 \ .
\eeq
The columns of $T$, plotted in $\IZ^5$, is then the toric diagram, and the number of vectors needed to characterize the toric diagram in $\IZ^5$ is $c$ which for our particular case is equal to 6. The $P$ matrix takes the form
\beq\label{PdP0}
P= {\tiny \left(
\begin{array}{llllll}
 1 & 0 & 0 & 1 & 0 & 0 \\
 0 & 1 & 0 & 1 & 0 & 0 \\
 0 & 0 & 1 & 1 & 0 & 0 \\
 1 & 0 & 0 & 0 & 1 & 0 \\
 0 & 1 & 0 & 0 & 1 & 0 \\
 0 & 0 & 1 & 0 & 1 & 0 \\
 1 & 0 & 0 & 0 & 0 & 1 \\
 0 & 1 & 0 & 0 & 0 & 1 \\
 0 & 0 & 1 & 0 & 0 & 1
\end{array}
\right) } .
\eeq

In fact, one can say much more about the product matrix $P$, of dimensions $E$ by $c$: it consists of only zeros and ones. In \cite{dimers}, it was shown that this matrix, which translates between the linear sigma model fields and space-times fields, also encodes perfect matchings of the dimer model description of the toric gauge theory. This provides a more efficient construction of the master space. We will return to a description of this in \sref{s:dimer}.

\paragraph{$\IC^3 / \IZ_2 \times \IZ_2$ Revisited: }
Next, let us construct the $K$-matrix for our $\IC^3 / \IZ_2 \times
\IZ_2$ example. We recall that the master space and its coherent component are of dimension 6. Using the superpotential
\eref{zkzm} to obtain the 12 F-terms, we can again readily solve the
system and obtain
\beq
{\scriptsize
K = \left( \begin{array}{cccccccccccc}
1 & 1 & 1 & 1 & 0 & 0 & 0 & 0 & 0 & 0 & 0 & 0 \cr 
   0 & 1 & 1 & 0 & 0 & 1 & 1 & 0 & 0 & 0 & 0 & 0 \cr 
   0 & -1 & 
    -1 & 0 & 1 & 0 & 0 & 1 & 0 & 0 & 0 & 0 \cr 1 & 1 & 
   0 & 0 & 0 & 0 & 0 & 0 & 1 & 1 & 0 & 0 \cr 0 & 0 & 
   0 & 0 & 1 & 0 & 1 & 0 & 1 & 0 & 1 & 0 \cr -1 & 
    -1 & 0 & 0 & -1 & 0 & -1 & 0 & 
    -1 & 0 & 0 & 1 \cr 
\end{array} \right) \ ,
}
\eeq
giving us the toric diagram with 9 vectors in 6-dimensions as
\beq\label{Tz2z2}
{\scriptsize
T = \left( \begin{array}{ccccccccc}
   0 & 0 & 1 & 0 & 0 & 1 &0 & 0 & 1   \cr 
   0 & 1& 0& 1 & 0 & 0 & 0 & 1 & 0   \cr 
0 & 1 & 0 & 1 & 0 & 0 &    0 & 0 & 1 \cr 
1 & 0 & 0 & 1 & 0 & 0 & 1 & 0 & 0   \cr 
   1 &0 & 0 & 0 & 1 & 1 & 0 & 0 & 0  
\cr 1 & 0 & 0 & 1 & 0  & 1 & 0 & 0 & 0  \cr  
\end{array} \right) \ .
}
\eeq

\subsubsection{Computing the Refined Hilbert Series} 
\label{s:hilb}
Let us now study the Hilbert series in the language of the $K$-matrix.
We mentioned in \sref{s:hilb1} that the Hilbert series should be refined. This is easily done and is central to the counting of gauge invariants in the plethystic programme. Recall that the master space $\f$ and its coherent component
$\firr{~}$ are given by a set of algebraic equations
in $\IC[X_1,...,X_E]$, where $E$ is the number of fields in the quiver.
Since we are dealing with a toric variety of dimension $g+2$ we have
an action of $(\IC^*)^{g+2}$ on $\f$ and $\firr{}$ and we can give
$g+2$ different weights to the variables $X_i$. 

What should these weights be? Now, all information about
the toric action is encoded in the matrix $K$. Therefore, a natural weight is to simply use the columns of $K$! There are $E$ columns, each being a vector of charges of length $g+2$, as needed, and we can assign the $i$-th column to the variable $X_i$ for $i=1,\ldots,E$. Since each weight is a vector of length $g+2$, we need a $g+2$-tuple for the counting which we can denote by $\underline{t}={t_1,...,t_{g+2}}$. Because the dummy variable $\underline{t}$ keeps track of the charge, we can think of the components as chemical potentials \cite{pleth,Butti:2007jv}. With this multi-index variable (chemical potential)
we can define the {\bf Refined Hilbert Series} of $\f$ as
the generating function for the dimension of the graded pieces:
\beq\label{refineHilb}
H(\underline{t}; \f) = \sum_{{\underline \alpha}} \dim_{\IC} \f_{{\underline \alpha}}\, {\underline t}^{\underline \alpha} \ ,
\eeq
where $\f_{\underline \alpha}$, the ${\underline \alpha}$-th multi-graded piece of $\f$, can be thought of as the number of independent multi-degree ${\underline \alpha}$ Laurent monomials on the variety $\f$. A similar expression
applies to $\firr{~}$. 

The refined Hilbert series for $\f$ and $\firr{~}$ can be computed
from the defining equations of the variety, using computer algebra program
and primary decomposition, as emphasized in \cite{Gray:2006jb}. 
In addition, for the coherent component $\firr{~}$, 
there exists an efficient algorithm \cite{M2book} 
for extracting the refined Hilbert series from the matrix $K$ that can be
implemented in Macaulay2 \cite{m2}. 
We give the actual code in Appendix \ref{ap:M2}.

A crucial step in the above analysis seems to rely upon our ability to
explicitly solve the F-terms in terms of a smaller independent set of
variables. This may be computationally intense. For Abelian orbifolds
the solutions may be written directly using the symmetries of the
group, as was done in \cite{Douglas:1997de,Beasley:1999uz}; 
in general, however, the $K$-matrix may not be immediately obtained.
We need, therefore, to resort to a more efficient method.

\subsubsection{The Symplectic Quotient Description}\label{s:Molien}
There is an alternative and useful description of the toric 
variety $\firr{~}$ as a symplectic quotient \cite{fulton}. In the math language this is also known as the {\bf Cox representation} of a toric variety \cite{cox} 
and in physics language, as a {\bf linear sigma model}. In this representation, we have a nice way of computing the refined Hilbert series using the Molien invariant. 

Now, the $c$ generators of the dual cone $T$ are not independent in $\mathbb{Z}^{g+2}$. The kernel of the matrix $T$
\begin{equation}
T \cdot Q =0
\end{equation}
or, equivalently, the kernel of the matrix $P = K^t \cdot T$
\begin{equation}
P \cdot Q =0
\end{equation}
defines a set of charges for the symplectic action. The $c-g-2$ rows of $Q^t$
define vectors of charges for the $c$ fields in the linear sigma model 
description of $\firr{~}$ \cite{Douglas:1997de,Beasley:1999uz,fulton}:
\begin{equation}
\firr{~} = \mathbb{C}^c//Q^t \ .
\end{equation}

A crucial observation is that if the rows of $Q^t$ sum to zero, then $\firr{~}$ is Calabi-Yau. 
In the following we will see that this is the case for all the examples we
will encounter. Indeed, it is possible to show this is general; for clarity
and emphasis we will leave the proof of the fact to the summary section of
\sref{MS} and first marvel at this fact for the detailed examples.

The refined Hilbert series for $\firr{~}$ can be computed using the
Molien formula \cite{Butti:2007jv,molien}, by projecting the trivial Hilbert series of $\mathbb{C}^c$ onto $(\mathbb{C}^{*})^{c-g-2}$ invariants. We will need
in the following the refined Hilbert series depending on some or all of the
$g+2$ chemical potentials $t_i$ and, therefore, we keep our
discussion as general as possible. The dependence on the full set of
parameters $t_i$ is given by using the Cox homogeneous coordinates for
the toric variety \cite{cox}. We introduce $c$ homogeneous variables $p_\alpha$ with chemical potentials 
$y_\alpha, \alpha=1,...,c$ acted on by $(\mathbb{C}^{*})^{c-g-2}$ with charges 
given by the rows of $Q^t$.
The Hilbert series for $\mathbb{C}^c$ is freely generated and is simply:
\begin{equation}
H(\underline{y},\IC^c) = 
H(\{y_{\alpha}\},\IC^c) = \prod_{\alpha=1}^c \frac{1}{1-y_\alpha} \ ,
\end{equation}
where we have written $\underline{y}$ as a vector, in the notation of \eref{refineHilb}, to indicate refinement, i.e., $H$ depends on all the 
$\{y_{\alpha}\}$'s.

Next, the vector of charges of $p_\alpha$ under the $(\mathbb{C}^{*})^{c-g-2}$ action is given by $\{Q_{1\alpha},...,Q_{c-g-2,\alpha}\}$. 
By introducing $c-g-2$ $U(1)$ chemical potentials $z_1,...,z_{c-g-2}$
we can write the Molien formula, which is a localisation formula of the Hilbert series from the ambient space to the embedded variety of interest, as
\begin{equation}\label{Molien}
H(\underline{y},~\firr{~}) =
\int \prod_{i=1}^{c-g-2} \frac{dz_i}{z_i} H(\{y_{\alpha} \ z_1^{Q_{1\alpha}} \ldots z_{c-g-2}^{Q_{c-g-2,\alpha}} \},\IC^c) =
\int \prod_{i=1}^{c-g-2} \frac{dz_i}{z_i} \prod_{\alpha=1}^c \frac{1}{1-y_\alpha z_1^{Q_{1\alpha}} \ldots z_{c-g-2}^{Q_{c-g-2,\alpha}}}
 \ .
\end{equation}
The effect of the integration on the $U(1)$ chemical potentials $z_i$ is
to project onto invariants of the $U(1)$'s. In this formula we integrate over the unit circles in $z_i$ and we should take $|y_\alpha|<1$.

Due to the integration on the $c-g-2$ variables $z_i$, the final result 
for the Hilbert series depends only on $g+2$ independent combinations
of the parameters $y_\alpha$, which can be set in correspondence with the
$g+2$ parameters $t_i$. We can convert the $y_\alpha$ variables to the set
of independent $g+2$ chemical potential $t_i$ for the toric action using the 
matrix $T$ as \cite{cox} 
\begin{equation}
t_i =\prod_{\alpha=1}^c y_\alpha^{T_{i\alpha}} \ .
\label{toricsympaction}
\end{equation}

Recall that the $g+2$  variables $t_i$ are the chemical potentials
of the $g+2$ elementary fields that have been chosen to parametrize $\f$. The weight of the $i$-th elementary field
$X_i$ for $i=1,...,E$ under this parameterization is given by 
the $i$-th column of the matrix $K$. Denoting with $q_i\equiv 
q_i(\underline{t})$
the chemical potential for the $i$-th field we thus have
\begin{equation}
q_i =   \prod_{\alpha=1}^c y_\alpha^{P_{i\alpha}}
\label{toricsympaction2}
\end{equation}
where we used $K^t\cdot T=P$. 

Formula (\ref{toricsympaction}), or equivalently (\ref{toricsympaction2}),
 allows us to
determine the parameters $y_i$ entering the Molien formula in terms of
the chemical potentials for the elementary fields of the quiver gauge theory.
This identification can be only done modulo an intrinsic $c-g-2$ 
dimensional ambiguity parameterized by the matrix $Q$: $y_\alpha$ 
are determined
by \eref{toricsympaction} up to vectors in the kernel of $P$.  
We will see in the next section that there is an efficient way of assigning
charges under the non-anomalous symmetries to the variables $y_\alpha$ using
perfect matchings. In particular, if we are interested in the Hilbert series
depending on a single parameter $t$, we can always assign charge $t$ to
the variables corresponding to external perfect matchings and charge one to
all the other variables. 
Let us now re-compute the refined Hilbert series for the two examples studied above, using \eref{Molien}. For simplicity, we compute the Hilbert series depending on one parameter $t$, referring to Appendix \ref{ap:ref} for an example
of computation of the refined Hilbert series depending on all parameters. 

\paragraph{Symplectic Quotient for $dP_0=\IC^3 / \IZ_3$: } 
The kernel of the matrix $T$, from \eref{TdP0},
can be easily computed to be the vector $Q$:
\begin{equation}
P \cdot Q = 0 \qquad \Rightarrow \qquad Q^t = \left(
\begin{array}{llllll}
- 1 & -1 & -1 & 1 & 1 & 1
\end{array}
\right) ,
\label{chargesdp0}
\end{equation}
which forms the vector of charges for the linear sigma model description of the master space for $dP_0$. In this description, therefore, we find that the master space, which we recall to be irreducible, is given by
\begin{equation}
\mathbb{C}^6//[-1,-1,-1,1,1,1] \ .
\label{MSdP0}
\end{equation}

We can compute the Hilbert series using the Molien formula \eref{Molien}. For 
simplicity, we consider the Hilbert series depending on a single chemical
potential $t_i\equiv t$. This is obtained by assigning 
chemical potential  $t$ to all fields of the linear sigma model with 
negative charge. This assignment of charges 
is consistent with formula (\ref{toricsympaction}) and, as we will see in the next section, is equivalent to assigning $t$ to the three external perfect 
matchings and $1$ to the three internal ones. 
We introduce a new chemical potential $z$ for the $U(1)$ charge
and integrate on a contour slightly smaller than the unit circle.
Using the residue technique outlined in Section 3.2 of \cite{Forcella:2007wk},
 we find that the contribution to the integral comes from the pole at $z=t$, whence
\begin{equation}
H(t; \f_{dP_0}) = \oint \frac{dz}{ 2 \pi i z(1-t/z)^3 (1-z)^3} = \frac {1+4t+t^2}{(1-t)^5} ,
\label{HSdP0}
\end{equation}
agreeing precisely with \eref{HSdP0-1}.

\paragraph{Symplectic Quotient for $\IC^3 / \IZ_2\times \IZ_2$: } 
In this case the kernel for the matrix $T$, from \eref{Tz2z2}, is three dimensional and it is encoded by the matrix:
\beq\label{chargesZ2}
Q^t= \left(
\begin{array}{lllllllll}
 -1 & -1 & 0 & 1 & 1 & 0 & 0 & 0 & 0 \\
 -1 & 0 & -1 & 0 & 0 & 1 & 1 & 0 & 0 \\
 0 & -1 & -1 & 0 & 0 & 0 & 0 & 1 & 1
\end{array}
\right) .
\eeq
The rows of $Q^t$ induce a $(\mathbb{C}^{*})^ 3$ action on $\mathbb{C}^9$ which allows us to represent the coherent component of $\f_{\IC^3 / \IZ_2\times \IZ_2}$ as a symplectic quotient:
\begin{equation}
\firr{\IC^3 / \IZ_2\times \IZ_2} = \mathbb{C}^9//(\mathbb{C}^*)^3 \ .
\end{equation}
We compute here the Hilbert series depending on a single parameter 
$t_i\equiv t$. Formula (\ref{toricsympaction}) is consistent with assigning
chemical potential $t$ to the fields of the sigma model with negative charges 
and chemical potential $1$ to all the others. As we will see in the next section, this corresponds to  the
natural choice which assigns $t$ to the external perfect matchings and
$1$ to the internal ones. The Molien formula reads
\begin{eqnarray}
H(t,\firr{\IC^3 / \IZ_2\times \IZ_2})&=&\int \frac{dr dw ds}{r w s}\frac{1}{(1-t/r w)(1- t/r s)(1- t/w s) (1-r)^2(1-w)^2(1-s)^2} \nonumber\\
& = & \frac{1+6 t+6 t^2+t^3}{(1-t)^6} \ ,
\end{eqnarray}
which agrees with \eref{Hz2z2} exactly. The computation of the refined Hilbert
series depending on all six parameters is deferred to Appendix \ref{ap:ref}.

\subsubsection{Dimer Models and Perfect Matchings}\label{s:dimer}
It was recently realized that the most convenient way of describing toric quiver gauge theories is that of dimers and brane-tilings. 
Let us re-examine our above analysis using the language of dimers and perfect matchings. The reader is referred to \cite{dimers} and for a comprehensive introduction, especially to \cite{Kennaway:2007tq}. We will focus on perfect matchings and the matrix $P$ defined in \eref{Pmat}.

Now, $K$ is of size $(g+2) \times E$ with $E$ the number of fields, and $g$ the number of gauge group factors. The matrix $T$ is of size $(g+2) \times c$, where $c$ is the number of generators of the dual cone. Thus, $P$ is a matrix of size $E \times c$. The number $c$ is, equivalently, the number of perfect matchings for the corresponding tiling (dimer model). In fact, the matrix $P$ contains entries which are either $0$ or $1$, encoding whether a field $X_i$ in the quiver is in the perfect matching $p_\alpha$:
\begin{equation}
P_{i \alpha} =
\begin{cases} 1 & \text{if $X_i \in p_\alpha$,} \\
0 &\text{if $X_i \not \in p_ \alpha $.}
\end{cases}
\end{equation}

\paragraph{Dimer Model for $dP_0$: }

\begin{figure}[t]
\begin{center}
\includegraphics[scale=0.4]{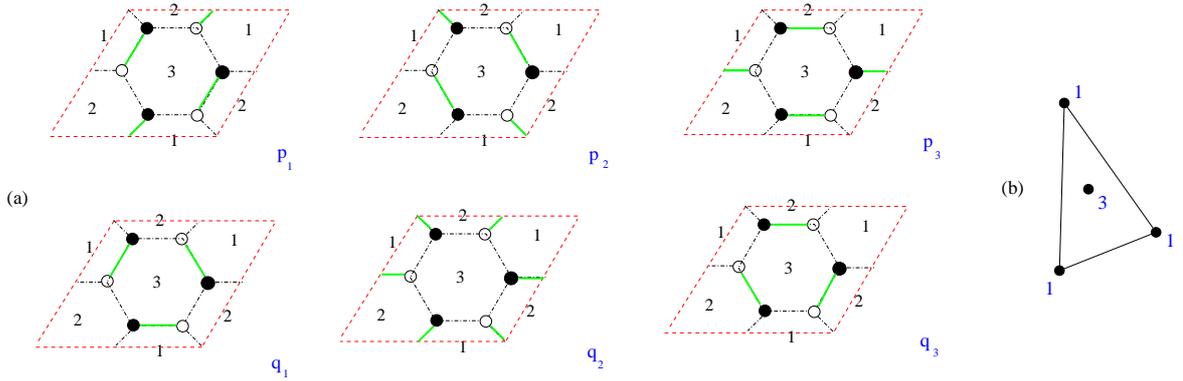} 
\caption{{\sf (a) The perfect matchings for the dimer model corresponding to $dP_0$, with $p_i$ the external matchings and $q_i$, the internal;
(b) The toric diagram, with the labeled multiplicity of GLSM fields, of $dP_0$.}}
\label{f:pmdP0}
\end{center}
\end{figure}

Let us first discuss in detail the example of $dP_0 = \IC^3/\IZ_3$.
Using the $P$ matrix in Equation (\ref{PdP0}) we can draw the 6 different perfect matchings. They are shown in Figure \ref{f:pmdP0}.
The first three perfect matchings are identified as the external perfect matchings $p_{1,2,3}$ while the last three are the internal perfect matchings $q_{1,2,3}$ associated with the internal point in the toric diagram of $dP_0$. For reference we have also drawn the toric diagram, together with the multiplicity of the gauged linear sigma model fields associated with the nodes, which we recall from \cite{Feng:2000mi}. In fact, it is this multiplicity that led to the formulation of dimers and brane tilings as originally discussed in the first reference of \cite{dimers}. Here we find another important application of this multiplicity.

Now, from Figure \ref{f:pmdP0} we notice that the collection of all external perfect matchings cover all edges in the tiling. Similarly, the collection of all internal perfect matchings cover all edges in the tiling, giving rise to a linear relation which formally states $p_1+p_2+p_3 = q_1+q_2+q_3$, or as a homogeneous relation, $-p_1-p_2-p_3+q_1+q_2+q_3=0$. Since the $P$ matrix encodes whether an edge is in a perfect matching, the linear combination of matchings encodes whether an edge is in that linear combination. Using the homogeneous form of the relation we in fact find that the vector $(-1,-1,-1,1,1,1)$ is in the kernel of $P$ and thus forms a row of the kernel matrix $Q^t$. Since the rank of the matrix $P$ is equal to the dimension of the master space, $g+2=5$, we conclude that this is the only element in the matrix $Q$. We have thus re-obtained the result
\eref{chargesdp0}.

\begin{figure}[t]
\begin{center}
\includegraphics[scale=0.4]{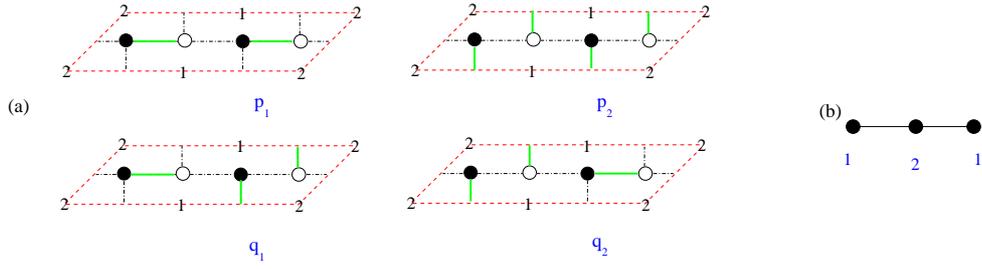} 
\caption{{\sf (a) The perfect matching for the dimer model corresponding to $\IC^2/\IZ_2$. The two upper perfect matchings are associated with the two external points in the toric diagram and the two lower perfect matchings are associated with the internal point in the toric diagram, drawn in (b).}}
\label{f:pmC2Z2}
\end{center}
\end{figure}

\paragraph{Dimer Model for $\IC^2/\IZ_2$: }
Next, let us look at the example of $\IC^2/\IZ_2$. The toric diagram, with multiplicity 1, 2, 1, and the corresponding perfect matchings are shown in \fref{f:pmC2Z2}, denoting the two external perfect matchings by $p_{1,2}$ and those of the internal point by $q_{1,2}$. 
A quick inspection of the perfect matchings shows a linear relation $-p_1-p_2+q_1+q_2=0$, leading to a charge matrix $(-1,-1,1,1)$ for the linear sigma model description of the master space for the orbifold $\IC^2/\IZ_2$. As is computed 
in \cite{Forcella:2007wk,Butti:2007jv} and as we shall later encounter in detail in \sref{C2Z2} we find that the master space is nothing but the conifold.

\begin{figure}[t]
\begin{center}
\includegraphics[scale=0.6]{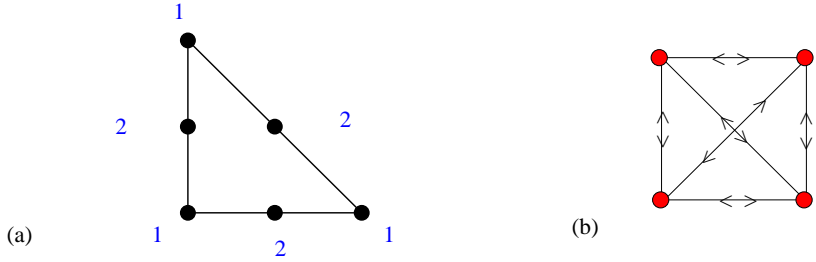} 
\caption{{\sf (a) The toric diagram for $\IC^3/\IZ_2\times \IZ_2$ together with the GLSM multiplicities/perfect matchings marked for the nodes. There is a total of 9 perfect matchings in the dimer model, which for sake of brevity we do not present here; (b) The associated quiver diagram.}}
\label{f:pmC3Z2Z2}
\end{center}
\end{figure}

\paragraph{Dimer Model for $\IC^3/\IZ_2\times \IZ_2$: }
The above arguments can also be used to compute the linear sigma model description of the orbifold $\IC^3/\IZ_2\times \IZ_2$. The toric diagram, shown in Figure \ref{f:pmC3Z2Z2}, consists of 6 points, 3 external with perfect matchings $p_{1,2,3}$, and 3 internal. 
The 3 internal points form a local $\IC^2/\IZ_2$ singularity and have multiplicity 2. We find 6 internal perfect matchings $q_{1,2,3,4,5,6}$. There are 3 sets of 3 points each, sitting on a line. Each such line is a $\IC^2/\IZ_2$ singularity and can therefore be used to write down a relation between the 4 perfect matchings, giving rise to 3 conifold like relations, $p_1 + p_2 = q_1+q_2, p_1+p_3 = q_3+q_4, p_2+p_3 = q_5+q_6$. Thus the master space of the orbifold $\IC^3/\IZ_2\times \IZ_2$ is an intersection of 3 conifold-like (quadric) relations in $\IC^9$, with a charge matrix
\beq\label{QC2Z2Z2}
Q^t= \left(
\begin{array}{lllllllll}
 -1 & -1 & 0 & 1 & 1 & 0 & 0 & 0 & 0 \\
 -1 & 0 & -1 & 0 & 0 & 1 & 1 & 0 & 0 \\
 0 & -1 & -1 & 0 & 0 & 0 & 0 & 1 & 1
\end{array}
\right) .
\eeq
which precisely agrees with Equation \eref{chargesZ2}.

We see that we can find a diagrammatic way, using dimers and perfect matchings, to find the charges of the matrix $Q$ and thereby the charges of the linear sigma model which describes the master space. This description is good for a relatively small number of perfect matchings and small number of fields in the quiver. When this number is large we will need to refer to the computation using the kernel of the $P$ matrix. We thus reach an important conclusion:
\begin{observation}
The coherent component of the master space of a toric quiver theory is generated by perfect matchings of the associated dimer model.
\end{observation}
This should be a corollary of the more general Birkhoff-Von Neumann Theorem\footnote{We thank Alastair King for first pointing this out to us. The precise relation between this theorem and perfect matching is now actively pursued by Alastair King and Nathan Broomhead and we look forward to their upcoming publication.} \cite{BvN}:
\begin{theorem}
An $n \times n$ doubly stochastic matrix (i.e., a square matrix of non-negative entries each of whose row and column sums to 1) is a convex linear combination of permutation matrices.
\end{theorem}

We can easily make contact between the perfect matching description and the
more mathematical description of the master space outlined in the previous 
section. As shown in
\cite{dimers}, the perfect matchings $p_\alpha$ parameterize the solutions of the
F-terms condition through the formula
\begin{equation}
X_i =\prod_{\alpha=1}^c p_\alpha^{P_{i \alpha}} \ .
\label{pmpar}
\end{equation}
This equation determines the charges of the perfect matchings (modulo
an ambiguity given by $Q^t$) in terms of the $g+2$ field theory charges.
In the previous section we introduced a homogeneous variable $y_\alpha$ for each perfect matching $p_\alpha$. We see that formula \eref{toricsympaction} for the chemical potential of the field $X_i$
\begin{equation}
q_i =   \prod_{\alpha =1}^c y_\alpha^{P_{i\alpha}} \, ,
\label{toricsympaction22}
\end{equation}
following from the
Cox description of the toric variety, nicely matches with \eref{pmpar} 
obtained from the dimer description.

Finally, there is a very simple way of determining the {\bf non-anomalous} 
charges of the perfect matchings, which is useful in computations 
based on the Molien formula.
The number of non-anomalous $U(1)$ symmetries of a toric 
quiver gauge theory is precisely the number of external perfect matchings,
or equivalently, the number $d$ of external points in the toric diagram.
This leads to a very simple parameterization for the non-anomalous 
charges \cite{Butti:2005vn,Butti:2005ps}: 
assign a different chemical potential $x_i$ for $i=1,...,d$ to each
external perfect matching and assign $1$ to the internal ones. An explicit
example is discussed in Appendix \ref{ap:ref}.
It follows
from \eref{pmpar} that this prescription is equivalent to the one discussed
in \cite{Butti:2005vn,Butti:2005ps}. 
In particular, in the computation of the Hilbert series depending on just
one parameter $t$, we can assign chemical potential $t$ to all the external
matchings and $1$ to the internal ones, as we did in section \ref{s:Molien}.


\subsection{Case Studies}\label{s:case}
Enriched by a conceptual grasp and armed with the computational techniques for describing the master space, we can now explore the wealth of examples of toric D-brane gauge theories which have bedecked the literature. We shall be reinforced with our lesson that $\f$ for the $U(1)^g$ quiver theory is of dimension $g+2$ and generically decomposes into several components the top dimensional one of which is also a Calabi-Yau variety of dimension $g+2$, as well as some trivial lower-dimensional linear pieces defined by the vanishing of combinations of the coordinates.

\subsubsection{The Suspended Pinched Point.}
\begin{figure}[t]
\begin{center}
\includegraphics[scale=0.6]{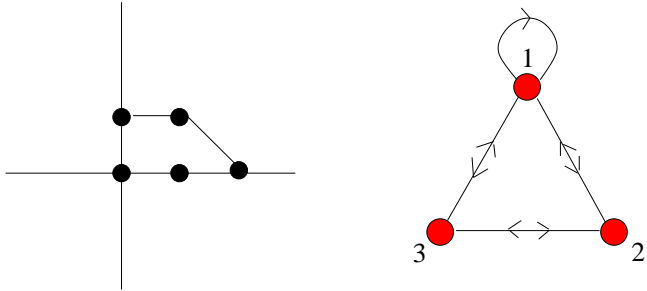} 
\caption{{\sf The toric diagram and the quiver for the $SPP$ singularity.}}
\label{f:SPPtq}
\end{center}
\end{figure}

We begin first with a non-orbifold type of singularity, the so-called suspended pinched point (SPP), first studied in \cite{Morrison:1998cs}. To remind the reader, the toric and quiver diagrams are presented in Figure~\ref{f:SPPtq} 
and the superpotential is
\begin{equation}
W_{SPP} = X_{11}(X_{12}X_{21} - X_{13} X_{31}) + X_{31}X_{13}X_{32}X_{23}- X_{21}X_{12}X_{23}X_{32} \ .
\end{equation}

The matrices $K$, $T$ and $P$ can readily found to be
\begin{equation}
K = \tmat{ 1 & 0 & 0 & 0 & 0 & 0 & 1 \cr 0 & 1 & 0 & 0 & 0 & 
   0 & 1 \cr 0 & 0 & 1 & 0 & 0 & 1 & 0 \cr 0 & 0 & 0 & 
   1 & 0 & 1 & 0 \cr 0 & 0 & 0 & 0 & 1 & -1 & 0 \cr}, \quad
T = \tmat{
0 & 0 & 0 & 0 & 0 & 1 \cr 0 & 0 & 0 & 0 & 1 & 0 \cr 
   0 & 0 & 1 & 1 & 0 & 0 \cr 1 & 1 & 0 & 0 & 0 & 0 \cr 
   0 & 1 & 0 & 1 & 0 & 0 \cr 
}, \quad
P = \tmat{
0 & 0 & 0 & 0 & 0 & 
    1 \cr 0 & 0 & 0 & 0 & 1 & 0 \cr 0 & 0 & 1 & 1 & 
   0 & 0 \cr 1 & 1 & 0 & 0 & 0 & 0 \cr 0 & 1 & 0 & 1 & 
   0 & 0 \cr 1 & 0 & 1 & 0 & 0 & 0 \cr 0 & 0 & 0 & 0 & 
   1 & 1 \cr 
} \ .
\end{equation}

In this example, we need to weight the variables appropriately by giving weight $t$ to all external points in the toric diagram, as discussed above:
\beq
\{X_{21},X_{12},X_{23},X_{32},X_{31},X_{13},X_{11} \} \rightarrow
\{1,1,1,1,1,1,2 \} \ .
\eeq
In the actual algebro-geometric computation this means that we weigh the variables of the polynomials with the above degrees and work in a weighted space.
\comment{
  \begin{equation}
    \hbox{R}(SPP)=\mathbb{C}[x21,x12,x23,x32,x31,x13,x11,
      Degrees=>{1,1,1,1,1,1,2}]
  \end{equation}
  \begin{eqnarray}
    \hbox{I}(SPP)&=&(x12*x23*x32-x12*x11,x21*x23*x32-x21*x11,\nonumber\\
    & & x21*x12*x32-x32*x31*x13,x21*x12*x23-x23*x31*x13,\nonumber\\
    & & x32*x23*x13-x13*x11,x32*x23*x31-x31*x11,x13*x31-x12*x21)\nonumber\\
  \end{eqnarray}
}
Now, we find that the moduli space is a reducible variety $\f_{SPP} = \firr{SPP} \cup L_{SPP}$ with Hilbert series:
\begin{equation}
H(t;\f_{SPP}) = \frac{1 + 2t + 2t^2 - 2t^3 + t^4}{(1- t)^4(1 -t^2)}
\end{equation}
and
\begin{eqnarray}
\firr{SPP} &=& \mathbb{V}(X_{23} X_{32} - X_{11}, X_{21} X_{12} - X_{31} X_{13}) \nonumber\\
L_{SPP} &=& \mathbb{V}(X_{13}, X_{31}, X_{12}, X_{21}) \ ,
\end{eqnarray}
where we have used the standard algebraic geometry notation that, given a set
$F=\{f_i\}$ of polynomials, $\IV(F)$ is the variety corresponding to the vanishing locus of $F$.
The top component $\firr{SPP}$ is a toric variety of complex dimension 5 which is the product of a conifold and a plane $\IC^2$; it has Hilbert series:
\begin{equation}
H(t;~\firr{SPP}) = \frac{1-t^2}{(1-t)^4} \frac{1}{(1-t)^2} = 
\frac{1+t}{(1-t)^5} \ ,
\end{equation}
with a palindromic numerator, as was with the \'etudes studied above.
The other component $L_{SPP}$ is a plane isomorphic $\IC^3$ with Hilbert series:
\begin{equation}
H(t;L_{SPP}) = \frac{1}{(1-t^2)(1-t)^2} \ .
\end{equation}
The two irreducible components intersect in a $\IC^2$ plane with Hilbert series:\begin{equation}
H(t;~\firr{SPP} \cap  L_{SPP}) = \frac{1}{(1-t)^2} \ .
\end{equation}
We observe that the Hilbert series of the various components satisfy the additive relation:
\begin{equation}
H(t;\f_{SPP}) = H(t;~\firr{SPP})+ H(t;L_{SPP})- H(t;~\firr{SPP} \cap  L_{SPP}) \ .
\end{equation}
This is, of course, the phenomenon of ``surgery'' discussed in \cite{Hanany:2006uc}. We will see this in all our subsequent examples.

In the symplectic quotient description, we find that the kernel of the $T$-matrix is $Q^t = \left( 1, -1, -1, 1, 0, 0 \right)$ and hence $\firr{SPP} \simeq 
\IC^6 // Q^t$.
The symmetry group of the coherent component is easily found to be $SU(2)^3\times U(1)^2$, a rank 5 group as expected from the toric property of this space. The non-Abelian part is realized as the repetition of the charges, $1, -1,$ and $0$, respectively.

\subsubsection{Cone over Zeroth Hirzebruch Surface}\label{s:F0}
\begin{figure}[t]
\begin{center}
  \epsfxsize = 11cm
  \centerline{\epsfbox{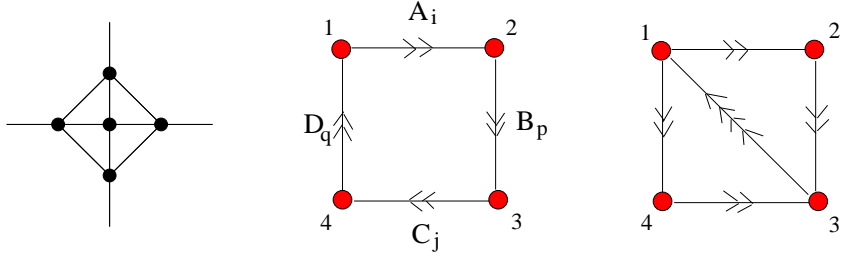}}
  \caption{{\sf The toric diagram and the quivers for phases I and II of $F_0$.}}
  \label{f:F0III}
\end{center}
\end{figure}

We continue with a simple toric threefold: the affine cone $F_0$ over the zeroth Hirzebruch surface, which is simply $\IP^1 \times \IP^1$. Indeed, we remark that the even simpler and more well-known example of the conifold was already studied in \cite{Butti:2006au,Forcella:2007wk}, the master space turns out to be just $\IC^4$; we will return to this in \sref{s:coni}. The toric diagram is drawn in \eref{f:F0III}.
There are two toric/Seiberg (see for example \cite{Feng:2001bn, Franco:2002mu}) dual phases, $(F_0)_I$ and $(F_0)_{II}$, of the gauge theory \cite{Feng:2000mi}, and the quivers and superpotentials are:
\begin{equation}\label{WF0}\ba{rcl}
W_{(F_0)_I} &=& \epsilon_{ij}\epsilon_{pq} {A}_i{B}_p{C}_j{D}_q ;
\\
W_{(F_0)_{II}} &=& \epsilon_{ij}\epsilon_{mn} X^i_{12}X^m_{23}X_{31}^{jn}- \epsilon_{ij}\epsilon_{mn} X^i_{14}X^m_{43}X_{31}^{jn} \ .
\ea\end{equation}
\comment{
  \begin{equation}
    R(F_0)=QQ[A1,A2,B1,B2,C1,C2,D1,D2,Degrees=>{1,1,1,1,1,1,1,1}]
  \end{equation}
  \begin{eqnarray}
    I(F_0) &=&(B1*C2*D2-B2*C2*D1,B2*C1*D1-B1*C1*D2,\nonumber\\
    & & A1*C2*D2-A2*C1*D2,A1*C2*D1-A2*C1*D1,\nonumber\\
    & & A2*B2*D1-A2*B1*D2,A1*B1*D2-A1*B2*D1,\nonumber\\
    & & A1*B2*C2-A2*B2*C1,A1*B1*C2-A2*B1*C1)
  \end{eqnarray}
}
\paragraph{Toric Phase I: }
We can readily find the F-terms from $W_{(F_0)_I}$ in \eref{WF0} and using the techniques outlined above, we can find the $K$-matrix to be
\begin{equation}
{\scriptsize
K=
\left(
\begin{array}{lllllllll}
    & A_1& A_2 & B_1 & B_2 & C_1 & D_1 & C_2 & D_2 \\
A_1 & 1 & 0 & 0 & 0 & 0 & 0 & -1 & 0 \\
A_2 & 0 & 1 & 0 & 0 & 0 & 0 & 1  & 0 \\
B_1 & 0 & 0 & 1 & 0 & 0 & 0 & 0  & -1 \\
B_2 & 0 & 0 & 0 & 1 & 0 & 0 & 0  &  1 \\
C_1 & 0 & 0 & 0 & 0 & 1 & 0 & 1  & 0 \\
D_1 & 0 & 0 & 0 & 0 & 0 & 1 & 0  & 1 \\
 \end{array}
\right) }
.
\end{equation}
Whence, primary decomposition gives that there are three irreducible pieces:
\beq\label{F0-I}\ba{rcl}
\f_{(F_0)_I} &=& \firr{(F_0)_I} \cup L^1_{(F_0)_I} \cup L^2_{(F_0)_I}, \\
\ea\eeq
with
\beq\label{FF0}
\ba{rcl}
\firr{(F_0)_I}  &=& \mathbb{V}(B_2 D_1 - B_1 D_2, A_2 C_1 - A_1 C_2)\\
L^1_{(F_0)_I} &=& \mathbb{V}(C_2, C_1, A_2, A_1)\\
L^2_{(F_0)_I} &=& \mathbb{V}(D_2,D_1, B_2, B_1) \ .
\ea\eeq
With weight $t$ to all 8 basic fields in the quiver the Hilbert series of the total space is given by
\beq
H(t;~\f_{(F_0)_I}) = \frac{1 + 2t + 3t^2 - 4t^3 + 2t^4}{(1-t)^6} \ .
\eeq

The top-dimensional component of $\f_{(F_0)_I}$ is toric Calabi-Yau, and here of dimension 6; this is consistent with the fact that the number of nodes in the quiver is 4. Specifically, $\firr{(F_0)_I}$ is the product of two conifolds and it has Hilbert series (again with palindromic numerator):
\begin{equation}
H(t;~\firr{(F_0)_I})=\frac{(1 + t)^2}{(1-t)^6} \ .
\end{equation} 
The two lower-dimensional components are simply $\IC^4$, with Hilbert series
\beq
H(t;~L^1_{(F_0)_I}) = H(t;~L^2_{(F_0)_I}) = \frac{1}{(1-t)^4} \ .
\eeq
These two hyperplanes are, as mentioned above, Coulomb branches of the moduli space, they intersect the $\firr{(F_0)_I}$ along one of the two three dimensional conifolds which have Hilbert series
\begin{equation}
H(t;~\firr{(F_0)_I} \cap L^1_{(F_0)_I}) = H(t; ~\firr{(F_0)_I} \cap L^2_{(F_0)_I}) = \frac{1+t}{(1-t)^3} \ .
\end{equation}
The Hilbert series of various components again satisfy the additive surgical relation of \cite{Hanany:2006uc}:
\begin{equation}
\hspace{-1cm}
H(t;~\f_{(F_0)_I} )
=H(t;~\firr{(F_0)_I})+ H(t;~L^1_{(F_0)_I}) + H(t;~L^2_{(F_0)_I}) - H(t;~\firr{(F_0)_I} \cap L^1_{(F_0)_I} ) - H(t;~\firr{(F_0)_I} \cap L^2_{(F_0)_I}) \ .
\end{equation}

For reference, the dual cone $T$-matrix and the perfect matching matrix $P = K^t \cdot T$ are:
\beq
T = \tmat{
0 & 0 & 0 & 0 & 0 & 0 & 1 & 1 \cr 0 & 0 & 0 & 0 & 0 & 1 & 0 & 1 \cr 0 & 0 & 0 & 1 & 1 & 0 & 0 & 0 \cr 0 & 0 & 1 & 0 & 1 & 
   0 & 0 & 0 \cr 0 & 1 & 0 & 0 & 0 & 0 & 1 & 0 \cr 1 & 0 & 0 & 1 & 0 & 0 & 0 & 0 \cr 
} \ , \qquad
P = \tmat{
 0 & 0 & 0 & 0 & 0 & 0 & 1 & 1 \cr 0 & 0 & 0 & 0 & 0 & 1 & 0 & 1 \cr 0 & 0 & 0 & 1 & 1 & 0 & 0 & 0 \cr 0 & 0 & 1 & 0 & 1 & 0 & 0 & 0 \cr 0 & 1 & 0 & 0 & 0 & 0 & 1 & 0 \cr 1 & 0 & 0 & 1 & 0 & 0 & 0 & 
 0 \cr 0 & 1 & 0 & 0 & 0 & 1 & 0 & 0 \cr 1 & 0 & 1 & 0 & 0 & 0 & 0 & 0 \cr 
} \ .
\eeq
Subsequently, their kernel is $Q^t$, giving us
\beq
Q^t = \tmat{ 0 & 1 & 0 & 0 & 0 & -1 & -1 & 1 \cr 1 & 0 & 
    -1 & -1 & 1 & 0 & 0 & 0 \cr  } \Rightarrow
\firr{(F_0)_I} \simeq \IC^8 // Q^t \ .
\eeq
The fact that the rows of $Q^t$ sum to 0 means that the toric variety is indeed Calabi-Yau.
The symmetry of the coherent component is $SU(2)^4\times U(1)^2$, suitable for a product of two conifolds. We note that the charge matrix $Q$ has 8 columns which are formed out of 4 pairs, each with two identical columns. This repetition of columns in the charge matrix is another way of determining the non-Abelian part of the symmetry group of the coherent component.

\paragraph{Toric Phase II: }
We can perform a similar analysis for the second toric phase which is a Seiberg dual of the first. Note from \eref{WF0} that the gauge-invariant terms in $W_{(F_0)_{II}}$ now have a different number of fields; correspondingly, we must thus assign different weights to the variables. The ones that are composed of Seiberg dual mesons of the fields of the first toric phase should be assigned twice the weight:
\begin{equation}
\{X^1_{12},X^1_{23},X^{22}_{31},X^2_{23},X^{21}_{31},X^2_{12},X^{12}_{31},X^{11}_{31},X^1_{14},X^1_{43},X^2_{43},X^2_{14} \} \rightarrow \{ {1,1,2,1,2,1,2,2,1,1,1,1} \} \ .
\end{equation}
In the actual algebro-geometric computation this means that we weight the variables of the polynomials with the above degrees and work in a weighted space.
\comment{
  \begin{equation}
    \hbox{R}(F_0 II) = \mathbb{C}[x12,x23,yy31,y23,yx31,y12,xy31,xx31,
      x14,x43,y43,y14,Degrees=>{1,1,2,1,2,1,2,2,1,1,1,1}]
  \end{equation}
  Observe the different weights for the various fields: the ones with weights $2$ are the ones that are composed mesons of the fields of the first toric phase.
  \begin{eqnarray}
    \hbox{I}(F_0 II)&=&(x23*yy31-y23*yx31,x12*yy31-y12*xy31,\nonumber\\
    & & x12*x23-x14*x43,x12*yx31-y12*xx31,x12*y23-x14*y43,\nonumber\\
    & & x23*xy31-y23*xx31,y12*x23-y14*x43,y12*y23-y14*y43,\nonumber\\
    & & x43*yy31-y43*yx31,x14*yy31-y14*xy31,x14*yx31-y14*xx31,\nonumber\\
    & & x43*xy31-y43*xx31) 
  \end{eqnarray}
}
Subsequently, we find that
\beq
{\tiny
K=
\left(
\begin{array}{lllllllllllll}
  & X^1_{43} & X^{11}_{31}&X^2_{14} &X^1_{23}&X^{21}_{31}& X^2_{23}& X^1_{12}& X^1_{14}& X^2_{12}& X^{12}_{31}& X^2_{43}& X^{22}_{31}\\
X^1_{43}   & 1 & 0 & 0 & 0 & 0 & 0 & 1  & 0 & 1 & 0 & 1 & 0 \\
X^{11}_{31} & 0 & 1 & 0 & 0 & 0 & 0 & 1  & 1 & 0 & 1 & 0 & 0 \\
X^2_{14}  &0 & 0 & 1 & 0 & 0 & 0 & 1  & 1 & 1 & 0 & 0 & 0 \\
X^1_{23}  &0 & 0 & 0 & 1 & 0 & 0 & -1 & 0 & -1&-1 &-1 &-1 \\
X^{21}_{31}& 0 & 0 & 0 & 0 & 1 & 0 & -1 & -1& 0 & 0 & 0 & 1 \\
X^2_{23}  & 0 & 0 & 0 & 0 & 0 & 1 & 0  & 0 & 0 & 1 & 1 & 1 \\
 \end{array}
\right)
} \ .
\eeq
The master space affords the primary decomposition
$\f_{(F_0)_{II}} = \firr{(F_0)_{II}} \cup L^1_{(F_0)_{II}} \cup L^2_{(F_0)_{II}} \cup L^3_{(F_0)_{II}}$ with
\beq\begin{array}{rcl}
\firr{(F_0)_{II}}  &=& \mathbb{V}(X^{12}_{31}X^1_{43} - X^{11}_{31} X^2_{43}, X^2_{23}X^1_{43} - X^1_{23}X^2_{43}, X^{22}_{31}X^1_{43} - X^{21}_{31} X^2_{43},
X^2_{12}X^1_{14} - X^1_{12}X^2_{14}, \\
&&
X^{21}_{31}X^1_{14} - X^{22}_{31}X^2_{14}, X^{22}_{31}X^1_{14} - X^{12}_{31}X^2_{14}, X^{21}_{31}X^{12}_{31} -  X^{22}_{31}X^{22}_{31}, X^1_{23}X^{12}_{31} - X^2_{23}X^{11}_{31}, \\ 
&&
X^2_{23}X^2_{12} - X^2_{43}X^2_{14}, X^1_{23}X^2_{12} - X^1_{43}X^2_{14}, X^1_{12}X^{21}_{31} - X^2_{12}X^{11}_{31},X^1_{12}X^2_{23} - X^1_{14}X^2_{43}, \\
&& 
X^1_{23}X^{22}_{31} - X^2_{23}X^{21}_{31}, X^1_{12}X^{22}_{31} - X^2_{12}X^{12}_{31}, X^1_{12}X^1_{23} - X^1_{14}X^1_{43})\\
L^1_{(F_0)_{II}} &=& \mathbb{V}(X^2_{14}, X^2_{43}, X^1_{43}, X^1_{14}, X^2_{12}, X^2_{23}, X^1_{23},X^1_{12})\\
L^2_{(F_0)_{II}} &=& \mathbb{V}(X^2_{43}, X^1_{43}, X^{11}_{31}, X^{12}_{31}, X^{21}_{31}, X^2_{23}, X^{22}_{31}, X^1_{23})\\
L^3_{(F_0)_{II}} &=& \mathbb{V}(X^2_{14}, X^1_{14},X^{11}_{31}, X^{12}_{31}, X^2_{12}, X^{21}_{31}, X^{22}_{31}, X^1_{12}) \ .
\end{array}
\eeq
The Hilbert series is
\beq
H(t;~\f_{(F_0)_{II}}) = \frac{1 + 6t + 17t^2 + 24t^3 + 14t^4 - 4t^5 + 4t^7 + 2t^8}{(1- t)^2(1 - t^2)^4} \ .
\eeq
We see that $\f_{(F_0)_{II}}$ is composed of four irreducible components: a six dimensional $\firr{(F_0)_{II}}$ which is the product of two conifolds; this is the biggest irreducible component and it actually has the same Hilbert series $\frac{(1 + t)^2}{(1-t)^6}$ as the $\firr{(F_0)_{I}}$ component of first toric phase.

The other three irreducible components are three four dimensional complex planes each defined by the vanishing of 8 coordinates out of the 12 total. These planes intersect only at the origin of the coordinate system and have the Hilbert series
\beq
H(t;~L^1_{(F_0)_{II}})=\frac{1}{(1-t^2)^4}, \quad
H(t;~L^2_{(F_0)_{II}})= H(t;~L^3_{(F_0)_{II}})=\frac{1}{(1-t)^4} \ .
\eeq
We see that $L^1_{(F_0)_{II}}$ has $t^2$ in the denominator instead of $t$ because it is parameterized precisely by the coordinates of twice the weight.

The three $\IC^4$ components and the $\firr{(F_0)_{II}}$ component intersect in a three dimensional conifold variety but with different grading of the coordinates:
\beq
H(t;~\firr{(F_0)_{II}} \cap L^1_{(F_0)_{II}}) = \frac{1+t^2}{(1-t^2)^3}, \quad
H(t;~\firr{(F_0)_{II}} \cap L^2_{(F_0)_{II}}) = H(t;~\firr{(F_0)_{II}} \cap L^3_{(F_0)_{II}}) = \frac{1+t}{(1-t)^3} \ .
\eeq
Once again, we have a surgery relation:
\beq\ba{rcl}
H(t; ~\f_{(F_0)_{II}}) & = & H(t;~\firr{(F_0)_{II}} )+H(t;~L^1_{(F_0)_{II}})
+ H(t;~L^2_{(F_0)_{II}}) + H(t;~L^3_{(F_0)_{II}} )\\
&& - H(t;~\firr{(F_0)_{II}} \cap L^1_{(F_0)_{II}} ) - H(t;~\firr{(F_0)_{II}} \cap L^2_{(F_0)_{II}} ) - H(t;~\firr{(F_0)_{II}} \cap L^3_{(F_0)_{II}}) \ .
\ea\eeq

The dual cone $T$-matrix and the perfect matching matrix $P = K^t \cdot T$ are:
\beq
T = \tmat{ 0 & 0 & 0 & 0 & 0 & 0 & 1 & 1 & 1 \cr 0 & 0 & 0 & 
   0 & 1 & 1 & 0 & 0 & 1 \cr 0 & 1 & 1 & 1 & 0 & 0 & 
   0 & 0 & 0 \cr 0 & 0 & 0 & 1 & 0 & 0 & 0 & 1 & 1 \cr
   0 & 0 & 1 & 0 & 0 & 1 & 0 & 0 & 1 \cr 1 & 0 & 0 & 
   1 & 0 & 0 & 0 & 1 & 0 \cr} \ , \qquad
P = \tmat{ 0 & 0 & 0 & 0 & 0 & 0 & 1 & 1 & 1 \cr 0 & 0 & 0 & 
   0 & 1 & 1 & 0 & 0 & 1 \cr 0 & 
    1 & 1 & 1 & 0 & 0 & 0 & 0 & 0 \cr 0 & 0 & 0 & 1 &
   0 & 0 & 0 & 1 & 1 \cr 0 & 0 & 1 & 0 & 0 & 1 & 0 & 
   0 & 1 \cr 1 & 0 & 0 & 1 & 0 & 0 & 0 & 1 & 0 \cr 0 &
   1 & 0 & 0 & 1 & 0 & 1 & 0 & 0 \cr 0 & 1 & 0 & 1 & 
   1 & 0 & 0 & 0 & 0 \cr 0 & 1 & 1 & 0 & 0 & 0 & 1 & 
   0 & 0 \cr 1 & 0 & 0 & 0 & 1 & 1 & 0 & 0 & 0 \cr 1 &
   0 & 0 & 0 & 0 & 0 & 1 & 1 & 0 \cr 1 & 0 & 1 & 0 & 
   0 & 1 & 0 & 0 & 0 \cr} \ .
\eeq
Hence, the kernel is $Q^t$ and we have
\beq
Q^t = \tmat{
 1 & 1 & 0 & -1 & 0 & -1 & 
    -1 & 0 & 1 \cr 0 & 1 & 0 & -1 & 0 & 0 & 
    -1 & 1 & 0 \cr 0 & 1 & -1 & 0 & 
    -1 & 1 & 0 & 0 & 0 \cr } \Rightarrow
\firr{(F_0)_{II}} \simeq \IC^9 // Q^t \ .
\eeq
Again, the rows of $Q^t$ sum to 0 and the toric variety is Calabi-Yau.
The second and third rows are conifold like relations and the first row is a relation which is found for the master space of $\IC^3/\IZ_3$ and is not in the first phase of $F_0$.

We see here that a manifestation of Seiberg duality is in the fact that the coherent component of each of the phases are the same. This feature is going to repeat itself. Different toric phases of a given Calabi-Yau singularity will exhibit the same coherent component of the master space. This is going to be our conjectured relation:
\begin{conjecture}
Quivers which are toric (Seiberg) duals have the same coherent component of the master space.
\end{conjecture}
It would be interesting to understand the fate of the linear components under Seiberg duality and this is left for future work.

\subsubsection{Cone over First del Pezzo Surface: $dP_1$}\label{s:dP1}
\begin{figure}[t]
\begin{center}
\includegraphics[scale=0.6]{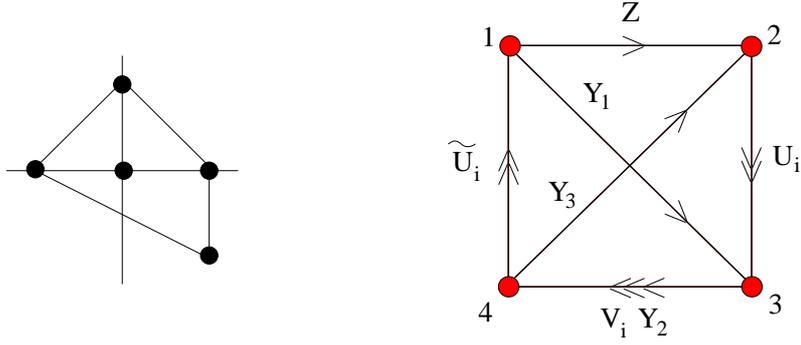} 
\caption{{\sf The toric diagram and the quiver for $dP_1$.}}
\label{f:dP1tq}
\end{center}
\end{figure}

In our analysis in \sref{s:irred} we initiated the example of $dP_0$ to which we will later return; for now, it is only natural that we continue to the higher cases. We know that $\IP^2$ blown up till 3 generic points admit toric description; the affine cones thereupon have been referred to $dP_{i=0,\ldots,3}$ theories \cite{Beasley:1999uz,Feng:2000mi}. Moreover, if we persisted with the blow-ups, now at {\it non-generic} points, we arrive at the so-called Pseudo-del Pezzo surfaces \cite{Feng:2002fv}; these continue to be toric and the corresponding theories are known as $PdP_{i \ge 4}$.
The toric and quiver diagrams for $dP_1$ are given in Figure~\ref{f:dP1tq}
and the superpotential is:
\begin{equation}
W_{dP_1} =\epsilon_{ab}Y_1 V_a \tilde U_b +\epsilon_{ab}Y_3  U_a V_b +\epsilon_{ab}Y_2 \tilde U_a Z U_b \ .
\end{equation}
\comment{
  \begin{equation}
    \hbox{R}(dP_1)=\mathbb{C}[u1,u2,uu1,uu2,v1,v2,y1,y2,y3,z,
      Degrees=>{1,1,1,1,2,2,1,1,1,1}]
  \end{equation}
  \begin{eqnarray}
    \hbox{I}(dP_1)&=&(v1*uu2-v2*uu1,y1*uu2-y3*u2,y1*v1-y2*z*u1,\nonumber\\
    & & y1*uu1-y3*u1,y1*v2-y2*z*u2,u1*v2-u2*v1,\nonumber\\
    & & y3*v2-y2*uu2*z,y3*v1-y2*uu1*z,uu1*z*u2-uu2*z*u1,\nonumber\\
    & & y2*uu1*u2-y2*uu2*u1]
  \end{eqnarray}
}
Now the $K$-matrix is
\begin{equation}
{\tiny
K=
\left(
\begin{array}{lllllllllll}
   & V_2 & \tilde U_1 & \tilde U_2 & U_2 & Y_3 & Y_2 & V_1 & Y_1 & U_1 & Z\\
 V_2 & 1 & 0 & 0 & 0 & 0 & 0 & 1 & 0 & 0 & 1 \\
 \tilde U_1 &0 & 1 & 0 & 0 & 0 & 0 & 1 & 0 & 1 & 0\\
 \tilde U_2 &0 & 0 & 1 & 0 & 0 & 0 & -1 & -1 & -1 & -1 \\
 U_2 & 0 & 0 & 0 & 1 & 0 & 0 & 0 & 1 & 1 & 0 \\
 Y_3 & 0 & 0 & 0 & 0 & 1 & 0 & 0 & 1 & 0 & 1 \\
 Y_2 & 0 & 0 & 0 & 0 & 0 & 1 & 0 & 0 & 0 & -1\\
 \end{array}
\right)}
\end{equation}
and the master space is the union of two components $\f_{dP_1} = \firr{dP_1} \cup L_{dP_1} $, a 6-dimensional $\firr{dP_1}$ piece and a plane $L_{dP_1}$ representing a $\IC^4$:
\begin{equation}\label{f-dP1}\ba{rcl}
\firr{dP_1} &=& \mathbb{V}(\tilde U_2 Y_1 - U_2 Y_3, \tilde U_1 Y_1 - U_1 Y_3, \tilde U_2 V_1 - \tilde U_1 V_2, U_2 V_1 - U_1 V_2, U_2 \tilde U_1 - U_1 \tilde U_2,\\
& & \tilde U_2 Y_2 Z - V_2 Y_3, \tilde U_1 Y_2 Z - V_1 Y_3, U_2 Y_2 Z - V_2 Y_1, U_1 Y_2 Z - V_1 Y_1) \\
L_{dP_1} &=& \mathbb{V}(Z, Y_3, Y_2, Y_1, V_2, V_1) \ .
\ea
\end{equation}
We need to weight the fields appropriately:
\begin{equation} 
\{ V_2 , \tilde U_1 , \tilde U_2 , U_2 , Y_3 , Y_2 , V_1 , Y_1 , U_1 , Z\}\rightarrow \{2,1,1,1,1,1,2,1,1,1\} \ .
\end{equation}
The Hilbert series of the total space is
\begin{equation}
H(t;~\f_{dP_1})=\frac{1 + 4 t + 8 t^2 + 4 t^3 - t^4 + t^6}{(1 - t)^6(1 + t)^2}
\end{equation}
and that of the $\firr{dP_1}$ component is
\begin{equation}
H(t;~\firr{dP_1})= \frac{1 + 4 t + 7 t^2 + 4 t^3 + t^4}{(1 - t)^6(1 + t)^2} \ .
\end{equation}
Note that the denominator is equal to $(1 - t)^4(1 - t^2)^2$, signifying that two variables are of weight 2. Hence, the above Hilbert series is in the Second form.

The two components intersect in the three dimensional conifold variety with Hilbert series:
\begin{equation}
H(t;\firr{dP_1} \cap L_{dP_1}) = \frac{1 + t}{(1 - t)^3}
\end{equation}
and we indeed have the surgery relation
\begin{equation}
H(t;\f_{dP_1})= H(t;~\firr{dP_1}) + H(t;~L_{dP_1}) - H(t;\firr{dP_1} \cap L_{dP_1}) \ .
\end{equation}

For the case of $dP_1$ the $6 \times 8$ dual-cone $T$-matrix and the
$10\times8$ perfect matching matrix $P$ take the form
\beq
T = \tmat{
0 & 0 & 0 & 0 & 0 & 1 & 1 & 1 \cr 0 & 0 & 0 & 1 & 
   1 & 0 & 0 & 0 \cr 0 & 0 & 0 & 0 & 1 & 0 & 0 & 1 \cr 
   0 & 0 & 1 & 0 & 0 & 0 & 0 & 1 \cr 1 & 1 & 0 & 0 & 
   1 & 0 & 0 & 0 \cr 0 & 1 & 0 & 0 & 0 & 0 & 1 & 0 \cr
} \ , \qquad
P = \tmat{
0 & 0 & 0 & 0 & 0 & 
    1 & 1 & 1 \cr 0 & 0 & 0 & 1 & 1 & 0 & 0 & 0 \cr 
   0 & 0 & 0 & 0 & 1 & 0 & 0 & 1 \cr 0 & 0 & 1 & 0 & 
   0 & 0 & 0 & 1 \cr 1 & 1 & 0 & 0 & 1 & 0 & 0 & 0 \cr 
   0 & 1 & 0 & 0 & 0 & 0 & 1 & 0 \cr 0 & 0 & 0 & 1 & 
   0 & 1 & 1 & 0 \cr 1 & 1 & 1 & 0 & 0 & 0 & 0 & 0 \cr 
   0 & 0 & 1 & 1 & 0 & 0 & 0 & 0 \cr 1 & 0 & 0 & 0 & 
   0 & 1 & 0 & 0 \cr  
} \ .
\eeq
The rank of this matrix is $g+2=6$ and we expect a 2 dimensional kernel. This can be easily computed to be the matrix $Q$,
which forms two vectors of charges for the linear sigma model description of the coherent component of the master space for $dP_1$. In summary,
\begin{equation}
T \cdot Q = 0 \quad \Rightarrow \quad Q^t =
\tmat{
1 & 0 & -1 & 1 & -1 & -1 & 0 & 1 \cr 1 & 
    -1 & 0 & 0 & 0 & -1 & 1 & 0 \cr 
}
\quad \Rightarrow
\firr{dP_1} \simeq \mathbb{C}^8//Q^t \ .
\label{MSdP1}
\end{equation}
The sum of charges (rows of $Q^t$) is zero, giving a Calabi Yau 6-fold. One relation is a relation found for $dP_0$, as appropriate for $dP_1$ is a blowup of $dP_0$ that manifests itself by Higgsing the $Z$ field of $dP_1$. The second relation is a conifold-like relation. The symmetry is $SU(2)\times SU(2)\times U(1)^4$. One $U(1)$ is the R-symmetry and the first $SU(2)$ is the natural one acting on the mesonic moduli space. The second $SU(2)$ is a ``hidden" symmetry coming from one of the two anomalous baryonic $U(1)$ symmetries. We will use the full symmetry to compute the refined Hilbert series for this space in section \ref{s:dp1rev}. 


\subsubsection{Cone over Second del Pezzo Surface: $dP_2$}\label{s:dP2}
\begin{figure}[t]
\begin{center}
\includegraphics[scale=0.6]{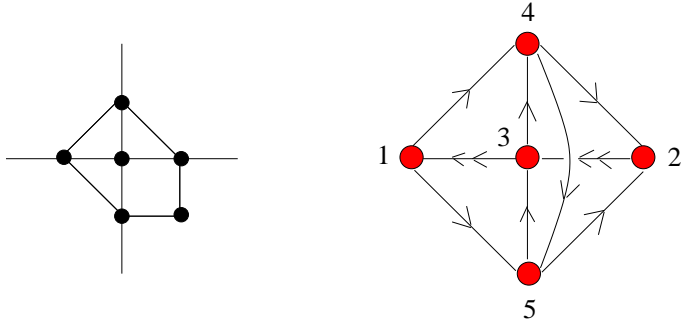} 
\caption{{\sf The toric diagram and the quiver for $dP_2$.}}
\label{f:dP2tq}
\end{center}
\end{figure}

Moving onto the next blowup, we have the $dP_2$ theory, whose quiver and toric diagrams are given in \fref{f:dP2tq}
and the superpotential is:
\begin{equation}
W_{(dP_2)} =X_{34} X_{45} X_{53} - X_{53}Y_{31} X_{15} - X_{34} X_{42} Y_{23} + Y_{23} X_{31} X_{15} X_{52} + X_{42} X_{23} Y_{31} X_{14}- X_{23} X_{31} X_{14} X_{45} X_{52} \ .
\end{equation}

\comment{
  \begin{equation}
    \hbox{R}(dP_2)=\mathbb{C}[x34,x45,x53,y31,x15,x42,y23,x31,x52,x23,x14,
      Degrees=>{2,1,2,2,1,1,2,1,1,1,1}]
  \end{equation}
  \begin{eqnarray}
    \hbox{I}(dP_2)&=&(x45*x53-x42*y23,x34*x53-x23*x31*x14*x52,
    x34*x45-y31*x15,\nonumber\\
    & & -x53*x15+x42*x23*x14,-x53*y31+y23*x31*x52,-x34*y23+x23*y31*x14,
    \nonumber\\
    & &-x34*x42+x31*x15*x52,y23*x15*x52-x23*x14*x45*x52,
    y23*x31*x15-x23*x31*x14*x45,\nonumber\\
    & &x42*y31*x14-x31*x14*x45*x52,x42*x23*y31-x23*x31*x45*x52)
  \end{eqnarray}
}
We point out that this model is one of the toric phases (phase II in the notation of \cite{Feng:2002zw}), and we shall only consider this one for now.

The master space has $K$-matrix 
\begin{equation}
{\tiny
K=
\left(
\begin{array}{llllllllllll}
     & X_{14} & X_{23} & X_{31} & X_{45} & X_{52} & X_{42} & Y_{23} & X_{34} & X_{53} & Y_{31} & X_{15}\\
 X_{14} & 1 & 0 & 0 & 0 & 0 & 0 & 0 & 1 & 0 & 0 & 1\\
 X_{23} & 0 & 1 & 0 & 0 & 0 & 0 & 0 & 1 & 0 & 0 & 1\\
 X_{31} &0 & 0 & 1 & 0 & 0 & 0 & 0 & 1 & 0 & 1 & 0 \\
 X_{45} &0 & 0 & 0 & 1 & 0 & 0 & 0 & 1 & -1 & 1 & 1 \\
 X_{52} &0 & 0 & 0 & 0 & 1 & 0 & 0 & 1 & 0 & 1 & 0 \\
 X_{42} &0 & 0 & 0 & 0 & 0 & 1 & 0 & -1 & 1 & -1 & 0\\
 Y_{23} &0 & 0 & 0 & 0 & 0 & 0 & 1 & -1 & 1 & 0 & -1\\ 
 \end{array}
\right) .
}
\end{equation}
and decomposes as $\f_{dP_2}=\firr{dP_2} \cup L^1_{dP_2} \cup L^2_{dP_2}$, with: 
\beq\label{f-dP2}\ba{rcl}
\firr{dP_2} &=& \mathbb{V}(X_{45}X_{53} - X_{42}Y_{23}, X_{34}X_{45} - Y_{31}X_{15}, X_{42}X_{23}X_{14} - X_{53}X_{15},\\
& & X_{45}X_{23}X_{14} - X_{15}Y_{23}, X_{15}X_{31}X_{52} - X_{34}X_{42}, X_{45}X_{31}X_{52} - Y_{31}X_{42},\\
& & Y_{31}X_{23}X_{14} - X_{34}Y_{23}, Y_{23}X_{31}X_{52} - X_{53}Y_{31}, X_{31}X_{52}X_{23}X_{14} - X_{34}X_{53})\\
L^1_{dP_2} &=& \mathbb{V}(X_{14}, X_{23}, X_{15}, Y_{23}, X_{53}, X_{34})\\ 
L^2_{dP_2} &=& \mathbb{V}(X_{52}, X_{31}, X_{42}, Y_{31}, X_{53},X_{34}) \ .
\ea\eeq
We weight the fields appropriately:
\begin{equation}
\{ X_{14} , X_{23} , X_{31} , X_{45} , X_{52} , X_{42} , Y_{23} , X_{34} , X_{53} , Y_{31} , X_{15}\} \rightarrow \{1,1,1,1,1,2,1,2,2,1,2\} \ .
\end{equation}
The Hilbert series of the total space is
\begin{equation}
H(t;~\f_{dP_2})=\frac{1 + 2 t + 5t^2 + 4 t^3 - t^4 - 2 t^5 + 2t^6}
{(1 - t)^7(1 + t)^2}
\end{equation}
while that of the top component, a 7-dimensional sub-variety, has
\begin{equation}
H(t;~\firr{dP_2})= \frac{1 + 2t + 5 t^2 + 2t^3 + t^4}{(1-t)^7(1 +t)^2} \ .
\end{equation}

The two planes $L^1_{dP_2}$ and $L^2_{dP_2}$ are simply $\IC^5$ (with appropriate graded coordinates) and their Hilbert series are
\begin{equation}
H(t;~L^1_{dP_2})=H(t;~L^2_{dP_2})=\frac{1}{(1- t)^4(1-t^2)} \ .
\end{equation}
They themselves intersect on a complex line $\mathbb{C}$, which is in fact their common intersection with the $\firr{dP_2}$ component:
\begin{equation}
H(t;~L^1_{dP_2} \cap L^2_{dP_2}) = H(t;~ (L^1_{dP_2} \cap L^2_{dP_2}) \cap \firr{dP_2} )=\frac{1}{1-t} \ .
\end{equation}
Each, however, intersects $\firr{dP_2}$ on a non trivial 4 dimensional variety with Hilbert series:
\begin{equation}
H(t;~\firr{dP_2} \cap L^1_{dP_2} )=H(t;~ \firr{dP_2} \cap L^2_{dP_2} )
=\frac{1 + t + t^2 }{(1- t)^3(1 - t^2)} \ .
\end{equation}

This is a manifold of complete intersection generated by 4 generators of degree 1 and 1 generator of degree 2 satisfying 1 relation of degree 3.
Again, we have a surgery relation among the Hilbert series, fully exhibiting the intersection structure of the components of the master space:
\beq\ba{rcl}
H(t;~\f_{dP_2}) &=& 
H(t;~\firr{dP_2} )+H(t;~ L^1_{dP_2} )+H(t;~ L^2_{dP_2} ) +H(t;~ (L^1_{dP_2} \cap L^2_{dP_2}) \cap \firr{dP_2} ) \\
& & -H(t;~\firr{dP_2} \cap L^1_{dP_2} )-H(t;~\firr{dP_2} \cap L^2_{dP_2} )-H(t;~L^1_{dP_2} \cap L^2_{dP_2} )
\ .
\ea\eeq

For reference, the dual cone $T$-matrix and the perfect matching matrix $P = K^t \cdot T$ are:
\beq
T = \tmat{
0 & 0 & 0 & 0 & 0 & 0 & 0 & 0 & 1 & 1 \cr 0 & 0 & 
   0 & 0 & 0 & 0 & 1 & 1 & 0 & 0 \cr 0 & 0 & 0 & 0 & 
   1 & 1 & 0 & 0 & 0 & 0 \cr 0 & 0 & 1 & 1 & 0 & 0 & 
   0 & 0 & 0 & 0 \cr 1 & 1 & 0 & 0 & 0 & 0 & 0 & 0 & 
   0 & 0 \cr 0 & 1 & 0 & 1 & 0 & 1 & 0 & 0 & 0 & 0 \cr
   0 & 0 & 1 & 0 & 0 & 0 & 0 & 1 & 0 & 1 \cr 
} \ , \qquad
P = \tmat{
0 & 0 & 0 & 0 & 0 & 0 & 0 & 0 & 1 & 
    1 \cr 0 & 0 & 0 & 0 & 0 & 0 & 1 & 1 & 0 & 0 \cr 
   0 & 0 & 0 & 0 & 1 & 1 & 0 & 0 & 0 & 0 \cr 0 & 0 & 
   1 & 1 & 0 & 0 & 0 & 0 & 0 & 0 \cr 1 & 1 & 0 & 0 & 
   0 & 0 & 0 & 0 & 0 & 0 \cr 0 & 1 & 0 & 1 & 0 & 1 & 
   0 & 0 & 0 & 0 \cr 0 & 0 & 1 & 0 & 0 & 0 & 0 & 1 & 
   0 & 1 \cr 1 & 0 & 0 & 0 & 1 & 0 & 1 & 0 & 1 & 0 \cr
   0 & 1 & 0 & 0 & 0 & 1 & 0 & 1 & 0 & 1 \cr 1 & 0 & 
   1 & 0 & 1 & 0 & 0 & 0 & 0 & 0 \cr 0 & 0 & 0 & 1 & 
   0 & 0 & 1 & 0 & 1 & 0 \cr 
} \ .
\eeq
Subsequently, their kernel is $Q^t$, giving us
\beq
Q^t = \tmat{ 
1 & -1 & -1 & 1 & 0 & 0 & 0 & 0 & 
    -1 & 1 \cr 1 & -1 & -1 & 1 & 0 & 0 & 
    -1 & 1 & 0 & 0 \cr 1 & -1 & 0 & 0 & 
    -1 & 1 & 0 & 0 & 0 & 0 \cr  
} \Rightarrow
\firr{dP_2} \simeq \IC^{10} // Q^t \ .
\eeq
Once again, the rows of $Q^t$ sum to 0 and the toric variety is Calabi-Yau.

\subsubsection{Cone over Third del Pezzo Surface: $dP_3$}\label{s:dP3}
\begin{figure}[t]
\begin{center}
\includegraphics[scale=0.6]{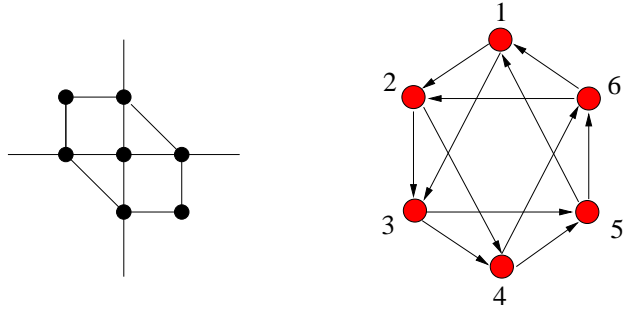} 
\caption{{\sf The toric diagram and the quiver for $dP_3$.}}
\label{f:dP3tq}
\end{center}
\end{figure}

Now for the last true toric del Pezzo, let us study $dP_3$, whose toric and quiver diagrams are given in \fref{f:dP3tq}
and superpotential:
\begin{eqnarray}
W_{(dP_3)} &=& X_{12}X_{23} X_{34} X_{45} X_{56} X_{61} + X_{13} X_{35} X_{51}+ X_{24} X_{46} X_{62} \nonumber \\
& & - X_{23} X_{35} X_{56} X_{62} - X_{13} X_{34} X_{46} X_{61} - X_{12} X_{24} X_{45} X_{51} \ .
\end{eqnarray}

Again, we need to weight the variables appropriately:
\beq\label{weightsdP3}
\{X_{12}, X_{23}, X_{34}, X_{45}, X_{56}, X_{61}, X_{13}, 
      X_{35}, X_{51}, X_{24}, X_{46}, X_{62} \} \rightarrow
\{1,1,1,1,1,1,2,2,2,2,2,2\} \ . 
\eeq
\comment{
  \begin{eqnarray}
    \nonumber
    \hbox{R}(dP_3)&=&\mathbb{C}[x12,x23,x34,x45,x56,x61,x13,
      x35,x51,x24,x46,x62],\\Degrees&=&[{1,1,1,1,1,1,2,2,2,2,2,2}]
    \label{weights}
  \end{eqnarray}
  \begin{eqnarray}
    \hbox{I}(dP_3)&=&(x23*x34*x45*x56*x61-x24*x45*x51,
    x12*x34*x45*x56*x61-x35*x56*x62,\nonumber\\
    & & x12*x23*x45*x56*x61-x13*x46*x61,x12*x23*x34*x56*x61-x12*x24*x51,\\
    & & x12*x23*x34*x45*x61-x23*x35*x62,x12*x23*x34*x45*x56-x13*x34*x46,\\
    & & x35*x51-x34*x46*x61,x13*x51-x23*x56*x62,x13*x35-x12*x24*x45,\nonumber\\
    & & x46*x62-x12*x45*x51,x24*x62-x13*x34*x61,x24*x46-x23*x35*x56)
  \end{eqnarray}
}
Again, this is only one of the 4 toric phases of the theory (Phase I in the notation of \cite{Feng:2002zw}). For now, we shall focus on this one.

The $K$-matrix for the master space is
\begin{equation}
{\tiny
K=
\left(
\begin{array}{lllllllllllll}
 & X_{12} & X_{23} & X_{45} & X_{56} & X_{46} & X_{34} & X_{61} & X_{51} & X_{13} & X_{24} & X_{62} & X_{35}\\
 X_{12} &1 & 0 & 0 & 0 & 0 & 0 & 0 & 0 & 1 & -1 & 1 & 1 \\
 X_{23} & 0 & 1 & 0 & 0 & 0 & 0 & 0 & 0 & 1 & 0 & 1  & 0 \\
 X_{45} & 0 & 0 & 1 & 0 & 0 & 0 & 0 & 0 & 1 & 0 & 0  & 1 \\
 X_{56} & 0 & 0 & 0 & 1 & 0 & 0 & 0 & 0 & 1 & -1 & 1 & 1 \\
 X_{46} & 0 & 0 & 0 & 0 & 1 & 0 & 0 & 0 & 1 & 0 & 1  & 0 \\
 X_{34} & 0 & 0 & 0 & 0 & 0 & 1 & 0 & 0 & 1 & 0 & 0  & 1 \\
 X_{61} & 0 & 0 & 0 & 0 & 0 & 0 & 1 & 0 & -1 & 1 & -1& 0 \\
 X_{51} & 0 & 0 & 0 & 0 & 0 & 0 & 0 & 1 & -1 & 1 & 0 & -1\\
 \end{array}
\right)
}
\end{equation}
and $\f_{dP_3}$ decomposes as $\firr{dP_3} \cup L^1_{dP_3} \cup L^2_{dP_3} \cup L^3_{dP_3}$ with
\begin{equation}\label{f-dP3}\ba{rcl}
\firr{dP_3} &=& \mathbb{V}(X_{23}X_{56}X_{62} - X_{13}X_{51}, X_{34}X_{61}X_{46} - X_{35}X_{51}, X_{12}X_{45}X_{24} - X_{13}X_{35},\\
& & X_{12}X_{45}X_{51} - X_{46}X_{62}, X_{23}X_{56}X_{35} - X_{24}X_{46}, X_{34}X_{61}X_{13} -X_{24}X_{62}, \\
& & X_{23}X_{34}X_{56}X_{61} - X_{51}X_{24}, X_{12}X_{34}X_{45}X_{61} - X_{35}X_{62},X_{12}X_{23}X_{45}X_{56} - X_{13}X_{46}),\\
L^1_{dP_3} &=& \mathbb{V}(X_{62}, X_{46}, X_{35}, X_{13}, X_{45}, X_{12}),\\
L^2_{dP_3} &=& \mathbb{V}(X_{46}, X_{24}, X_{51}, X_{13}, X_{56}, X_{23}),\\
L^3_{dP_3} &=& \mathbb{V}(X_{62}, X_{24}, X_{51}, X_{35}, X_{61}, X_{34}) \ .
\ea\eeq
We see that there are four irreducible sub-varieties: the 8 dimensional Calabi-Yau cone and three 6 dimensional planes. The various Hilbert series are:
\beq\label{dP3CY}
\ba{l}
H(t;~\f_{dP_3})=\frac{1 + 4t^2 + 4t^4 - 6t^5 + 3t^6}{(1 - t)^8(1 + t)^2} \ ;
\quad
H(t;~\firr{dP_3})= \frac{1 + 4t^2 + t^4}{(1 - t)^8(1 +t)^2}, \\
H(t;~L^1_{dP_3})=H(t;~L^2_{dP_3})=H(t;~L^3_{dP_3})=\frac{1}{(1-t)^4(1 - t^2)^2} \ .
\ea
\eeq

The planes $L^1_{dP_3}$, $L^2_{dP_3}$ and $L^3_{dP_3}$ intersect the $\firr{dP_3}$ in a non-trivial 5-dimensional variety with Hilbert series
\begin{equation}
H(t;~\firr{dP_3} \cap L^1_{dP_3})=H(t;~ \firr{dP_3} \cap L^2_{dP_3})=H(t;~ \firr{dP_3} \cap L^3_{dP_3})=
\frac{1 + t^2}{(1-t)^4(1- t^2)}
\end{equation}
and they intersect themselves in a complex plane $\mathbb{C}^2$, which is also their common intersection with $\firr{dP_3}$; this intersection has Hilbert series
\begin{equation}\ba{l}
H(t;~L^1_{dP_3}\cap L^2_{dP_3})=H(t;~ L^1_{dP_3} \cap L^3_{dP_3})=H(t;~ L^2_{dP_3} \cap L^3_{dP_3})=\\
H(t;~\firr{dP_3} \cap L^1_{dP_3}\cap L^2_{dP_3})=H(t;~ \firr{dP_3} \cap L^1_{dP_3} \cap L^3_{dP_3})=
H(t;~ \firr{dP_3} \cap L^2_{dP_3} \cap L^3_{dP_3})=\frac{1}{(1-t)^2} \ .
\ea\end{equation}

The common intersection among the three planes and among all the four irreducible components of $\f_{dP_3}$ is just the origin of the embedding space. Once more, we have the surgery relation
\begin{equation}\ba{rcl}
H(t;~ \f_{dP_3})&=&H(t;~\firr{dP_3} )+H(t;~  L^1_{dP_3})+H(t;~  L^2_{dP_3})+H(t;~ L^3_{dP_3})
-H(t;~ \firr{dP_3} \cap L^1_{dP_3})\\
&-& H(t;~ \firr{dP_3} \cap L^2_{dP_3})-H(t;~\firr{dP_3} \cap L^3_{dP_3})-H(t;~ L^1_{dP_3} \cap L^2_{dP_3})\\
&-&H(t;~ L^1_{dP_3} \cap L^3_{dP_3})-H(t;~L^2_{dP_3}\cap L^3_{dP_3})+H(t;~ \firr{dP_3} \cap L^1_{dP_3} \cap L^2_{dP_3})+\nonumber\\& & H(t;~ \firr{dP_3} \cap L^1_{dP_3} \cap L^3_{dP_3})+H(t;~ \firr{dP_3} \cap L^2_{dP_3} \cap L^3_{dP_3}) \ .
\ea\eeq

Finally, the dual cone $T$-matrix and the perfect matching matrix $P = K^t \cdot T$ are:
\beq
T = {\tiny \left( \ba{cccccccccccc}
 0 & 0 & 0 & 0 & 0 & 0 & 0 & 0 & 0 & 0 & 1 & 1 \cr 
   0 & 0 & 0 & 0 & 0 & 0 & 0 & 0 & 1 & 1 & 0 & 0 \cr 
   0 & 0 & 0 & 0 & 0 & 0 & 1 & 1 & 0 & 0 & 0 & 0 \cr 
   0 & 0 & 0 & 0 & 1 & 1 & 0 & 0 & 0 & 0 & 0 & 0 \cr 
   0 & 0 & 1 & 1 & 0 & 0 & 0 & 0 & 0 & 0 & 0 & 0 \cr 
   1 & 1 & 0 & 0 & 0 & 0 & 0 & 0 & 0 & 0 & 0 & 0 \cr 
   0 & 0 & 0 & 1 & 0 & 1 & 0 & 0 & 0 & 1 & 0 & 1 \cr 
   0 & 1 & 0 & 0 & 1 & 0 & 0 & 1 & 0 & 0 & 1 & 0 \cr
\ea \right)} \ , \quad
P ={\tiny \left( \ba{cccccccccccc}
 0 & 0 & 0 & 0 & 0 & 0 & 0 & 0 & 0 & 0 & 1 & 1 \cr 
   0 & 0 & 0 & 0 & 0 & 0 & 0 & 0 & 1 & 1 & 0 & 0 \cr 
   0 & 0 & 0 & 0 & 0 & 0 & 1 & 1 & 0 & 0 & 0 & 0 \cr 
   0 & 0 & 0 & 0 & 1 & 1 & 0 & 0 & 0 & 0 & 0 & 0 \cr 
   0 & 0 & 1 & 1 & 0 & 0 & 0 & 0 & 0 & 0 & 0 & 0 \cr 
   1 & 1 & 0 & 0 & 0 & 0 & 0 & 0 & 0 & 0 & 0 & 0 \cr 
   0 & 0 & 0 & 1 & 0 & 1 & 0 & 0 & 0 & 1 & 0 & 1 \cr 
   0 & 1 & 0 & 0 & 1 & 0 & 0 & 1 & 0 & 0 & 1 & 0 \cr 
   1 & 0 & 1 & 0 & 0 & 0 & 1 & 0 & 1 & 0 & 0 & 0 \cr 
   0 & 1 & 0 & 1 & 0 & 0 & 0 & 1 & 0 & 1 & 0 & 0 \cr 
   0 & 0 & 1 & 0 & 1 & 0 & 0 & 0 & 1 & 0 & 1 & 0 \cr 
   1 & 0 & 0 & 0 & 0 & 1 & 1 & 0 & 0 & 0 & 0 & 1 
\ea \right)} \ .
\eeq
Subsequently, their kernel is $Q^t$, giving us
\beq
Q^t =  {\tiny \left( \ba{cccccccccccc}
-1 & 1 & 1 & -1 & 0 & 0 & 0 & 0 & 0 & 0 & 
    -1 & 1 \cr 0 & 0 & 1 & -1 & 0 & 0 & 0 & 0 & 
    -1 & 1 & 0 & 0 \cr 1 & -1 & 0 & 0 & 0 & 0 & 
    -1 & 1 & 0 & 0 & 0 & 0 \cr -1 & 1 & 1 & -1 & 
    -1 & 1 & 0 & 0 & 0 & 0 & 0 & 0
\ea \right) } \Rightarrow
\firr{dP_3} \simeq \IC^{12} // Q^t \ .
\eeq
We see that the rows of $Q^t$ sum to 0 and the toric variety is Calabi-Yau.

\subsubsection{Cone over Fourth Pseudo del Pezzo Surface: $PdP_4$}
By blowing $dP_3$ up at a non-generic point we can obtain another toric variety, referred to as a pseudo del Pezzo \cite{Feng:2002fv}. The toric and quiver diagrams are given in \fref{f:dP4tq}
and the superpotential is
\beq\ba{rcl}
W_{PdP_4}&=&X_{61} X_{17} X_{74} X_{46} + X_{21} X_{13} X_{35} X_{52} + X_{27} X_{73} X_{36} X_{62} + X_{14} X_{45} X_{51} \nonumber\\
& & - X_{51} X_{17} X_{73} X_{35} - X_{21} X_{14} X_{46} X_{62} - X_{27} X_{74} X_{45} X_{52} - X_{13} X_{36} X_{61} \ .
\ea\eeq
\begin{figure}[t]
\begin{center}
\includegraphics[scale=0.6]{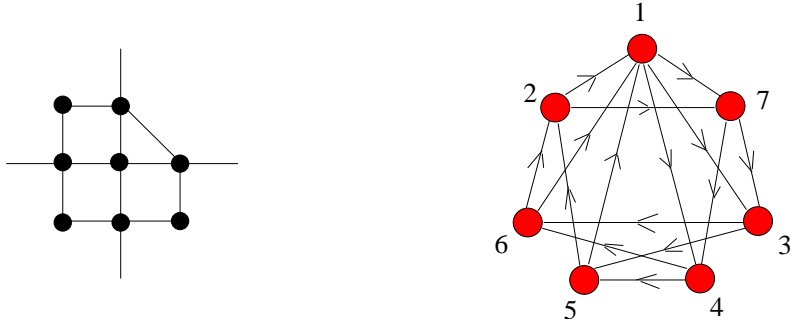} 
\caption{{\sf The toric diagram and the quiver for $PdP_4$.}}
\label{f:dP4tq}
\end{center}
\end{figure}

Once again, we need to assign appropriate weights to the variables:
\begin{equation}\ba{l}
\{ X_{45}, X_{73}, X_{46}, X_{21}, X_{36}, X_{74}, X_{52}, X_{35}, X_{17}, X_{62}, X_{51}, X_{27}, X_{61}, X_{13}, X_{14} \} \\
\rightarrow \{2,2,2,2,2,2,2,2,2,2,4,4,4,4,4 \} \ .
\ea
\end{equation}
Subsequently, the $K$-matrix is given as
\begin{equation}
{\tiny
K=
\left(
\begin{array}{llllllllllllllll}
 & X_{13} & X_{17} & X_{35} & X_{61} & X_{14} & X_{27} & X_{62} & X_{51} & X_{74} &  X_{45}  & X_{36}  &  X_{46} & X_{21}  & X_{73}  & X_{52} \\
X_{13} &1 & 0 & 0 & 0 & 0 & 0 & 0 & 0 & 0 &  1  & 0  &  1 & 0  & 1  & 0 \\
X_{17} &0 & 1 & 0 & 0 & 0 & 0 & 0 & 0 & 0 &  1  & 1  &  0 &  1  & 0  & 0 \\
X_{35} &0 & 0 & 1 & 0 & 0 & 0 & 0 & 0 & 0 &  1  & 1  &  1 &  0  & 0  & 0 \\
X_{61} &0 & 0 & 0 & 1 & 0 & 0 & 0 & 0 & 0 &  1 & 0 &  0 &  1  & 1  & 0   \\
X_{14} &0 & 0 & 0 & 0 & 1 & 0 & 0 & 0 & 0 &  -1  & 0  &  0 &  -1  & 0 &  1 \\
X_{27} &0 & 0 & 0 & 0 & 0 & 1 & 0 & 0 & 0 &  -1 & -1  & -1 &  0  & -1 & -1 \\
X_{62} &0 & 0 & 0 & 0 & 0 & 0 & 1 & 0 & 0 &  -1  & -1 &  -1 &  -1 & -1  & 0 \\
X_{51} &0 & 0 & 0 & 0 & 0 & 0 & 0 & 1 & 0 &  0  & 1  & 1 &  0  & 0 & 1\\
X_{74} &0 & 0 & 0 & 0 & 0 & 0 & 0 & 0 & 1 &  0  & 0  & -1 &  1  & 0 & -1\\ 
\end{array}
\right) .
}
\end{equation}
and the master space has Hilbert series
{\scriptsize
\begin{equation}
H(t;~\f_{PdP_4})=\frac{1 - 5t^6 - 10t^8 + 8t^{10} + 26t^{12} + 67t^{14} - 162t^{16} - 
      56t^{18} + 276t^{20} - 111t^{22} - 122t^{24} + 112t^{26} - 14
      t^{28} - 15t^{30} + 5t^{32}}{(1 - t^4)^5(1 - t^2)^{10}} \ .
\end{equation}
}
Now, $\f_{PdP_4}$ has a top $\firr{PdP_4}$ component of dimension 9:
\begin{equation}
{\scriptsize
\ba{rl}
\firr{PdP_4}&=\mathbb{V}(
X_{46}X_{21}X_{62} - X_{45}X_{51}, X_{73}X_{35}X_{17} - X_{45}X_{14},X_{46}X_{74}X_{17} - X_{36}X_{13}, X_{21}X_{52}X_{35} - X_{36}X_{61} ,  \\ 
&
X_{45}X_{74}X_{52} - X_{73}X_{36}X_{62} , X_{52}X_{35}X_{13} - X_{46}X_{62}X_{14}, X_{74}X_{17}X_{61} - X_{21}X_{62}X_{14} , X_{73}X_{62}X_{27} - X_{61}X_{13} ,\\
&
X_{74}X_{52}X_{27} - X_{51}X_{14},X_{45}X_{52}X_{27} - X_{46}X_{17}X_{61} , X_{45}X_{74}X_{27} - X_{21}X_{35}X_{13}, X_{73}X_{36}X_{27} - X_{46}X_{21}X_{14},\\
&
X_{35}X_{17}X_{51} - X_{36}X_{62}X_{27} , X_{73}X_{17}X_{51} - X_{21}X_{52}X_{13}, X_{73}X_{35}X_{51} - X_{46}X_{74}X_{61},X_{73}X_{46}X_{17}X_{62} - X_{45}X_{52}X_{13}, \\
&
X_{73}X_{21}X_{35}X_{62} - X_{45}X_{74}X_{61},X_{74}X_{52}X_{35}X_{17} - X_{36}X_{62}X_{14}, X_{46}X_{21}X_{35}X_{17} - X_{45}X_{36}X_{27}, X_{46}X_{21}X_{74}X_{52} - X_{73}X_{36}X_{51} , \\
&
X_{21}X_{52}X_{62}X_{27} - X_{17}X_{51}X_{61}, X_{46}X_{74}X_{62}X_{27} - X_{35}X_{51}X_{13}, X_{73}X_{74}X_{17}X_{27} - X_{21}X_{13}X_{14},
X_{73}X_{52}X_{35}X_{27} - X_{46}X_{61}X_{14}
) \ ,
\ea}
\end{equation}
with Hilbert series
\begin{equation}
H(t;~\firr{PdP_4}) = \frac{1 + t^2 + 6t^4 + t^6 + t^8}{(1 - t^2)^9} \ .
\end{equation}

\comment{
  \begin{eqnarray}\nonumber
  \hbox{R}(PdP_4)&=&\mathbb{C}[x56,x67,x72,x25,x73,x35,x57,
    x46,x61,x12,x24,x13,x34,x45,x51],
  Degrees&=&>[{3,1,1,2,2,2,3,2,1,2,2,1,2,1,3}]
  \end{eqnarray}
  \begin{eqnarray}
    \hbox{I}(PdP_4)&=&(x67*x72*x25-x13*x35*x61,x56*x72*x25-x46*x73*x34,\\
    & & x56*x67*x25-x57*x24*x45,x56*x67*x72-x51*x12,x35*x57-x46*x67*x34,\\
    & & x73*x57-x13*x56*x61,x73*x35-x72*x24*x45,x61*x12*x24-x67*x73*x34,\\
    & & x46*x12*x24-x13*x35*x56,x46*x61*x24-x51*x25,x46*x61*x12-x57*x72*x45,\\
    & & x34*x45*x51-x35*x56*x61,x13*x45*x51-x46*x67*x73,x13*x34*x51-x57*x72*x24,\\
    & & x13*x34*x45-x12*x25)
  \end{eqnarray}
}

The dual cone $T$-matrix and the perfect matching matrix $P = K^t \cdot T$ are:
\beq\ba{l}
T = {\tiny \left( \ba{cccccccccccccccccc}
0 & 0 & 0 & 0 & 0 & 0 & 0 & 0 & 0 & 0 & 0 & 0 & 1 &
   1 & 1 & 1 & 1 & 1 \cr 0 & 0 & 0 & 0 & 0 & 0 & 0 & 
   0 & 0 & 0 & 1 & 1 & 0 & 0 & 0 & 0 & 1 & 1 \cr 0 & 
   0 & 0 & 0 & 0 & 0 & 1 & 1 & 1 & 1 & 0 & 0 & 0 & 0 &
   0 & 0 & 0 & 0 \cr 0 & 0 & 1 & 1 & 1 & 1 & 0 & 0 & 
   1 & 1 & 0 & 0 & 0 & 0 & 0 & 0 & 0 & 0 \cr 0 & 0 & 
   0 & 0 & 0 & 1 & 0 & 1 & 0 & 1 & 0 & 1 & 0 & 0 & 0 &
   1 & 0 & 1 \cr 0 & 0 & 0 & 0 & 1 & 0 & 0 & 0 & 0 & 
   1 & 0 & 0 & 0 & 0 & 1 & 0 & 0 & 1 \cr 0 & 0 & 0 & 
   1 & 0 & 0 & 0 & 0 & 1 & 0 & 0 & 0 & 0 & 1 & 0 & 0 &
   1 & 0 \cr 1 & 1 & 0 & 1 & 1 & 0 & 0 & 0 & 0 & 0 & 
   0 & 0 & 0 & 1 & 1 & 0 & 0 & 0 \cr 0 & 1 & 0 & 0 & 
   0 & 0 & 0 & 1 & 0 & 0 & 0 & 0 & 0 & 1 & 0 & 1 & 0 &
   0 \cr 
\ea \right)} \ , \\ \\
P ={\tiny \left( \ba{cccccccccccccccccc}
 0 & 0 & 0 & 0 & 0 & 0 & 0 & 0 & 0 & 0 & 0 & 0 & 1 &
   1 & 1 & 1 & 1 & 1 \cr 0 & 0 & 0 & 0 & 0 & 0 & 0 & 
   0 & 0 & 0 & 1 & 1 & 0 & 0 & 0 & 0 & 1 & 1 \cr 0 & 
   0 & 0 & 0 & 0 & 0 & 1 & 1 & 1 & 1 & 0 & 0 & 0 & 0 &
   0 & 0 & 0 & 0 \cr 0 & 0 & 1 & 1 & 1 & 1 & 0 & 0 & 
   1 & 1 & 0 & 0 & 0 & 0 & 0 & 0 & 0 & 0 \cr 0 & 0 & 
   0 & 0 & 0 & 1 & 0 & 1 & 0 & 1 & 0 & 1 & 0 & 0 & 0 &
   1 & 0 & 1 \cr 0 & 0 & 0 & 0 & 1 & 0 & 0 & 0 & 0 & 
   1 & 0 & 0 & 0 & 0 & 1 & 0 & 0 & 1 \cr 0 & 0 & 0 & 
   1 & 0 & 0 & 0 & 0 & 1 & 0 & 0 & 0 & 0 & 1 & 0 & 0 &
   1 & 0 \cr 1 & 1 & 0 & 1 & 1 & 0 & 0 & 0 & 0 & 0 & 
   0 & 0 & 0 & 1 & 1 & 0 & 0 & 0 \cr 0 & 1 & 0 & 0 & 
   0 & 0 & 0 & 1 & 0 & 0 & 0 & 0 & 0 & 1 & 0 & 1 & 0 &
   0 \cr 0 & 0 & 1 & 0 & 0 & 0 & 1 & 0 & 1 & 0 & 1 & 
   0 & 1 & 0 & 0 & 0 & 1 & 0 \cr 1 & 1 & 0 & 0 & 0 & 
   0 & 1 & 1 & 0 & 0 & 1 & 1 & 0 & 0 & 0 & 0 & 0 & 
   0 \cr 1 & 0 & 0 & 0 & 0 & 0 & 1 & 0 & 0 & 0 & 0 & 
   0 & 1 & 0 & 1 & 0 & 0 & 0 \cr 0 & 1 & 1 & 0 & 1 & 
   0 & 0 & 0 & 0 & 0 & 1 & 0 & 0 & 0 & 0 & 0 & 0 & 
   0 \cr 0 & 0 & 1 & 0 & 0 & 1 & 0 & 0 & 0 & 0 & 0 & 
   0 & 1 & 0 & 0 & 1 & 0 & 0 \cr 1 & 0 & 0 & 1 & 0 & 
   1 & 0 & 0 & 0 & 0 & 0 & 1 & 0 & 0 & 0 & 0 & 0 & 
   0 \cr  
\ea \right)} \ .
\ea\eeq
Subsequently, their kernel is $Q^t$, giving us
\beq
Q^t =  {\tiny \left( \ba{cccccccccccccccccc}
0 & 0 & 1 & 1 & -1 & -1 & 0 & 0 & 0 & 0 & 
    0 & 0 & 0 & 0 & 0 & 0 & -1 & 1 \cr 1 & 0 & 1 & 
    -1 & 0 & 0 & 0 & 0 & 0 & 0 & -1 & 0 & 
    -1 & 0 & 0 & 0 & 1 & 0 \cr 1 & -1 & 1 & 0 & 0 & 
    -1 & 0 & 0 & 0 & 0 & 0 & 0 & 
    -1 & 0 & 0 & 1 & 0 & 0 \cr 0 & 0 & 1 & 0 & 
    -1 & 0 & 0 & 0 & 0 & 0 & 0 & 0 & 
    -1 & 0 & 1 & 0 & 0 & 0 \cr 1 & -1 & 1 & 
    -1 & 0 & 0 & 0 & 0 & 0 & 0 & 0 & 0 & 
    -1 & 1 & 0 & 0 & 0 & 0 \cr 0 & 0 & 1 & 0 & 0 & 
    -1 & 0 & 0 & 0 & 0 & 
    -1 & 1 & 0 & 0 & 0 & 0 & 0 & 0 \cr 1 & 0 & 1 & 0 &
   -1 & -1 & 
    -1 & 0 & 0 & 1 & 0 & 0 & 0 & 0 & 0 & 0 & 0 & 0 \cr
   1 & 0 & 0 & -1 & 0 & 0 & 
    -1 & 0 & 1 & 0 & 0 & 0 & 0 & 0 & 0 & 0 & 0 & 0 \cr
   1 & -1 & 1 & 0 & 0 & -1 & 
    -1 & 1 & 0 & 0 & 0 & 0 & 0 & 0 & 0 & 0 & 0 & 0 \cr
\ea \right) } \Rightarrow
\firr{PdP_4} \simeq \IC^{18} // Q^t \ .
\eeq
Indeed, the rows of $Q^t$ sum to 0 and the toric variety is Calabi-Yau.

\subsubsection{Cone over Fifth Pseudo del Pezzo Surface: $PdP_5$}
Trudging on, we can also study one more blowup, the $PdP_5$ theory.
The toric and quiver diagrams are given in \fref{f:dP5tq}
and the superpotential is
\begin{equation}\ba{rcl}
W_{PdP_5} &=& -X_{13} X_{35} X_{58} X_{81} + X_{14} X_{46} X_{68} X_{81} + X_{35} X_{57} X_{72} X_{23} - X_{46} X_{67} X_{72} X_{24} \\
& &+ X_{67}X_{71} X_{13} X_{36} - X_{57} X_{71} X_{14} X_{45} + X_{58} X_{82} X_{24} X_{45} - X_{68} X_{82} X_{23} X_{36} \ . \\
\ea\end{equation}
\begin{figure}[t]
\begin{center}
\includegraphics[scale=0.6]{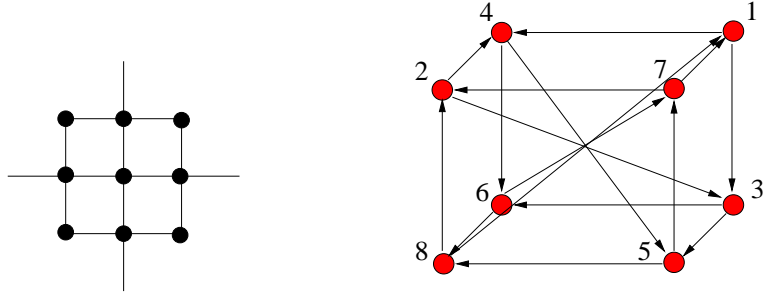} 
\caption{{\sf The toric diagram and the quiver for $PdP_5$.}}
\label{f:dP5tq}
\end{center}
\end{figure}
\comment{
  \begin{equation}
    \hbox{R}(PdP_5)=\mathbb{C}[x13,x35,x58,x81,x14,x46,x68,x57,x72,x23,x67,x24,x71,x36,x45,x82,Degrees=>{1,1,1,1,1,1,1,1,1,1,1,1,1,1,1,1}]
  \end{equation}
  \begin{eqnarray}
    \hbox{I}(PdP_5)&=&(-35*x58*x81+x67*x71*x36,-x13*x58*x81+x57*x72*x23,\nonumber\\
    & & -x13*x35*x81+x82*x24*x45,-x13*x35*x58+x14*x46*x68,x46*x68*x81-x57*x71*x45,\nonumber\\
    & & x14*x68*x81-x67*x72*x24,x14*x46*x81-x82*x23*x36,x35*x72*x23-x71*x14*x45,\nonumber\\
    & & x35*x57*x23-x46*x67*x24,x35*x57*x72-x68*x82*x36,-x46*x72*x24+x71*x13*x36,\nonumber\\
    & & -x46*x67*x72+x58*x82*x45,x67*x13*x36-x57*x14*x45,x67*x71*x13-x68*x82*x23,\nonumber\\
    & & -x57*x71*x14+x58*x82*x24,x58*x24*x45-x68*x23*x36)
  \end{eqnarray}
 (x13*x67*x36 - x14*x57*x45, x68*x23*x36 - x58*x24*x45, x13*x67*x71
      -------------------------------------------------------------------------
      - x68*x23*x82, x14*x57*x71 - x58*x24*x82, x46*x72*x24 - x13*x71*x36,
      -------------------------------------------------------------------------
      x46*x72*x67 - x58*x45*x82, x35*x72*x23 - x14*x71*x45, x35*x57*x23 -
      -------------------------------------------------------------------------
      x46*x67*x24, x35*x57*x72 - x68*x36*x82, x81*x46*x68 - x57*x71*x45,
      -------------------------------------------------------------------------
      x81*x14*x68 - x72*x67*x24, x81*x14*x46 - x23*x36*x82, x35*x58*x81 -
      -------------------------------------------------------------------------
      x67*x71*x36, x13*x58*x81 - x57*x72*x23, x13*x35*x81 - x24*x45*x82,
      -------------------------------------------------------------------------
      x13*x35*x58 - x14*x46*x68, x72*x23*x67*x36 - x58*x81*x14*x45,
      -------------------------------------------------------------------------
      x46*x68*x67*x36 - x35*x58*x57*x45, x13*x35*x23*x36 - x14*x46*x24*x45,
      -------------------------------------------------------------------------
      x13*x81*x68*x36 - x57*x72*x24*x45, x57*x72*x67*x71 - x58*x81*x68*x82,
      -------------------------------------------------------------------------
      x14*x46*x67*x71 - x35*x58*x23*x82, x13*x35*x57*x71 - x46*x68*x24*x82,
      -------------------------------------------------------------------------
      x13*x81*x14*x71 - x72*x23*x24*x82, x35*x58*x72*x24 - x14*x68*x71*x36,
      -------------------------------------------------------------------------
      x58*x81*x46*x24 - x57*x23*x71*x36, x13*x35*x72*x67 - x14*x68*x45*x82,
      -------------------------------------------------------------------------
      x13*x81*x46*x67 - x57*x23*x45*x82, x46*x68*x72*x23 - x13*x58*x71*x45,
      -------------------------------------------------------------------------
      x14*x68*x57*x23 - x13*x58*x67*x24, x35*x81*x68*x23 - x67*x24*x71*x45,
      -------------------------------------------------------------------------
      x14*x46*x57*x72 - x13*x58*x36*x82, x35*x81*x46*x72 - x71*x36*x45*x82,
      -------------------------------------------------------------------------
      x35*x81*x14*x57 - x67*x24*x36*x82)
}
Now, all the variables have the same weight and the  master space has $K$-matrix
\begin{equation}
{\tiny
K=
\left(
\begin{array}{lllllllllllllllll}
      & X_{24} & X_{36} & X_{67} & X_{82} & X_{35} & X_{57} & X_{81} & X_{45} & X_{46} & X_{68} & X_{14} & X_{13} & X_{23} & X_{72} & X_{58} & X_{71}\\
X_{24}& 1 & 0 & 0 & 0 & 0 & 0 & 0 & 0 & 0 & 0 & 1  & 1  &  1 & 0  & 0  & 0 \\
X_{36}& 0 & 1 & 0 & 0 & 0 & 0 & 0 & 0 & 0 & 0 & 1  & 0  &  0 & 1  & 1  & 0 \\
X_{67}& 0 & 0 & 1 & 0 & 0 & 0 & 0 & 0 & 0 & 0 & 1  & 0  &  1 & 0  & 1  & 0 \\
X_{82}& 0 & 0 & 0 & 1 & 0 & 0 & 0 & 0 & 0 & 0 & 1  & 1  &  0 & 1  & 0  & 0 \\
X_{35}& 0 & 0 & 0 & 0 & 1 & 0 & 0 & 0 & 0 & 0 & -1 & -1  & -1 & -1 & -1 & 0 \\
X_{57}& 0 & 0 & 0 & 0 & 0 & 1 & 0 & 0 & 0 & 0 & -1 & 0 & -1 & -1 & -1 & -1 \\
X_{81}& 0 & 0 & 0 & 0 & 0 & 0 & 1 & 0 & 0 & 0 & -1 & -1 &  0 & 0  & 0  & 1 \\
X_{45}& 0 & 0 & 0 & 0 & 0 & 0 & 0 & 1 & 0 & 0 & 0  & 1  &  0 & 0  & -1 & -1\\
X_{46}& 0 & 0 & 0 & 0 & 0 & 0 & 0 & 0 & 1 & 0 & 0  & 0  &  1 & 0  &  1 & 1\\
X_{68}& 0 & 0 & 0 & 0 & 0 & 0 & 0 & 0 & 0 & 1 & 0  & 0  &  0 & 1  &  1 & 1\\
 \end{array}
\right) \ ,
}
\end{equation}
as well as the Hilbert series
\begin{equation}
H(t;~\f_{PdP_5})=\frac{1 + 6 t + 21 t^2 + 40 t^3 + 39 t^4 - 30 t^5 + 19t^6}
{(-1 +t)^{10}} \ .
\end{equation}
The top component is Calabi-Yau of dimension 10 given by the intersection:
\begin{equation}
{\scriptsize
\ba{rcl}
\firr{PdP_5} &=& \mathbb{V}(
X_{13}X_{67}X_{36} - X_{14}X_{57}X_{45}, X_{68}X_{23}X_{36} - X_{58}X_{24}X_{45}, X_{13}X_{67}X_{71} - X_{68}X_{23}X_{82}, X_{14}X_{57}X_{71} - X_{58}X_{24}X_{82}, \\
&&
X_{46}X_{72}X_{24} - X_{13}X_{71}X_{36},X_{46}X_{72}X_{67} - X_{58}X_{45}X_{82}, X_{35}X_{72}X_{23} - X_{14}X_{71}X_{45}, X_{35}X_{57}X_{23} -X_{46}X_{67}X_{24}, \\
&&
X_{35}X_{57}X_{72} - X_{68}X_{36}X_{82}, X_{81}X_{46}X_{68} - X_{57}X_{71}X_{45}X_{81}X_{14}X_{68} - X_{72}X_{67}X_{24}, X_{81}X_{14}X_{46} - X_{23}X_{36}X_{82}, \\
&&
X_{35}X_{58}X_{81} - X_{67}X_{71}X_{36}, X_{13}X_{58}X_{81} - X_{57}X_{72}X_{23}, X_{13}X_{35}X_{81} - X_{24}X_{45}X_{82}, X_{13}X_{35}X_{58} - X_{14}X_{46}X_{68}, \\
&&
X_{72}X_{23}X_{67}X_{36} - X_{58}X_{81}X_{14}X_{45},
X_{46}X_{68}X_{67}X_{36} - X_{35}X_{58}X_{57}X_{45}, 
X_{13}X_{35}X_{23}X_{36} - X_{14}X_{46}X_{24}X_{45}, \\
&&
X_{13}X_{81}X_{68}X_{36} - X_{57}X_{72}X_{24}X_{45},
X_{57}X_{72}X_{67}X_{71} - X_{58}X_{81}X_{68}X_{82},
X_{14}X_{46}X_{67}X_{71} - X_{35}X_{58}X_{23}X_{82}, \\
&&
X_{13}X_{35}X_{57}X_{71} - X_{46}X_{68}X_{24}X_{82},
X_{13}X_{81}X_{14}X_{71} - X_{72}X_{23}X_{24}X_{82}, 
X_{35}X_{58}X_{72}X_{24} - X_{14}X_{68}X_{71}X_{36}, \\
& &  
X_{58}X_{81}X_{46}X_{24} - X_{57}X_{23}X_{71}X_{36}, 
X_{13}X_{35}X_{72}X_{67} - X_{14}X_{68}X_{45}X_{82},
X_{13}X_{81}X_{46}X_{67} - X_{57}X_{23}X_{45}X_{82}, \\
&&
X_{46}X_{68}X_{72}X_{23} - X_{13}X_{58}X_{71}X_{45},
X_{14}X_{68}X_{57}X_{23} - X_{13}X_{58}X_{67}X_{24}, 
X_{35}X_{81}X_{68}X_{23} - X_{67}X_{24}X_{71}X_{45}, \\
&&  
X_{14}X_{46}X_{57}X_{72} - X_{13}X_{58}X_{36}X_{82}, 
X_{35}X_{81}X_{46}X_{72} - X_{71}X_{36}X_{45}X_{82},
X_{35}X_{81}X_{14}X_{57} - X_{67}X_{24}X_{36}X_{82})
\ea}\end{equation}
and has Hilbert series
\begin{equation}
H(t;~\firr{PdP_5}) = \frac{1 + 6 t + 21t^2 + 40t^3 + 21t^4 + 6t^5 + t^6}{(1- t)^{10}}
\ .
\end{equation}

Finally, the dual cone $T$-matrix and the perfect matching matrix $P = K^t \cdot T$ are:
\beq\ba{l}
T = {\tiny \left( \ba{cccccccccccccccccccccccc}
0 & 0 & 0 & 0 & 0 & 0 & 0 & 0 & 0 & 0 & 0 & 0 & 0 &
   0 & 0 & 0 & 0 & 0 & 1 & 1 & 1 & 1 & 1 & 1 \cr 0 & 
   0 & 0 & 0 & 0 & 0 & 0 & 0 & 0 & 0 & 0 & 0 & 0 & 0 &
   1 & 1 & 1 & 1 & 0 & 0 & 0 & 0 & 1 & 1 \cr 0 & 0 & 
   0 & 0 & 0 & 0 & 0 & 0 & 1 & 1 & 1 & 1 & 1 & 1 & 0 &
   0 & 0 & 0 & 0 & 0 & 0 & 0 & 0 & 0 \cr 0 & 0 & 0 & 
   0 & 1 & 1 & 1 & 1 & 0 & 0 & 0 & 0 & 1 & 1 & 0 & 0 &
   0 & 0 & 0 & 0 & 0 & 0 & 0 & 0 \cr 0 & 0 & 0 & 0 & 
   0 & 0 & 0 & 1 & 0 & 0 & 0 & 1 & 0 & 1 & 0 & 0 & 0 &
   1 & 0 & 0 & 0 & 1 & 0 & 1 \cr 0 & 0 & 0 & 0 & 0 & 
   0 & 1 & 0 & 0 & 0 & 1 & 0 & 1 & 0 & 0 & 0 & 1 & 0 &
   0 & 0 & 1 & 0 & 1 & 0 \cr 0 & 0 & 0 & 0 & 0 & 1 & 
   0 & 0 & 0 & 1 & 0 & 0 & 1 & 0 & 0 & 1 & 0 & 0 & 0 &
   1 & 0 & 0 & 1 & 0 \cr 0 & 0 & 1 & 1 & 0 & 0 & 0 & 
   0 & 0 & 1 & 0 & 1 & 0 & 0 & 0 & 1 & 0 & 1 & 0 & 0 &
   0 & 0 & 0 & 0 \cr 0 & 1 & 0 & 1 & 0 & 0 & 1 & 1 & 
   0 & 0 & 0 & 0 & 0 & 0 & 0 & 0 & 1 & 1 & 0 & 0 & 0 &
   0 & 0 & 0 \cr 1 & 0 & 1 & 0 & 0 & 0 & 0 & 0 & 0 & 
   0 & 1 & 1 & 0 & 0 & 0 & 0 & 0 & 0 & 0 & 0 & 1 & 1 &
   0 & 0 \cr 
\ea \right)} \ , \\ \\
P ={\tiny \left( \ba{cccccccccccccccccccccccc}
 0 & 0 & 0 & 0 & 0 & 0 & 0 & 0 & 0 & 0 & 0 & 0 & 0 &
   0 & 0 & 0 & 0 & 0 & 1 & 1 & 1 & 1 & 1 & 1 \cr 0 & 
   0 & 0 & 0 & 0 & 0 & 0 & 0 & 0 & 0 & 0 & 0 & 0 & 0 &
   1 & 1 & 1 & 1 & 0 & 0 & 0 & 0 & 1 & 1 \cr 0 & 0 & 
   0 & 0 & 0 & 0 & 0 & 0 & 1 & 1 & 1 & 1 & 1 & 1 & 0 &
   0 & 0 & 0 & 0 & 0 & 0 & 0 & 0 & 0 \cr 0 & 0 & 0 & 
   0 & 1 & 1 & 1 & 1 & 0 & 0 & 0 & 0 & 1 & 1 & 0 & 0 &
   0 & 0 & 0 & 0 & 0 & 0 & 0 & 0 \cr 0 & 0 & 0 & 0 & 
   0 & 0 & 0 & 1 & 0 & 0 & 0 & 1 & 0 & 1 & 0 & 0 & 0 &
   1 & 0 & 0 & 0 & 1 & 0 & 1 \cr 0 & 0 & 0 & 0 & 0 & 
   0 & 1 & 0 & 0 & 0 & 1 & 0 & 1 & 0 & 0 & 0 & 1 & 0 &
   0 & 0 & 1 & 0 & 1 & 0 \cr 0 & 0 & 0 & 0 & 0 & 1 & 
   0 & 0 & 0 & 1 & 0 & 0 & 1 & 0 & 0 & 1 & 0 & 0 & 0 &
   1 & 0 & 0 & 1 & 0 \cr 0 & 0 & 1 & 1 & 0 & 0 & 0 & 
   0 & 0 & 1 & 0 & 1 & 0 & 0 & 0 & 1 & 0 & 1 & 0 & 0 &
   0 & 0 & 0 & 0 \cr 0 & 1 & 0 & 1 & 0 & 0 & 1 & 1 & 
   0 & 0 & 0 & 0 & 0 & 0 & 0 & 0 & 1 & 1 & 0 & 0 & 0 &
   0 & 0 & 0 \cr 1 & 0 & 1 & 0 & 0 & 0 & 0 & 0 & 0 & 
   0 & 1 & 1 & 0 & 0 & 0 & 0 & 0 & 0 & 0 & 0 & 1 & 1 &
   0 & 0 \cr 0 & 0 & 0 & 0 & 1 & 0 & 0 & 0 & 1 & 0 & 
   0 & 0 & 0 & 1 & 1 & 0 & 0 & 0 & 1 & 0 & 0 & 0 & 0 &
   1 \cr 0 & 0 & 1 & 1 & 1 & 0 & 1 & 0 & 0 & 0 & 0 & 
   0 & 0 & 0 & 0 & 0 & 0 & 0 & 1 & 0 & 1 & 0 & 0 & 
   0 \cr 0 & 1 & 0 & 1 & 0 & 0 & 0 & 0 & 1 & 1 & 0 & 
   0 & 0 & 0 & 0 & 0 & 0 & 0 & 1 & 1 & 0 & 0 & 0 & 
   0 \cr 1 & 0 & 1 & 0 & 1 & 1 & 0 & 0 & 0 & 0 & 0 & 
   0 & 0 & 0 & 1 & 1 & 0 & 0 & 0 & 0 & 0 & 0 & 0 & 
   0 \cr 1 & 1 & 0 & 0 & 0 & 0 & 0 & 0 & 1 & 0 & 1 & 
   0 & 0 & 0 & 1 & 0 & 1 & 0 & 0 & 0 & 0 & 0 & 0 & 
   0 \cr 1 & 1 & 0 & 0 & 0 & 1 & 0 & 1 & 0 & 0 & 0 & 
   0 & 0 & 0 & 0 & 0 & 0 & 0 & 0 & 1 & 0 & 1 & 0 & 
   0 \cr  
\ea \right)} \ .
\ea\eeq
Subsequently, their kernel is $Q^t$, giving us
\beq\ba{r}
Q^t =  {\tiny \left( \ba{cccccccccccccccccccccccc}
0 & 1 & 0 & 0 & 1 & 0 & 0 & 
    -1 & 0 & 0 & 0 & 0 & 0 & 0 & -1 & 0 & 0 & 0 & 
    -1 & 0 & 0 & 0 & 0 & 1 \cr 0 & 0 & 0 & 0 & 1 & 
    -1 & -1 & 1 & 0 & 0 & 0 & 0 & 0 & 0 & 
    0 & 0 & 0 & 0 & 0 & 0 & 0 & 0 & 1 & -1 \cr 
    -1 & 1 & 0 & 0 & 1 & 0 & 0 & 
    -1 & 0 & 0 & 0 & 0 & 0 & 0 & 0 & 0 & 0 & 0 & 
    -1 & 0 & 0 & 1 & 0 & 0 \cr 
    -1 & 1 & 0 & 0 & 1 & 0 & 
    -1 & 0 & 0 & 0 & 0 & 0 & 0 & 0 & 0 & 0 & 0 & 0 & 
    -1 & 0 & 1 & 0 & 0 & 0 \cr 0 & 0 & 0 & 0 & 1 & 
    -1 & 0 & 0 & 0 & 0 & 0 & 0 & 0 & 0 & 0 & 0 & 0 & 
   0 & -1 & 1 & 0 & 0 & 0 & 0 \cr 1 & 0 & 
    -1 & 0 & 1 & 0 & 0 & 
    -1 & 0 & 0 & 0 & 0 & 0 & 0 & 
    -1 & 0 & 0 & 1 & 0 & 0 & 0 & 0 & 0 & 0 \cr 0 & 0 &
   0 & 0 & 1 & 0 & -1 & 0 & 0 & 0 & 0 & 0 & 0 & 0 & 
    -1 & 0 & 1 & 0 & 0 & 0 & 0 & 0 & 0 & 0 \cr 1 & 0 &
   -1 & 0 & 1 & -1 & 0 & 0 & 0 & 0 & 0 & 0 & 0 & 0 & 
    -1 & 1 & 0 & 0 & 0 & 0 & 0 & 0 & 0 & 0 \cr 0 & 1 &
   0 & 0 & 0 & 0 & 0 & -1 & 
    -1 & 0 & 0 & 0 & 0 & 1 & 0 & 0 & 0 & 0 & 0 & 0 & 
   0 & 0 & 0 & 0 \cr 0 & 1 & 0 & 0 & 1 & -1 & 
    -1 & 0 & 
    -1 & 0 & 0 & 0 & 1 & 0 & 0 & 0 & 0 & 0 & 0 & 0 & 
   0 & 0 & 0 & 0 \cr 0 & 1 & -1 & 0 & 1 & 0 & 0 & 
    -1 & 
    -1 & 0 & 0 & 1 & 0 & 0 & 0 & 0 & 0 & 0 & 0 & 0 & 
   0 & 0 & 0 & 0 \cr -1 & 1 & 0 & 0 & 1 & 0 & 
    -1 & 0 & 
    -1 & 0 & 1 & 0 & 0 & 0 & 0 & 0 & 0 & 0 & 0 & 0 & 
   0 & 0 & 0 & 0 \cr 1 & 0 & -1 & 0 & 1 & 
    -1 & 0 & 0 & 
    -1 & 1 & 0 & 0 & 0 & 0 & 0 & 0 & 0 & 0 & 0 & 0 & 
   0 & 0 & 0 & 0 \cr 1 & -1 & 
    -1 & 1 & 0 & 0 & 0 & 0 & 0 & 0 & 0 & 0 & 0 & 0 & 
   0 & 0 & 0 & 0 & 0 & 0 & 0 & 0 & 0 & 0 \cr
\ea \right) } \\
\Rightarrow
\firr{PdP_5} \simeq \IC^{24} // Q^t \ .
\ea \eeq
Again, the rows of $Q^t$ sum to 0 and the toric variety is Calabi-Yau.

\subsection{The Master Space: Recapitulation}
\label{MS}
Having warmed up with an \'{e}tude and some developmental case studies let us recapitulate with our theme in $\f$. We have studied, with extensive examples for toric theories at number of D3-branes equaling to $N=1$, the algebraic variety formed by the F-flatness equations, we have seen that in general this is a reducible variety, with a top Calabi-Yau component of dimension $g+2$.
In this section we recapitulate and deepen the properties of the master space
$\f$ for such toric quiver gauge theories and will discuss what happens for $N >1$.

\subsubsection{The Toric $N=1$ Case}\label{recaptoric}
As we have seen, $\f$ is a toric $g+2$ dimensional variety, not necessarily
irreducible. The coherent component $\firr{~}$, the
largest irreducible component of $\f$, also of dimension $g+2$,
can be quite explicitly characterized. 
Denoting, as usual, with $E$ the number of fields in the quiver, $g$ the number of nodes in the quiver and with $c$ the number of perfect matchings in the dimer realization of the gauge theory, we have defined three matrices:
\begin{itemize}
\item{A $(g+2) \times E$ matrix $K$ obtained by solving the F-terms 
in terms of a set of independent fields. The columns give the 
charges of the elementary fields under the $(\IC^*)^{g+2}$ action;
in a more mathematical language, they give the
semi-group generators of the cone $\sigma_K^{\vee}$
in the toric presentation for $\firr{~}$:
\beq
\firr{~} \simeq \mbox{Spec}_{\IC}[\sigma_K^{\vee} \cap \IZ^{g+2}]
\eeq
}
\item{A $(g+2) \times c$ matrix $T$, defined by $K^t \cdot T \ge 0$, and
representing the dual cone $\sigma_K$. The columns are 
the $c$ toric vectors of the $g+2$ dimensional variety $\firr{~}$.
We see that the number of perfect matchings in the dimer realization of the
quiver theory is the number of external points in the 
toric diagram for $\firr{~}$. This generalizes the fact that the
{\it external} perfect matchings are related to the external points of the 
toric diagram for the three dimensional transverse Calabi-Yau space $\cX$.}
\item{A $E \times c$ matrix $P=K^t \cdot T$ which defines the perfect matchings as collections of elementary fields.}
\end{itemize}

The variety $\firr{~}$ also has a linear sigma model, or symplectic quotient, description as
\begin{equation}
\firr{~} = \mathbb{C}^c//Q^t \ ,
\end{equation}
where $Q$ is the kernel of the matrices $P$ and $T$.
From all these descriptions we can extract some general properties of the
master space and of its Hilbert series.

\paragraph{As a Higher-Dimensional Calabi-Yau: }
The most suggestive property is that $\firr{~}$ is always a $g+2$
dimensional {\em Calabi-Yau manifold}. This has been explicitly checked
in all the examples we have discussed. It is simple to check the
Calabi-Yau condition in the linear sigma model description, since
it corresponds there to the fact that the vectors of charges in $Q^t$
are traceless, or equivalently, the toric diagram has all the vectors ending on the same hyper-plane. 

There is a remarkably simple proof that $\firr{~}$ is Calabi-Yau
which emphasizes the r\^{o}le of perfect matchings. Recall from \cite{dimers} and
\sref{s:dimer} that
the quiver gauge theory has a description in terms of a dimer model,
this is a  bi-partite tiling
of the two torus, with $V/2$ white vertices and $V/2$ black vertices,
where $V$ is the number of superpotential terms. The elementary fields
of the quiver correspond to the edges in the tiling, each of which connects
a black vertex to a white one. Now, by definition, a perfect matching
is a choice of fields/dimers  that cover each vertex
precisely once. In particular, a perfect matching contains exactly $V/2$
dimers, connecting the $V/2$ black vertices to the $V/2$ white ones.
Since the columns of the matrix $P$, of size $E \times c$,
tell us which fields/dimers occur in a given perfect matching we have
the nice identity
\begin{equation}
\underbrace{(1,1,....,1)}_{E} \cdot P = \frac{V}{2}
\underbrace{(1,1,....,1)}_{c} ,
\end{equation}
which basically counts the number of edges in each perfect matching.

By multiplying this equation on the right by the matrix
kernel $Q$ of $P$, $P\cdot Q=0$, we obtain
\begin{equation}
0 = \underbrace{(1,1,....,1)}_{c} \cdot Q
\end{equation}
and we conclude that the vector of charges of the linear sigma model,
which are the rows of the matrix $Q^t$, are traceless. This proves that
$\firr{~}$ is Calabi-Yau.

We again see that the prefect matchings description is crucial in our understanding of the properties of the master space.
As explained in \sref{s:dimer}, the perfect matchings generates the
coherent component of the master space as a consequence of the Birkhoff-Von Neumann Theorem.

\paragraph{Seiberg Duality: }
From our study of the toric dual phases, exemplified above by the
two phases of the cone over the Hirzebruch surface, we conjecture that
\begin{quote}
For a pair of Seiberg dual theories, $\firr{}$ is the same for both.
\end{quote}

\paragraph{Palindromic Hilbert Series: }
An intriguing property of the Hilbert series for $\firr{~}$
is its symmetry. As manifest from all our examples, the numerator of
the Hilbert series (in second form) for $\firr{~}$ is a polynomial in $t$
\begin{equation}
P(t)=\sum_{k=0}^N a_k t^k
\end{equation}
with symmetric coefficients $a_{N-k}=a_k$. This means that
there is a remarkable symmetry of the Hilbert series for $\firr{~}$ under 
$t\rightarrow 1/t$,
\begin{equation}
H(1/t;\firr{~})= t^w H(t;\firr{~})
\end{equation}
where the modular weight $w$ depends on $\firr{~}$. 

A polynomial with such a symmetry between its consecutively highest and lowest coefficients $a_{N-k} \leftrightarrow a_k$ is known as a {\bf palindromic polynomial}. A beautiful theorem due to Stanley \cite{stanley} states the following
\begin{theorem}
The numerator to the Hilbert series of a graded Cohen-Macaulay domain $R$ is palindromic iff $R$ is Gorenstein.
\end{theorem}
What this means, for us, is that the coordinate ring of the affine variety 
$\firr{~}$ must be Gorenstein. However, an affine toric variety is Gorenstein
precisely if its toric diagram is co-planar, i.e., it is Calabi-Yau 
(cf.~e.g.~\S~4.2 of \cite{Morrison:1998cs}). Thus we have killed two birds with one stone: proving that $\firr{~}$ is affine toric Calabi-Yau above from perfect matchings also shows that the numerator of its Hilbert series has the palindromic symmetry.

As we will see, this symmetry extends to the refined Hilbert series written
in terms of the R-charge parameter $t$ and chemical potentials for a global
symmetry $G$. Although $G$ has been up to now Abelian, we will see that
in some special cases of theories with hidden symmetries $G$ becomes non
Abelian. Introducing chemical potentials $z$ for Cartan sub-algebra of $G$, 
we will write the refined Hilbert series as a sum over $G$-characters
in the general form,
\begin{equation}
H(t,G) = \left (\sum_{k=0}^N \chi_k(z) t^k\right ) PE\left [\sum_{i=1}^G \chi_i(z) t^i\right ]
\end{equation} 
where PE is the plethystic exponential to be reviewed in \sref{s:plet}, 
computing symmetric products on
a set of generators of the coherent component. The refined Hilbert series
is now invariant under the the combined action of $t\rightarrow 1/t$ and
charge conjugation, $\chi_{N-k}=\chi_k^*$. 
We will see many examples in Section 4.

\subsubsection{General $N$}
The case for an arbitrary number $N$ of D3-branes is much more subtle and less understood in the mathematical literature \footnote{We thank Balazs Szendr\"oi for discussions on this point.}, even though it is clear from the gauge theory perspective. We know that the world-volume theory for $N$ D3-branes is a quiver theory with product $U(N_i)$ gauge groups and in the IR, the $U(1)$ factors decouple since only the special unitary groups exhibit asymptotic freedom and are strongly coupled in the IR. Thus the moduli space  of interest is the space of solutions to the F-flatness, quotiented out by a non-Abelian gauge group 
\beq
\CM_N = \f_N / (SU(N_1) \times \ldots \times SU(N_g)).
\eeq 
where the index $N$ recalls that we are dealing with $N$ branes. The moduli 
space $\CM_N$ is of difficult characterization since the quotient is
fully non-Abelian and it can not be described by toric methods, as in the $N=1$
case.

The more familiar mesonic moduli space is obtained by performing a
further quotient by the Abelian symmetries. Even for $N$ branes, the Abelian group will be constituted of the decoupled $U(1)$ factors, and hence will be the same as in the toric, $N=1$ case. 
Once again, we expect to have some symplectic quotient structure, in analogy with \eref{mesmod}, for this mesonic moduli space:
\beq\label{symM-N}
{}^{{\rm mes}}\!{\cal M}_N    \simeq \CM_N // U(1)^{g-1} \ .
\eeq
Hence, a toric symplectic quotient still persists for \eref{symM-N}, even though the moduli space in question is not necessarily toric.

Moreover, our plethystic techniques, which we will shortly review, will illuminate us with much physical intuition. First, the mesonic moduli space for $N$ branes is the $N$-th symmetrized product of that of $N=1$:
\begin{equation}\label{symM}
{}^{{\rm mes}}\!{\cal M}_N \simeq {\rm Sym}^N \cX := \cX^N / \Sigma_N \ ,
\end{equation}
where $\Sigma_N$ is the symmetric group on $N$ elements.
We see that the mesonic moduli space, for $\cX$ a Calabi-Yau threefold, is of dimension $3N$ by \eref{symM}.
The dimension of the moduli space $\CM_N$ is thus $3N + g - 1$ for general $N$.

\subsubsection{The Plethystic Programme Reviewed}\label{s:plet}
We cannot embark upon a study of general $N$ without delving into the plethystic programme, whose one key purpose of design was for this treatment. Here let us review some essentials, partly to set the notation for some extensive usage in \sref{s:hidden} later. The realisation in \cite{pleth} is that the mesonic generating function, with all baryonic numbers fixed to be zero, $g_{1,0}(t;~\cX) = f_\infty(t;~\cX)$ for the single-trace mesonic operators for D3-branes probing a Calabi-Yau threefold $\cX$ at $N \rightarrow \infty$ is the Hilbert series of $\cX$. Let us define the {\bf Plethystic Exponential} of a multi-variable function $g(t_1, \ldots, t_n)$ that vanishes at the origin, $g(0, \ldots, 0) = 0,$ to be

\beq\label{defPE}
PE \left [ g (t_1, \ldots, t_n) \right ] := \exp\left( \sum\limits_{k=1}^\infty\frac{g (t_1^k, \ldots, t_n^k)}{k}\right)  \ .
\eeq

Then the multi-trace mesonic operators at $N \rightarrow \infty$ are counted by the plethystic exponential\footnote{Note that in order to avoid an infinity the PE is defined with respect to a function that vanishes when all chemical potentials are set to zero.}
\beq\label{defPE1}
g_{\infty,0}(t;~\cX) = PE[f_\infty(t;~\cX)-1] := \exp\left( \sum\limits_{r=1}^\infty\frac{f_\infty (t^r)-1}{r}\right)  \ .
\eeq
The inverse, $f_1(t;~\cX) = PE^{-1}[f_\infty(t;~\cX)]$, is counting objects in the defining equation (syzygy) of the threefold. The mesonic multi-trace generating function $g_N$ at finite $N$ is found by the series expansion of the $\nu$-inserted plethystic exponential\footnote{Note that the $\nu$ insertion satisfies the condition that the argument of PE vanishes when all chemical potentials are set to zero. Any attempt to subtract something from this function leads to incorrect results.} as 

\beq
PE[ \nu f_\infty(t;~\cX)] = \exp\left( \sum\limits_{r=1}^\infty\frac{\nu^r f_\infty(t^r)}{r}\right) = \sum\limits_{N=0}^\infty g_N(t) \nu^N.
\eeq

In general \cite{Forcella:2007wk}, for the combined mesonic and baryonic branches of the moduli space\footnote{To be Strict, we should say here the mesonic branch together with given FI-parameters, since at $N=1$, there are no baryons. Nevertheless, we can still generate the counting for the baryons for $N>1$ using PE of the $N=1$ case.}, the $N=1$ operators are counted by the Hilbert series of the master space. The plethystic program can be efficiently applied to the study
of the coherent component of the moduli space \cite{Butti:2007jv}.
With the generating function for the coherent component of the master space,
which we denote $g_1(t;~\cX)\equiv H(t; ~\firr{~}) $, we can proceed with the
plethystic program and find the result for $g_N(t;~\cX)$, counting the combined baryonic and mesonic gauge invariant operators at finite $N$. 

The implementation of the plethystic program requires a decomposition
of the $g_1(t;~\cX)$ generating function in sectors of definite baryonic 
charge, to which the plethystic exponential is applied. An interesting 
connection of this decomposition is found in \cite{Butti:2007jv,Forcella:2007wk} with K\"ahler moduli. This also enables  a different computation of $g_1(t;~\cX)$. Though our current techniques compute this quantity using the Hilbert series of the master space, we will later check that this indeed agrees with the formalism of \cite{Butti:2007jv}.

The decomposition of $g_1(t;~\cX)$ requires the knowledge
of two sets of partition functions, the geometrical ones, obtained by 
localization, and the auxiliary one, obtained from dimer combinatorics.
Specifically, it was realized that 
\begin{equation}\label{GKZ1}
g_1(t;~\cX) = \sum\limits_{P \in GKZ} m(P) g_{1,P}(t;~\cX)
\end{equation} 
where the summation is extended over the lattice points $P$, of multiplicity $m(P)$, of a so-called {\bf GKZ (or secondary) fan} of the Calabi-Yau threefold and $g_{1,P}$ is a much more manageable object obtained from a localisation computation, as given in Eq (4.18) of \cite{Butti:2007jv}. The GKZ fan, to remind the reader, is the fan of an auxiliary toric variety, which is the space of K\"ahler parameters of the original toric threefold $\cX$. This space is of dimension $I - 3 + d$, where $I$ is the number of internal points and $d$, the number of vertices, of the toric diagram of $\cX$. 

The multiplicity $m(P)$ of points in this GKZ lattice fan is counted by an {\bf auxiliary partition function}, so-named $Z_{\rm aux}$. This is simply the (refined) Hilbert series of the following space: take the simpler quiver than the original by neglecting any repeated arrows and then form the space of open but not closed loops in this simplified quiver. The expansion of $Z_{\rm aux}$ can then be used to compute the generating function for one D3-brane. 

As a brief reminder to the reader, the procedure to determine the refined generating function $g_1$ in \eref{GKZ1} is to (1) obtain the generating function $g_{1,\beta_1, ... \beta_K}$ in terms of a set of K\"ahler parameters $\beta_1, ..., \beta_K$ using the localisation formula (4.18) of \cite{Butti:2007jv}; (2) obtain $Z_{\rm aux}(t_1,...,t_K)$ as above, and (3) replace a term $t_1^{\beta_1} ...  t_K^{\beta_K}$ in $Z_{\rm aux}$ by an expression for $g_{1,\beta_1,...\beta_K}$. 
We will not enter in the details of this construction and we refer the reader
to \cite{Butti:2007jv}.
The important point for our ensuing
discussions is that the plethystic program
can be applied to the $N=1$ partition functions at each point of the GKZ fan 
in order to obtain the finite  $N$ generating function 
\begin{equation}\label{GKZN}
g(t;~\cX) := \sum_{N=0}^\infty \nu^N g_N(t;~\cX) = \sum\limits_{P \in GKZ} m(P) PE \left [\nu  g_{1,P}(t;~\cX)\right ] \ .
\end{equation}

\section{Linear Branches of Moduli Spaces}\label{s:branch}
\setall

From the previous section we learned that the master space $\f$ of an $\mathcal{N}=1$ quiver gauge theory on a toric singularity $\cX$ generically presents a reducible moduli space. This fact could appear surprising. Indeed the $\mathcal{N}=2$ supersymmetric gauge theories are the classical examples of theories with reducible moduli spaces. These theories present two well separated branches: the Higgs branch and the Coulomb branch. Moving along these different branches has a well defined characterization both in terms of the geometry of the moduli space and in terms of the VEV structure of the gauge theory. It would be interesting to have a similar interpretation in this more generic setup\footnote{One can clearly look for other interpretations of the reducibility of $\f$. An interesting line of research would be to connect the geometrical structure of $\f$ to the pattern of BPS branes in $\cX$. We will leave this topic for future work.}.

In the case of just $N=1$ brane on the tip of the conical singularity $\cX$, the reducibility of $\f$ follows easily from the toric condition. Indeed, the equations defining $\f$ are of the form ``monomial $=$ monomial'', whereby giving us toric ideals as discussed in \sref{s:toric}. Let us embed this variety into $\mathbb{C}^d$ with coordinates $\{x_1,...,x_d\}$, then its algebraic equations have the form:
\begin{equation}\label{fget}
x_1^{i_{j_1}}...x_d^{i_{j_d}}\Big( M_{1_j}(x)-M_{2_j}(x) \Big)=0 \ ,
\end{equation}
where $j=1,...,k$ runs over the number of polynomials defining the variety and the polynomials $M_{1_j}(x)-M_{2_j}(x)$ are irreducible. If some $i_{j_p}$ is different from zero the variety is reducible and is given by the union of the zero locus of $M_{1_j}(x)-M_{2_j}(x)$ together with a set of planes (linear components) $L^l$ given by the zero locus of the factorized part in (\ref{fget}).

Hence, the master space $\f$ could be the union of a generically complicated $g+2$ dimensional variety with a set of smaller dimensional linear varieties $\{\mathbb{C}^{d_1},...,\mathbb{C}^{d_n}\}$ parameterized by combinations of the coordinates\footnote{The reducibility of the moduli space could persist for $N > 1$, the simplest example being the conifold with $N=2$ (see \cite{Butti:2007jv} with other examples) and it would be interesting to have a clearer geometric picture even in these more complicated cases. In this section we will concentrate on the $N=1$ case.}; 
this is certainly true for the extensive case studies seen above in \sref{s:case}. It would be nice to give a gauge theory interpretation to these smaller dimensional linear branches. 

In the following subsections we will give a nice picture for these planes, for some selected examples. The lesson we will learn is that {\it these planes of the master space could parametrize flows in the gauge theory}. Specifically, there may be chains of flows from one irreducible component of the master space of a theory to another. The archetypal example which will be the terminus of many of these flows will be the gauge theory on a D3-brane probing the Calabi-Yau singularity $\mathbb{C}^2/\mathbb{Z}_2\times \mathbb{C}$, to which we alluded in \sref{s:dimer}. Let us first briefly review this theory, continuing along the same vein as our discussions in \sref{s:irred}.

\subsection{The $(\mathbb{C}^2/\mathbb{Z}_2) \times \mathbb{C}$ Singularity}
\label{C2Z2}
The quiver gauge theory for $(\mathbb{C}^2/\mathbb{Z}_2) \times \mathbb{C}$ has ${\cal N}=2$ supersymmetry with two vector multiplets and two bi-fundamental hyper-multiplets. In ${\cal N}=1$ notation we have six chiral multiplets denoted as $\phi_1, \phi_2, A_1, A_2, B_1, B_2$, with a superpotential
\begin{equation}
W = \phi_1 (A_1 B_1 - A_2 B_2) + {\phi}_2 (B_2 A_2 - B_1 A_1)
\end{equation}
and the quiver and toric diagrams are given in \fref{f:c2z2quiver}.
\begin{figure}[t]
\begin{center}
  \epsfxsize = 11cm
  \centerline{\epsfbox{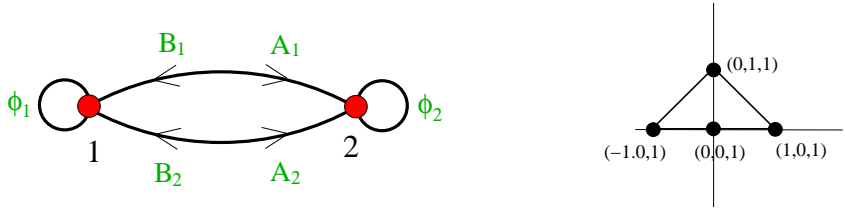}}
  \caption{{\sf Quiver and toric diagrams for $(\mathbb{C}^2/\mathbb{Z}_2) \times \mathbb{C}$.}}
  \label{f:c2z2quiver}
\end{center}
\end{figure}
The master space $\f_{(\mathbb{C}^2/\mathbb{Z}_2)\times \mathbb{C}}$ of the theory can be easily found to be
\begin{equation}\label{N2id}
\f_{(\mathbb{C}^2/\mathbb{Z}_2)\times \mathbb{C}}=\mathbb{V}(A_1B_1-A_2B_2,\phi_1A_1-\phi_2A_1,\phi_1A_2-\phi_2A_2,\phi_1B_1-\phi_2B_1,\phi_1B_2-\phi_2B_2) \ .
\end{equation}
Now, $\f_{(\mathbb{C}^2/\mathbb{Z}_2)\times \mathbb{C}}$ is clearly reducible and decomposes into the following two irreducible components as $\f_{(\mathbb{C}^2/\mathbb{Z}_2)\times \mathbb{C}}= \firr{(\mathbb{C}^2/\mathbb{Z}_2)\times \mathbb{C}} \hbox{  } \cup \hbox{  }L$ with
\beq\label{N2red}
\firr{(\mathbb{C}^2/\mathbb{Z}_2)\times \mathbb{C}}=\mathbb{V}(\phi_1 - \phi_2, A_1B_1 - A_2B_2) \ , \qquad
L =\mathbb{V}(A_1, A_2, B_1, B_2) \ .
\eeq
Specifically, $\firr{(\mathbb{C}^2/\mathbb{Z}_2)\times \mathbb{C}}$ is $\mathbb{C} \times \mathcal{C}$, where the $\mathbb{C}$ is defined by $\phi_1 = \phi_2$ and the conifold singularity $\mathcal{C}$ is described by the chiral fields $\{A_1,A_2,B_1,B_2\}$ with the constraint $A_1 B_1 = A_2 B_2 $. The component $L = \mathbb{C}^2$ is parametrized by the fields $\{\phi_1,\phi_2 \}$. These two branches meet on the complex line parametrized by $\phi_1=\phi_2$. 

The field theory interpretation of these two branches is standard: moving in $L$ we are giving VEV to the scalars in the vector multiplet and hence we call $L$ the Coulomb branch; while moving in $\firr{(\mathbb{C}^2/\mathbb{Z}_2)\times \mathbb{C}}$ we are giving VEV to the scalars in the hyper-multiplets and hence we call $\firr{(\mathbb{C}^2/\mathbb{Z}_2)\times \mathbb{C}}$ the Higgs branch.
Let us go on and revisit the reducibility of some of the master spaces studied in the previous section trying to give them a gauge theory interpretation.

\subsection{Case Studies Re-examined}

\paragraph{First Toric Phase of $F_0$: }
Let us start by re-examining $(F_0)_I$, encountered in \sref{s:F0}. We recall from \eref{F0-I} that the master space $\f_{F_0}$ is the union of three branches: the biggest one is six dimensional and is the set product of two conifold singularities, i.e., $\mathcal{C} \times \mathcal{C}$, and the two smallest ones are two copies of $\mathbb{C}^4$, parametrized respectively by the VEV of $\{B_1, B_2, D_1, D_2\}$ and $\{A_1, A_2, C_1, C_2\}$.

\begin{figure}[t]
\begin{center}
\includegraphics[scale=0.5]{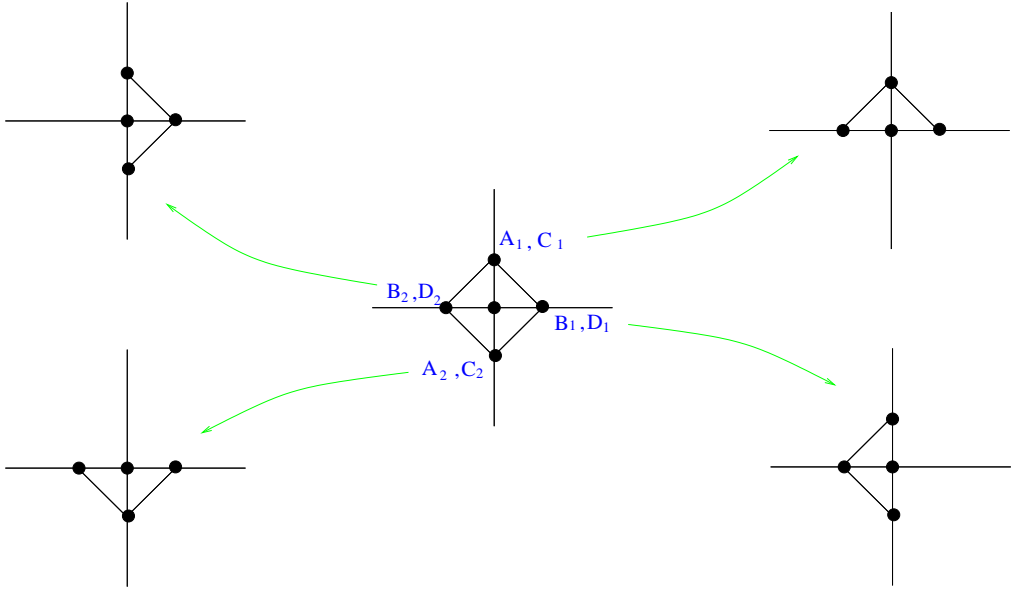} 
\caption{{\sf The four possible flows from $F_0$ to $(\mathbb{C}^2/\mathbb{Z}_2) \times \mathbb{C}$ as exhibited in the toric diagrams.}}
\label{flowF0}
\end{center}
\end{figure}

Looking at the toric diagram\footnote{We remind the reader that the prescription is as follows \cite{Beasley:1999uz,Feng:2000mi,Feng:2002fv}. 
If one toric diagram is derived by deleting external vertices of the other, then the associated gauge theories can flow one into the other. Geometrically this is blowing up the singularity, while in the gauge theory a FI term is turned on and as a
consequence gives VEV to some chiral fields, for a product of $U(N_i)$ gauge groups; or a baryonic VEV is turned on in the case of product of $SU(N_i)$ gauge groups. It is indeed possible to associate fields of the gauge theory to each
vertex in the toric diagram, as shown in Figure \ref{flowF0}. 
The deleted vertex
can be associated with the fields that are getting a VEV.} one can easily understand that the gauge theory can flow to the one associated to the $\mathbb{C}^2/\mathbb{Z}_2 \times \mathbb{C}$ singularity in four different ways by giving VEV to one out of the four possible sets of fields: $\{A_1,C_1\}$, $\{A_2,C_2\}$, $\{B_1,D_1\}$, $\{B_2,D_2\}$. Once arrived at the fixed point one can move along the Coulomb branch of the IR theory giving VEV respectively to the set of fields: $\{A_2,C_2\}$, $\{A_1,C_1\}$, $\{B_2,D_2\}$, $\{B_1,D_1\}$.
Hence, along the two smaller branches of the moduli space the theory admits an accidental $\mathcal{N}=2$ supersymmetry and the theory moves along the Coulomb branch of the resulting gauge theory.

\paragraph{First del Pezzo: }
In \eref{f-dP1} we saw that $\f_{dP_1}$ decomposes into two irreducible parts: $\firr{dP_1}$ and $L$. The former is a six dimensional variety, while the existence of the latter four dimensional linear variety $L$ parametrized by $\{U_1,U_2,\tilde U_1, \tilde U_2 \}$ can be interpreted as the effect of the presence of an accidental $\mathcal{N}=2$ sector. 

Indeed, looking at the toric diagram (see figure \ref{f:dP1tq}) it is easy to realize that the theory flows to the $\mathcal{N}=2$ theory of $(\mathbb{C}^2/\mathbb{Z}_2)\times \mathbb{C}$ in two different ways: giving VEV to $\{U_1,\tilde U_1\}$ or to $\{U_2,\tilde U_2\}$, and the Coulomb branch of these ones is parametrized respectively by $\{U_2,\tilde U_2\}$ and $\{U_1,\tilde U_1\}$. Hence, as in the case of $F_0$ the linear component in the reducible master space parameterizes the Coulomb branch of the IR $\mathcal{N}=2$ theory.

\subsubsection{Multiple Flows}
At this point we have seen that the reducible part of the master space of $F_0$ and $dP_1$ can be interpreted as parameterizing the flows of these theories to the Coulomb branch of $(\mathbb{C}^2/\mathbb{Z}_2)\times \mathbb{C}$. Indeed, $\f_{F_0}$ contains two 4 dimensional planes: along $L^1_{F_0}$ the theory can flow in two different ways to $(\mathbb{C}^2/\mathbb{Z}_2)\times \mathbb{C}$ and along $L^2_{F_0}$ the theory can flow in two other different ways to $(\mathbb{C}^2/\mathbb{Z}_2)\times \mathbb{C}$. In all the $F_0$ theory has 4 ways to flow to an  $\mathcal{N}=2$ theory and these various flows are parametrized by the two linear irreducible components of the moduli space. On the other hand, $\f_{dP_1}$ contains just one linear four dimensional irreducible component which parameterizes the two different ways in which this theory can flow to the $(\mathbb{C}^2/\mathbb{Z}_2)\times \mathbb{C}$ theory. It is important to note that the linear components of the master spaces of $F_0$ and $dP_1$ do not have non trivial intersections among themselves.

It is a well-known fact that the $dP_3$ theory can flow to the $dP_2$ and thence further to $dP_1$ or $F_0$ (cf.~e.g.~\cite{Feng:2002fv}).
The linear irreducible components of the master space $\f_{dP_3}$ and $\f_{dP_2}$ intersect in a non trivial way and it would be interesting to understand the link between the non trivial linear structure of the moduli space of these theories and the possible flows.

\paragraph{Flow Through $dP_2$: }
As already explained in \eref{f-dP2}, $\f_{dP_2}$ contains two five dimensional planes that are parameterized by the following set of coordinates in the moduli space:
\beq\ba{rcl}
L^1_{dP_2}&=&\{X_{45},Y_{31},X_{42},X_{31},X_{52} \}\\
L^2_{dP_2}&=&\{X_{45},X_{15},Y_{23},X_{23},X_{14} \} \ .
\ea\eeq
The two hyperplanes intersect along a complex line parametrized by $X_{45}$:
\begin{equation}
L^1_{dP_2}\cap L^2_{dP_2}=\{ X_{45} \} \ .
\end{equation}
Inspecting the toric diagram (see figure \ref{f:dP2tq}) it is easy to see that the $dP_2$ theory can flow to $F_0$ just in one way and this flow is parametrized in the moduli space by the linear variety $L^1_{dP_2}\cap L^2_{dP_2}$. Along this flow the two linear five dimensional components of $dP_2$ become the two linear four dimensional components of the moduli space of $F_0$:
\beq
\vev{X_{45}} \neq 0: \hbox{  } L^1_{dP_2} \rightarrow L^1_{F_0} \hbox{ , }  L^2_{dP_2} \rightarrow L^2_{F_0} \ .
\eeq
Whence the $F_0$ theory can flow to the Coulomb phase of $(\mathbb{C}^2/\mathbb{Z}_2)\times \mathbb{C}$ in the four different ways previously explained.

The $dP_2$ theory has another interesting flow to $dP_1$. Again, looking at the blowing up structure of the two toric diagrams one can see that there are two possible ways to flow to $dP_1$: one is to give VEV to the field $X_{52}$ in $L^1_{dP_2}$ in which case this five dimensional linear space flows to the four dimensional linear irreducible component $L_{dP_1}$ of the master space of $dP_1$; the other is to give a VEV to the field $X_{14}$ in $L^2_{dP_2}$ in which case this five dimensional linear space flows to $L_{dP_1}$. In summary, 
\beq\ba{rcl}
\vev{X_{52}} \neq 0 &:& L^1_{dP_2} \rightarrow L_{dP_1} \nonumber \\
\vev{X_{14}} \neq 0 &:& L^2_{dP_2} \rightarrow L_{dP_1} \ .
\ea\eeq
Whence the theory can flow to the Coulomb branch of $(\mathbb{C}^2/\mathbb{Z}_2)\times \mathbb{C}$ along $L_{dP_1}$ in the two ways previously explained.

\paragraph{Third del Pezzo: }
From \eref{f-dP3} we know that $\f_{dP_3} $ contains three six dimensional hyperplanes and these are parameterized by the following set of coordinates in the moduli space:
\begin{eqnarray}\label{HH}
L^1_{dP_3}&=&\{X_{56}, X_{23}, X_{34}, X_{51}, X_{61}, X_{24} \} \nonumber \\
L^2_{dP_3}&=&\{X_{61}, X_{34}, X_{12}, X_{45}, X_{35}, X_{62} \} \nonumber \\
L^3_{dP_3}&=&\{X_{12}, X_{45}, X_{23}, X_{56}, X_{13}, X_{46}  \} \ ;
\end{eqnarray}
every pair of hyperplanes intersect in a $\mathbb{C}^2$ parameterized by the following coordinates:
\begin{eqnarray}\label{intH}
L^1_{dP_3}\cap L^2_{dP_3}&=&\{ X_{34},X_{61} \} \nonumber \\
L^2_{dP_3}\cap L^3_{dP_3}&=&\{ X_{12},X_{45} \} \nonumber \\ 
L^1_{dP_3}\cap L^3_{dP_3}&=&\{ X_{56},X_{23} \} \ .
\end{eqnarray}
From the toric diagram (see figure \ref{f:dP3tq}) one can see that $dP_3$ can flow to $dP_2$ in six different ways. These are indeed the six different $\mathbb{C}$ contained in the various pairwise intersections in (\ref{intH}). The six possible different VEVs one can give to flow to $dP_2$ are:
\begin{equation}
\vev{X_{12}} \hbox{ , } \vev{X_{23}} \hbox{ , }
\vev{X_{34}} \hbox{ , } \vev{X_{45}} \hbox{ , } \vev{X_{56}}
\hbox{ , } \vev{X_{61}} \ .
\end{equation} 
Observe that every coordinate parameterizing the six different complex lines are contained just in two of the three six dimensional planes in (\ref{HH}). Indeed along these flows the two planes $L^i_{dP_3}$, $L^j_{dP_3}$ containing the field $F_{i,j}^k$, $k=1,2$ with non trivial VEV, flow to the two five dimensional planes $L^1_{dP_2}$, $L^2_{dP_2}$ of the moduli space of $dP_2$:
\beq
\vev{F_{i,j}^k} \neq 0 : L^i_{dP_3}, L^j_{dP_3} 
\rightarrow L^1_{dP_2}, L^2_{dP_2} \ .
\eeq
Whence, giving VEV to the field parameterizing the second $\mathbb{C}$ in the pairwise intersections in (\ref{intH}) the theory flows to $F_0$ and the two five dimensional planes in the master space of $dP_2$ flows to the two four dimensional planes in the master space of $F_0$. We can summarize these flows, observing that $dP_3$ can flow in three different ways to $F_0$ along the three intersections in (\ref{intH}), giving VEV to the two fields in the $\mathbb{C}^2$:
\beq
\vev{L^i_{dP_3}\cap L^j_{dP_3}} \neq 0 :  L^i_{dP_3}, L^j_{dP_3} 
\rightarrow L^1_{F_0}, L^2_{F_0} \ .
\eeq
As explained above $dP_2$ can also flow to $dP_1$.
Summarizing the two main different flows, we have that
\beq\ba{rcl}
dP_3 \rightarrow dP_2 \rightarrow dP_1 \rightarrow (\mathbb{C}^2/\mathbb{Z}_2)\times \mathbb{C} \rightarrow \hbox{Coulomb branch}\\
dP_3 \rightarrow dP_2 \rightarrow F_0 \rightarrow (\mathbb{C}^2/\mathbb{Z}_2)\times \mathbb{C} \rightarrow \hbox{Coulomb branch} \ ;
\ea\eeq
these can be geometrically interpreted as flows along the various irreducible linear components of the complete reducible master space of $dP_3$.

In this section we have proposed a simple field theory interpretation for the linear irreducible components of the master space. It seems to work nicely for the case examined. But there are other cases in which the correspondence between flows 
and linear spaces is less straightforward and need further analysis.

\comment{
\paragraph{Second Toric Phase of $ F_0$)}
The linear red components of the moduli space seems to parametrize flux of the gauge theories...this fact seems very well understood in the cases of the minimal toric phases of dP and F0...indeed as reported in the coming section the linear reducible components of moduli space parameterize exactly the flows of these theory till the Coulomb branch of the  $(\mathbb{C}^2/\mathbb{Z}_2)\times \mathbb{C}$theory...In the case of the SPP the situation is similar but not completely equal: indeed two of the coordinate of the three dim hyperplane $H$ $\{ X_{23},X_{32} \}$ work easily: giving VEV to one of the two the theory flows to the Coulomb branch of $(\mathbb{C}^2/\mathbb{Z}_2)\times \mathbb{C}$ parametrized by the other two component of the 3 dim plane $\{ X_{32},X_{11}\}$ and giving VEV to the other component the theory flows again to the coulomb branch of $(\mathbb{C}^2/\mathbb{Z}_2)\times \mathbb{C}$ but this time parametrized by  $\{ X_{23},X_{11}\}$...if we give VEV to the third component of the 3 dim plane ($X_{11}$) instead the theory clearly do not flow to the Coulomb branch of $(\mathbb{C}^2/\mathbb{Z}_2)\times \mathbb{C}$...
The case of the II phase of $F_0$ the situation is even more involved...the second phase of $F_0$ clearly flows to the Coulomb branch of $(\mathbb{C}^2/\mathbb{Z}_2)\times \mathbb{C}$ but these flows are parametrized by fields that are not in the same linear irreducible components...for example one flows is obtained giving VEV to $X_{12}$ and $X_{43}$ and the coulomb branch of the theory obtained is parametrized by $Y_{12}$ and $Y_{43}$.
SUMMARIZING: the linear irreducible components seems very strictly related to flows to the Coulomb branch of $(\mathbb{C}^2/\mathbb{Z}_2)\times \mathbb{C}$ and this fact is nicely realized in the case of the minimal phase of dP and $F_0$, but the situation is generically more involved and probably more important not invariant under Seiberg duality! 
}

\section{Hidden Global Symmetries}\label{s:hidden}\setall
The moduli space of a field theory may possess symmetries beyond gauge symmetry or reparametrisation. Searching for hidden symmetries of a given supersymmetric field theory often leads to insight of the structure of the theory and may even provide selection rules for operators of high mass dimension. For example, seeking unexpected algebro-geometric signatures of the (supersymmetric) standard model is the subject of \cite{Gray:2006jb}. For D-brane quiver theories, an underlying symmetry is ever-present: the symmetry of the Calabi-Yau space $\cX$ is visible in the UV as a global flavor symmetry in the Lagrangian, while the symmetry of the full moduli space $\CM$ can reveal a new {\bf hidden global symmetry} which develops as the theory flows to the IR. 

In particular, in \cite{Franco:2004rt}, the basic fields of the quiver for $dP_n$ theories were reordered into multiplets of a proposed $E_n$ symmetry and consequently the superpotential terms into singlets of this symmetry. The exceptional Lie group $E_n$ acts geometrically on the divisors of the del Pezzo 
surfaces and is realized in the quantum field theory as a hidden symmetry
enhancing the $n$ non-anomalous baryonic $U(1)$'s of the $dP_n$ quiver.

Do the symmetries of the master space, whose geometrical significance we have learned to appreciate in the foregoing discussions, manifest themselves in the full moduli space $\CM$ of the gauge theory? Phrased another way, do these symmetries survive the symplectic quotient of \eref{M} and manifest themselves also
at finite $N$? This is indubitably a natural question. 

In this section, 
equipped with the new notion of the master space we can revisit this problem and recast all operators into irreducible representations of the symmetry of $\f$.
We will show that these symmetries are encoded in a subtle and beautiful way by the fundamental invariant of the plethystic programme \cite{pleth} for $\CM$, viz., the Hilbert series.
Moreover, we will demonstrate many other examples
of hidden symmetries in orbifolds and toric quivers, enhancing anomalous and 
non-anomalous Abelian symmetries.

Since we will always deal with the coherent component of the moduli space in the following and there is no source for ambiguity, we will adopt a simplified notation for the Hilbert series which was already used in \cite{Butti:2007jv}:
\begin{equation}
g_1(t;\cX) \equiv H(t; \firr{\cX})
\end{equation}

\subsection{Character Expansion: A Warm-up with $\mathbb{C}^3$}
The ${\cal N}=4$ supersymmetric gauge theory does not have a baryonic branch and therefore the master space $\f$ coincides with the Calabi-Yau manifold $\cX = \mathbb{C}^3$.
We therefore do not expect any new symmetries but instead can use this example as a warm-up example for expanding in terms of characters of the global symmetry. For a single D3-brane, $\f \simeq \CM \simeq \cX \simeq \IC^3$ and there is a $U(3)$ symmetry.
Now, there is an $SU(4)_R$ symmetry for which $U(3)$ is a maximal subgroup, however, we shall see below that the slightly smaller $U(3)$ suffices to keep the structure of the BPS operators.

The generating function for $\mathbb{C}^3$ is well known and was computed in various places (cf.~e.g.~\cite{pleth}). It takes the form
\begin{equation}\label{C3}
g(\nu; t_1, t_2, t_3; \mathbb{C}^3) = \prod_{n_1=0}^\infty \prod_{n_2=0}^\infty \prod_{n_3=0}^\infty \frac{1}{1-\nu t_1^{n_1} t_2^{n_2} t_3^{n_3}} \ ,
\end{equation}
which coincides with the grand canonical partition function of the three dimensional harmonic oscillator.
This form is perhaps the simplest one can write down for the exact answer and from this extract the generating function for any fixed $N$.
In this subsection, we will rewrite it in terms of characters of the $U(3)$ global symmetry.
The expansion demonstrates how one can explicitly represent this function in terms of characters.
This will help in analyzing the next few examples in which expansion in terms of characters are done but for more complicated cases.

Equation (\ref{C3}) admits a plethystic exponential form,
\begin{equation}\label{C3PE}
g(\nu; t_1, t_2, t_3; \mathbb{C}^3) = PE [ \nu g_1 ] , \quad 
g_1(t_1, t_2, t_3; \mathbb{C}^3) = \frac{1}{(1-t_1)(1-t_2)(1-t_3)} = 
PE[t_1 + t_2 + t_3] \ .
\end{equation}
We recall that $g_1$, the generating function for a single D3-brane, $N=1$, is none other than the refined Hilbert series for $\IC^3$, itself being the $PE$ of the defining equations (syzygies) for $\IC^3$, here just the 3 variables. Furthermore, $PE [ \nu  g_1 ]$ encodes all the generators for the multi-trace operators (symmetric product) at general number $N$ of D-branes.

We can now introduce $SU(3)$ weights $f_1, f_2$ which reflect the fact that the chemical potentials $t_1, t_2, t_3$ are in the fundamental representation of $SU(3)$, and a chemical potential $t$ for the $U(1)_R$ charge,
\begin{equation}
(t_1, t_2, t_3) = t \left (f_1, \frac{f_2}{f_1}, \frac{1}{f_2} \right ) \ .
\end{equation}
We can define the character of the fundamental representation with the symbol
\begin{equation}
[1, 0 ] = f_1+ \frac{f_2}{f_1}+ \frac{1}{f_2} \ ,
\end{equation}
and get
\begin{equation}
g_1(t_1, t_2, t_3; \mathbb{C}^3) = PE \left [ [1, 0] t  \right ] = \sum_{n=0}^\infty [n,0] t^n \ .
\end{equation}
The second equality follows from the basic property of the plethystic exponential which produces all possible symmetric products of the function on which it acts. The full generating function is now rewritten as
\begin{equation}\label{C3PE-1}
g(\nu; t_1, t_2, t_3; \mathbb{C}^3) = PE \left [ \nu \sum_{n=0}^\infty [n,0] t^n \right ] ,
\end{equation}
giving an explicit representation as characters of $SU(3)$.
We can expand $g(\nu; t_1, t_2, t_3; \mathbb{C}^3) = \sum\limits_{N=0}^\infty g_N(t_1, t_2, t_3; \mathbb{C}^3) \nu^N$ and find, for example,
\begin{equation}\label{C3N2}
g_2(t_1, t_2, t_3; \mathbb{C}^3) = \left ( 1 - [0,2] t^4 + [1,1] t^6 - [0,1] t^8 \right ) PE \left [ [1,0] t + [2,0] t^2 \right ] \ .
\end{equation}
Alternatively we can write down an explicit power expansion for $g_2$ as
\begin{equation}\label{C3N2ser}
g_2(t_1, t_2, t_3; \mathbb{C}^3) = \sum_{n=0}^\infty \sum_{k=0}^{\lfloor \frac{n}{4}\rfloor + \lfloor \frac{n+1}{4}\rfloor} m(n,k)[n-2k, k] t^{n} , 
\qquad
m(n,k) 
= 
\left\{ \ba{lcl}
\lfloor \frac{n}{2}\rfloor -k+1 && n \mbox{ odd} \\ 
\lfloor \frac{n}{2}\rfloor - 2 \lfloor \frac{k+1}{2}\rfloor +1 &&
n \mbox{ even} 
\ea
\right. 
\end{equation}

Note that $g_2$ is not palindromic, indicating that the moduli space of 2 D-branes on $\mathbb{C}^3$ is not Calabi-Yau \footnote{We thank David Berenstein for an enlightening discussion on this point.}. 
Armed with this character expansion let us now turn to more involved cases where there is a baryonic branch.

\subsection{Conifold Revisited}\label{s:coni}
Having warmed up with $\IC^3$, let us
begin with our most familiar example, the conifold $\cX = \cC$.
The master space for the conifold is simply $\f = \mathbb{C}^4$ \cite{Butti:2006au,Forcella:2007wk}. The symmetry of this space is $SU(4)\times U(1)$ where the $U(1)$ is the R-symmetry while the $SU(4)$ symmetry is not visible at the level of the Lagrangian and therefore will be called ``hidden''. One should stress that at the IR the two $U(1)$ gauge fields become free and decouple and one is left with 4 non-interacting fields, which obviously transform as fundamental representation of this $SU(4)$ global symmetry. Now, there is a mesonic $SU(2)\times SU(2)$ symmetry and a baryonic $U(1)_B$ symmetry, this $SU(4)$ {\bf Hidden Symmetry} is an enhancement of both. 

\begin{figure}[t]\begin{center}
$\begin{array}{cc}
\begin{array}{c}
\epsfxsize = 8cm \epsfbox{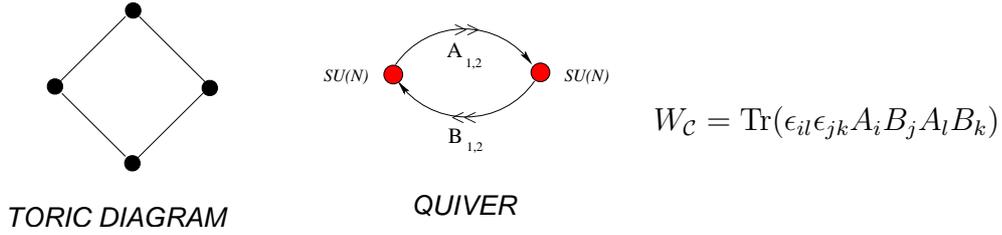}
\end{array}
&
W_{\cC} =  \tr(\epsilon_{il} \epsilon_{jk} A_i B_j A_l B_k)
\end{array}$
\caption{{\sf The quiver and toric diagrams, as well as the
    superpotential for the conifold $\cC$.}}
\label{f:coni}
\end{center}\end{figure}
To start let us recall the theory in \fref{f:coni}.
Indeed, we see that when the number of branes $N=1$, we have a $U(1)^2$ theory with $W = 0$. The vanishing of the superpotential in this case means that we have four free variables $A_{1,2}$ and $B_{1,2}$ and the master space should be $\IC^4$. The gauge theory has an explicit global symmetry $SU(2)_1\times SU(2)_2\times U(1)_R\times U(1)_B$ and the four fields transform under these symmetries according to \fref{weightsConi}.
\begin{table}[t]\begin{center}
\begin{tabular}{|c||c|c|c|c||c|}
\hline
 & $SU(2)_1$ $(j_1,m_1)$& $SU(2)_2$ $(j_2,m_2)$& $U(1)_R$ & $U(1)_B$ & monomial\\ \hline \hline
$A_1$ & $(\frac{1}{2}, +\frac{1}{2})$ & $(0,0)$ & $\frac{1}{2}$ & 1& $t_1 x$\\ \hline
$A_2$ & $(\frac{1}{2},-\frac{1}{2})$ & $(0,0)$ & $\frac{1}{2}$ & 1& $\frac{t_1}{x}$ \\ \hline
$B_1$ & $(0,0)$& $(\frac{1}{2},+\frac{1}{2})$ & $\frac{1}{2}$ & -1& $t_2 y$ \\ \hline
$B_2$ & $(0,0)$ & $(\frac{1}{2},-\frac{1}{2})$ & $\frac{1}{2}$ & -1& $\frac{t_2}{y}$ \\ \hline
\end{tabular}
\caption{{\sf The transformation, under the explicit global symmetry group 
$SU(2)_1\times SU(2)_2\times U(1)_R\times U(1)_B$, of the 4 fields in the conifold theory. The monomials indicate the associated chemical potentials in the Plethystic programme.}}
\label{weightsConi}
\end{center}\end{table}
We have marked the monomials for the counting of baryonic operators whose
generating function for $N=1$, in the notation of the plethystic programme, was given in Equation (3.3) of \cite{Forcella:2007wk}; this is also simply the refined Hilbert series for the master space $\f_{\cC}$:
\begin{equation}\label{g1coni}
\ba{rcl}
g_1(t_1,t_2,x,y; {\cal C}) &=& H(t_1,t_2,x,y; \f_{\cC} = \IC^4)
= \frac{1}{(1-t_1 x) (1-\frac{t_1}{x})(1-t_2 y) (1-\frac{t_2}{y})} \\
&=& PE[t_1 x + \frac{t_1}{x} + t_2 y + \frac{t_2}{y}] \ .
\ea
\end{equation}
In the above, we recall from \sref{s:plet} that the Hilbert series is itself the PE of the defining equations. Moreover, as in \cite{Forcella:2007wk} we can define $b$ which counts (i.e., is the chemical potential associated to) baryon number and $t$ which counts the total R-charge; then $t_1 = t b$ and $(t_2 = t/b)$ would respectively count the number of $A$ and $B$ fields appearing in the baryonic operator. Furthermore, $x$ and $y$ keep track of the first and second $SU(2)$ weights respectively. Indeed, if we unrefined by setting $t_1=t,t_2=t,x=1,y=1$, we would obtain the familiar Hilbert series for $\IC^4$ which is $g_1(t;~\IC^4) = (1-t)^{-4}$.

Now, the fields in \eref{weightsConi} are in the representations of the explicit $SU(2)\times SU(2)\times U(1)_R \times U(1)_B$ symmetry, while our hidden symmetry is $SU(4) \times U(1)$. We will therefore introduce $SU(4)$ weights, $h_1, h_2, h_3$ and map them to the three weights of the $SU(2)\times SU(2)\times U(1)_B$ global symmetry. This is done simply by first taking the four weights of the fundamental representation of $SU(4)$, whose character we will denote as $[1,0,0]$, and which we recall can be written in terms of the weights as
\begin{equation}
[1,0,0] = h_1 + \frac{h_2}{h_1} + \frac{h_3}{h_2} + \frac{1}{h_3} \ .
\end{equation}
Then we multiply by $t$ to obtain a $SU(4) \times U(1)$ representation, which should be mapped them to the four weights of $SU(2)\times SU(2)\times U(1)_R \times U(1)_B$ in the rightmost column of \eref{weightsConi} above:
\begin{equation}
t \left ( h_1, \frac{h_2}{h_1}, \frac{h_3}{h_2}, \frac{1}{h_3} \right ) = \left (t_1 x, \frac{t_1}{x}, t_2 y, \frac{t_2}{y} \right ) \ .
\end{equation}
This has a solution $t = \sqrt{t_1 t_2}, h_1 = b x, h_2 = b^2, h_3 = b y$.

In analogy with \eref{C3PE}, we can now write \eref{g1coni} in terms of a plethystic exponential:
\begin{equation}\label{g1conif}
g_1(t, h_1, h_2, h_3; {\cal C}) = PE \left [ [1,0,0] t \right ] = \sum_{n=0}^\infty [n, 0, 0] t^n \ ,
\end{equation}
where $[n, 0, 0]$ is the completely symmetric tensor of rank $n$ and dimension $n+3 \choose 3$. The first equality writes the 4 generators of $\IC^4$, viz., $t_1 x, \frac{t_1}{x}, t_2 y, \frac{t_2}{y}$, in the weights of $SU(4) \times U(1)$ and the second equality follows from the definition of $PE$ in \eref{defPE} and the fact that expansion of the plethystic exponential in power series of $t$ will compose the $n$-th symmetrized product. Equation \eref{g1conif} is a trivial and obvious demonstration that the $N=1$ generating function is decomposed into irreducible representations of $SU(4)$, precisely one copy of the irreducible representation $[n, 0, 0]$ at R-charge $n$. To be precise we are taking here the R-charge to be $n$ times that of the basic field $A$.

\subsubsection{Hidden Symmetries for Higher Plethystics}
We have now seen that the basic invariant, i.e., $g_1$, the Hilbert series, of $\f_{\cC}$, can be written explicitly as the plethystic exponential of the fundamental representation of $SU(4)$. Now, the hidden symmetry mixes baryon number with meson number and therefore we do not expect this symmetry to hold for general number $N$ of D3-branes. For the case of $N=2$, however, baryons and mesons have the same R-charge and therefore we may expect the global symmetry to be enhanced. 

Actually for $N=2$, $SU(4)$ becomes a symmetry of the Lagrangian of the conifold theory, which is an $SO(4)$ theory with four flavors in the vector
representation and an $SU(4)$ invariant superpotential.\footnote{We thank Vadim Kaplunovsky for a discussion on this point.} Writing the 4 flavors in the vector representation of $SO(4)$ as a $4\times4$ matrix $Q$ we find that the superpotential is $W = \det Q $. Using the formalism which is developed to count gauge invariant operators for the conifold \cite{Forcella:2007wk} we find that the $N=2$ generating function does indeed decompose into characters of irreducible representations of $SU(4)$.

The generating function for $N=2$ was computed in Equation (3.47) and its predecessor in \cite{Forcella:2007wk}. The expression is given as a function of all 4 chemical potentials and is quite lengthy. Here, we will take this expression and recast it into characters of the global symmetry $SU(4)$. The first point to note is that the generators form the $[2, 0, 0]$ representation of $SU(4)$. It is natural to expect this since this representation is the second rank symmetric product of the generators for $N=1$. The other terms are less obvious and need explicit computation. A short computation yields
\begin{equation}\label{g2coni}
g_2(t, h_1, h_2, h_3; \cC) = \left (1 - [0,0,2] t^6 + [1,0,1] t^8 - [0,1,0] t^{10} \right ) PE \left [ [2,0,0] t^2 \right ] \ .
\end{equation}
This can be seen if we write the explicit expressions for the characters of the irreducible representations \cite{rep} of $SU(4)$ which we can write in terms of the weights as
\begin{equation}\ba{rcl}
[2,0,0] &=& h_1^2+\frac{h_3
   h_1}{h_2}+\frac{h_1}{h_3}+\frac
   {h_3^2}{h_2^2}+h_2+\frac{1}{h_2
   }+\frac{1}{h_3^2}+\frac{h_3}{h_1}+\frac{h_2}{h_3
   h_1}+\frac{h_2^2}{h_1^2}  \\
\left [0,0,2\right ] &=& \frac{h_1^2}{h_2^2}+\frac{h_3
   h_1}{h_2}+\frac{h_1}{h_3}+h_3^2+h_2+\frac{1}{h_2}+\frac{h_2^2}
   {h_3^2}+\frac{h_3}{h_1}+\frac{h_2}{h_3 h_1}+\frac{1}{h_1^2}  \\
\left [1,0,1 \right ] &=& \frac{h_1^2}{h_2}+\frac{h_3
   h_1}{h_2^2}+h_3
   h_1+\frac{h_2
   h_1}{h_3}+\frac{h_1}{h_2
   h_3}+\frac{h_3^2}{h_2}+\frac{h_2}{h_3^2}+3+\frac{h_2
   h_3}{h_1}
   +\frac{h_3}{h_2
   h_1}+\frac{h_2^2}{h_3
   h_1}+\frac{1}{h_3
   h_1}+\frac{h_2}{h_1^2}  \\
\left [0,1,0 \right ] &=& \frac{h_3
   h_1}{h_2}+\frac{h_1}{h_3}+h_2+\frac{1}{h_2}+\frac{h_3}{h_1}+\frac{h_2}{h_3 h_1} \ .
\ea
\end{equation}
As a check, \eref{g2coni} can be expanded to first few orders in $t$,
\[\ba{rcl}
g_2(t, h_1, h_2, h_3; \cal C) &=&
1 + [2,0,0] t^2 + \left ( [4, 0, 0] + [0,2,0] \right ) t^4
+ \left ( [6,0,0] + [2,2,0] \right ) t^6  \\
&&+ \left ( [8,0,0] + [4,2,0] + [0,4,0] \right ) t^8 
+ \left ( [10,0,0] + [6,2,0] + [2,4,0] \right ) t^{10} + \ldots
\ea\]
and in series form it is
\begin{equation}
g_2(t, h_1, h_2, h_3; {\cal C}) = \sum_{n=0}^\infty \left ( \sum_{k=0}^{\lfloor \frac {n}{2} \rfloor } [2n -4k, 2k, 0] \right ) t^{ 2 n } .
\end{equation}

Now, using the formula for the dimension of a generic irreducible representation of $SU(4)$,
\begin{equation}
{\rm dim} [n_1, n_2, n_3] = \frac{ (n_1+n_2+n_3+3) (n_1+n_2+2) (n_2+n_3+2) (n_1+1) (n_2+1) (n_3+1) } {12} ,
\end{equation}
we find that the unrefined generating function for $N=2$ sums to
\begin{equation}
g_2(t, 1, 1, 1; {\cal C}) = \frac{ 1 + 3 t^2 +6 t^4 } { (1-t^2 )^7} ,
\end{equation}
as expected and in agreement with Equation (3.47) of \cite{Forcella:2007wk}. Taking the Plethystic Logarithm we find that the 7 dimensional manifold is generated by 10 operators of order 2 transforming in the $[2, 0, 0]$ representation of $SU(4)$ subject to 10 cubic relations transforming in the $[0, 0, 2]$ representation of $SU(4)$. Since $g_2$ is not palindromic we expect the moduli space of 2 D-branes on the conifold to be not Calabi-Yau.

This confirms the expansion in terms of characters of $SU(4)$ for the case of $N=2$. Unfortunately this symmetry does not extend to $N=3$ and is not a symmetry for higher values of $N$. In the next example we are going to see how a hidden symmetry extends to all values of $N$, simply because the hidden symmetry does not mix baryonic numbers with mesonic numbers as it does for the conifold. The symmetry structure then persists to all orders in $N$.

\subsection{$F_0$ Revisited}\label{s:F0rev}
Let us move on to the $F_0$ theory and focus on the first toric phase, whose master space we studied in \sref{s:F0}. We recall that for a single D3-brane, $N=1$, it is a six dimensional reducible variety composed by a set of coordinate planes and an irreducible six dimensional Calabi-Yau piece. This top piece, being toric, admits a symmetry group which is at least $U(1)^6$. Re-examining \eref{FF0} we see that it is actually the set product of two three dimensional conifold singularities:
\begin{equation}\label{dopcon}
B_2 D_1 - B_1 D_2 = 0 \ , \  A_2 C_1 - A_1 C_2 = 0 \ ;
\end{equation}
hence the group of symmetries is $SU(2)^4 \times U(1)^2$, twice of that in \sref{s:coni}. The first $SU(2)^2$ is the non-Abelian symmetry group of the variety $F_0$ and the second $SU(2)^2$ is a hidden symmetry related to the two anomalous baryonic symmetries of the gauge theory. The chiral spectrum of the theory is summarized by the Hilbert series:
\begin{equation}\label{g1f0}
g_1(t_1,t_2; F_0) = \frac {(1 - t_1^2)(1-t_2^2)}{(1-t_1)^4(1-t_2)^4} 
= PE[4 t_1 - t_1^2 + 4 t_2 - t_2^2]
\ ,
\end{equation}
where the chemical potential $t_1$ counts the fields $A_i,C_j$ while $t_2$ counts the fields $B_i,D_j$. 

Let us define the representation $[n]\times[m]\times[p]\times[q]=[n,m,p,q]$ for the $SU(2)^4$ group. Equation \eref{dopcon} and Table 3 of \cite{Butti:2007jv} then implies that $A,C$ are in the $[1,0,1,0]$ representation and $B,D$ are in the $[0,1,0,1]$ representation. The refinement for the Hilbert series (\ref{g1f0}) is then
\begin{equation}\label{g1f0s}
g_1(t_1,t_2,x,y,a_1,a_2; F_0) = (1 - t_1^2)(1-t_2^2) PE \left [ [1,0,1,0]t_1 + [0,1,0,1] t_2 \right ] .
\end{equation}
Amazingly, the group $SU(2)^4 \times U(1)^2$ is the symmetry for the chiral ring for generic $N$, not just for $N=1$, and hence the moduli space of the non-Abelian theory on $N$ D3-branes has this group as a symmetry group. 

The generating function for finite $N$ can be computed using the plethystic
exponential in each sector of the GKZ decomposition of the $N=1$ partition
function. The reader is referred to \cite{Butti:2007jv} for details and
to \sref{s:plet} for a short account of the general philosophy.
In particular, the implementation of the plethystic program goes through the
formulae \eref{GKZ1} and \eref{GKZN} and requires the computation of
the generating functions for fixed K\"ahler moduli and the auxiliary partition
function.

Recall from \cite{Butti:2007jv} that the generating function for fixed integral K\"ahler moduli, which in this case can be parameterized by two integers  $\beta$, $\beta'$,  is equal to
\begin{equation}
g_{1, \beta, \beta ' } (t_1,t_2; F_0) = \sum_{n=0}^\infty (2 n + \beta +1 )(2 n + \beta ' + 1 ) t_1^{2 n + \beta}t_2^{2 n + \beta '} \ .
\end{equation}
This can be easily refined in terms of representations of the global symmetry as
\begin{equation}\label{g1betadP0}
g_{1, \beta, \beta '} \left (t_1,t_2, x, y, a_1, a_2; F_0 \right ) = \sum_{n=0}^\infty [2 n + \beta, 2 n + \beta ', 0, 0] t_1^{2 n + \beta}t_2^{2 n + \beta '}
\ .
\end{equation}
The auxiliary partition function in Equation (5.56) of \cite{Butti:2007jv} also admits an expression in terms of representations of the global symmetry ,
\begin{equation}
Z_{\rm aux}(t_1, t_2, a_1, a_2; F_0 ) = (1 - t_1^2 t_2^2) PE \left [ [0, 0, 1, 0] t_1 + [0, 0, 0, 1] t_2 \right ] ,
\end{equation}
which has an expansion as
\begin{equation}
Z_{\rm aux}(t_1, t_2, a_1, a_2; F_0 ) = \sum_{\beta, \beta ' = 0}^\infty [0, 0, \beta , \beta ' ] t_1 ^\beta t_2^{\beta '} - \sum_{ \beta, \beta' = 2 }^\infty  [0, 0, \beta -2 , \beta' - 2 ] t_1^\beta t_2^{\beta ' } \ .
\end{equation}

Recalling from \sref{s:plet}, once we have $g_{1, \beta, \beta '}$ and $Z_{\rm aux}$ we can do the replacement $t_1^\beta t_2^{\beta'}$ in $Z_{\rm aux}$ by $g_{1,\beta,\beta'}$ (as done in line 2 below and using the fact that by our definition $[a,b,0,0] \times [0,0,c,d] = [a,b,c,d]$) to obtain the generating function for a single brane:
\begin{equation}
\hspace{-0.5in}
\ba{rcl}
&&g_{1} \left (t_1,t_2, x, y, a_1, a_2; F_0 \right )\\
&=& \sum\limits_{\beta, \beta' = 0}^\infty \sum\limits_{n=0}^\infty [2 n +\beta, 2n + \beta', \beta, \beta' ] t_1^{2 n + \beta} t_2^{2 n + \beta'} - \sum\limits_{\beta, \beta' = 2}^\infty \sum\limits_{n=0}^\infty [2 n + \beta,2n + \beta', \beta -2 , \beta' -2 ] t_1^{2 n + \beta}t_2^{2 n + \beta'}   \\
&=& \sum\limits_{\beta, \beta' = 0}^\infty \sum\limits_{n=0}^\infty [2 n + \beta, 2n + \beta', \beta, \beta' ] t_1^{2 n + \beta}t_2^{2n+\beta'} - \sum\limits_{\beta, \beta' = 0}^\infty \sum\limits_{n=1}^\infty [2 n + \beta, 2n + \beta', \beta, \beta' ] t_1^{2 n + \beta}t_2^{2n+ \beta'} \\
&=&  \sum\limits_{\beta, \beta' = 0}^\infty [\beta, \beta' , \beta, \beta' ] t_1^{\beta} t_2^{\beta'} ,
\ea\label{charg1F0}\end{equation}
leaving only positive coefficients and meaning that $SU(2)^4 \times U(1)^2$ is indeed a symmetry of the $N=1$ moduli space. The second equality follows by shifting $n$ by $-1$ and the $\beta$'s by $2$. The third equality is the remaining $n=0$ term from both contributions. It is important to note that Equation \eref{charg1F0} factorizes into two conifold generating functions,
\begin{equation}\ba{rcl}
g_{1} \left (t_1,t_2, x, y, a_1, a_2; F_0 \right ) &=& \left ( \sum\limits_{\beta = 0}^\infty [\beta, 0 , \beta, 0 ] t_1^{\beta} \right ) \left ( \sum\limits_{\beta' = 0}^\infty [0, \beta' , 0, \beta' ] t_2^{\beta'} \right ) \\
&=& \left( (1-t_1^2) \frac{}{} PE[ [1,0,1,0] t_1] \right)
\left( (1-t_2^2) \frac{}{} PE[ [0,1,0,1] t_2] \right) ,
\ea\end{equation}
which can be easily checked to equal Equation \eref{g1f0s}.

Next, using \eref{GKZN}, we can write a generic expression for any $N$.
\begin{equation}\ba{rcl}
g \left ( \nu; t_1,t_2, x, y, a_1, a_2; F_0 \right ) &= &
\sum\limits_{\beta,\beta' = 0}^\infty [0, 0, \beta, \beta'] \left ( PE \left [ \nu \sum\limits_{n=0}^\infty [2 n + \beta, 2 n + \beta', 0, 0 ] t_1^{2 n + \beta} t_2^{2 n +\beta'} \right ] - \right.\\
&& \left.
 - PE \left [ \nu \sum\limits_{n=1}^\infty [2 n + \beta, 2n +\beta', 0, 0] t_1^{2 n +  \beta} t_2^{2 n + \beta'} \right ] \right ) .
\ea\end{equation}
Note that the first $PE$ contains all the terms in the second $PE$ and hence all the coefficients in the expansion are positive. This is the explicit demonstration that for generic $N$ the chiral spectrum organizes into representations of $SU(2)^4 \times U(1)^2$ and hence the moduli space of the non-Abelian theory with generic rank $N$ has symmetry $SU(2)^4 \times U(1)^2$.

\subsection{$dP_0$ Revisited}
The master space for $dP_0$ is calculated above in \sref{s:toric} and is found to be irreducible.
\comment{
In fact, it can also be computed using dimer technology. 
Now, for the case of $dP_0$ the matrix $P$, defined in \eref{Pmat}, takes the form
{\tiny
\[ P= \left(
\begin{array}{llllll}
 1 & 0 & 0 & 1 & 0 & 0 \\
 0 & 1 & 0 & 1 & 0 & 0 \\
 0 & 0 & 1 & 1 & 0 & 0 \\
 1 & 0 & 0 & 0 & 1 & 0 \\
 0 & 1 & 0 & 0 & 1 & 0 \\
 0 & 0 & 1 & 0 & 1 & 0 \\
 1 & 0 & 0 & 0 & 0 & 1 \\
 0 & 1 & 0 & 0 & 0 & 1 \\
 0 & 0 & 1 & 0 & 0 & 1
\end{array}
\right) . \] 
}
The rank of this matrix should be the dimension of the master space (as encode in the cone $K$), i.e., it should be $G+2$ where $G$ is the number of gauge groups for this model. Hence, the rank is 5 and since there are 6 columns we expect a 1 dimensional kernel for $P$. This kernel can be easily computed to be the vector $Q$,
\begin{equation}
P \cdot Q = 0 \qquad \Rightarrow \qquad Q^t = \left(
\begin{array}{llllll}
 1 & 1 & 1 & -1 & -1 & -1
\end{array}
\right) ,
\end{equation}
which forms the vector of charges for the linear sigma model description of the master space for $dP_0$. In this description, therefore, we find that the master space is 
}
Its symplectic quotient description is
$\IC^6//\{-1,-1,-1,1,1,1\}$.
We note that the sum of charges is zero, implying that this 5-dimensional variety is Calabi-Yau, in agreement with \eref{FdP0}. This space is a natural generalization of the conifold and has the description of a cone over a 9 real dimensional Sasaki Einstein manifold given by a circle bundle over $\IP^2\times \IP^2$. This structure reveals a symmetry of the form $SU(3)\times SU(3)\times U(1)$. 

Indeed, we note that this construction is toric and therefore we would expect a symmetry which is at least $U(1)^5$ since the master space has dimension 5. However, due to the special symmetries of this space the symmetry is larger. The $U(1)$ is the R-symmetry and the first $SU(3)$ is the natural one acting on the mesonic moduli space $\cX = dP_0 = \IC^3 / \IZ_3$. The second $SU(3)$ symmetry is a ``hidden'' symmetry, and is related to the two anomalous baryonic $U(1)$ symmetries that play a r\^{o}le as the Cartan subgroup of this symmetry. We can use the full symmetry to compute the refined Hilbert series for this space. 



The Hilbert series for just one charge was computed with the Molien formula in 
\eref{HSdP0}
\begin{equation}
g_1(t; dP_0) = \oint \frac{dw}{ 2 \pi i w(1-t/w)^3 (1-w)^3} = \frac {1+4t+t^2}{(1-t)^5} .
\label{g1dP0sec}
\end{equation}
Taking the plethystic logarithm of this expression we find 9 generators at order $t$ subject to 9 relations at order $t^2$,
\begin{equation}
PE^{-1}[ g_1(t; dP_0) ] = 9 t - 9 t^2 + \dots
\end{equation}
This agrees exactly with the content of \eref{FdP0} which says that $\f_{dP_0}$ should be the incomplete (since the plethystic logarithm does not terminate) intersection of 9 quadrics in 9 variables.

Now, we would like to refine the Hilbert series to include all the 5 global charges. This can be done using the Molien formula or any other of the methods discussed in section 2. Here we find a shorter way of determining it.
To do this we recognize the 9 quiver fields as transforming in the $[1,0]\times [0,1]$ representation of $SU(3)\times SU(3)$. For short we will write an irreducible representation of this group as a collection of 4 non-negative integer numbers, here $[1,0,0,1]$ and with obvious extension to other representations. These are all the generators of the variety \eref{MSdP0}. The relations are derived from a superpotential of weight $t^3$ so we expect 9 relations at order $t^2$ transforming in the conjugate representation to the generators, $[0,1,1,0]$. To get this into effect we rewrite \eref{g1dP0sec} into a form which allows generalization to include characters (multiplying top and bottom by $(1-t)^4$):
\begin{equation}
g_1(t; dP_0) = \frac {1 - 9 t^2 + 16 t^3 - 9 t^4 + t^6 }{(1-t)^9} \ .
\end{equation}

The coefficient of the $t^2$ in the numerator can now be identified with the 9 relations, whose transformation rules are already determined to be $[0,1,1,0]$. Being irreducible we expect the Hilbert series to be palindromic. This gives as the $t^4$ term as $[1,0,0,1]$. The same property implies that the coefficient of the $t^3$ term is self conjugate and hence uniquely becomes the adjoint representation for $SU(3)\times SU(3)$, $[1,1,0,0] + [0,0,1,1]$. Finally, the denominator can be simply expressed as a plethystic exponential of the representation for the generators, $[1,0,0,1]$. In other words, $(1-t)^{-9} = PE[9 t]$. In summary, we end up with the refinement of the Hilbert series for $\f_{dP_0}$ as:
\begin{equation}\label{refg1dP0}
\ba{rcl}
&&g_1(t, f_1, f_2, a_1, a_2; dP_0) \\
&=& \left(1 - [0,1,1,0] t^2 + ( [1,1,0,0] + [0,0,1,1] ) t^3 
 - [1,0,0,1] t^4 + t^6 \right) PE \left [ [1,0,0,1] t \right ] \ .
\ea\end{equation}

For completeness, we list here the explicit expressions for the characters of the representations, using weights $f_1, f_2$ for the first, mesonic $SU(3)$ and $a_1, a_2$ for the second, hidden $SU(3)$:
\begin{equation}\ba{lll}
[1, 0, 0, 1] &=& \left ( f_1 + \frac{f_2}{f_1} + \frac{1}{f_2} \right ) \left ( \frac{1}{a_1} + \frac{a_1}{a_2} + a_2 \right ),  \\
\left [ 0, 1, 1, 0 \right ] &=& \left ( \frac{1}{f_1} + \frac{f_1}{f_2} + f_2 \right ) \left ( a_1 + \frac{a_2}{a_1} + \frac{1}{a_2} \right ) ,  \\
\left [ 1, 1, 0, 0 \right ] &=& \frac{f_1^2}{f_2}+f_1 f_2+\frac{f_1}{f_2^2}+2+\frac{f_2^2}{f_1}+\frac{1}{f_1 f_2}+\frac{f_2}{f_1^2} ,  \\
\left [ 0, 0, 1, 1 \right ] &=& \frac{a_1^2}{a_2}+a_1 a_2+\frac{a_1}{a_2^2}+2+\frac{a_2^2}{a_1}+\frac{1}{a_1 a_2}+\frac{a_2}{a_1^2} .
\ea\eeq
In terms of the above weights, being generated by the representation $[1, 0, 0, 1]$, the Hilbert series, \eref{refg1dP0}, for the case of $N=1$ D3-brane admits a simple and natural series expansion of the form
\begin{equation}
\label{g1dP0-2}
g_1 \left (t, f_1, f_2, a_1, a_2; dP_0 \right ) = \sum_{n=0}^\infty \left [n, 0, 0, n \right ] t^{ n } \ .
\end{equation}

\subsubsection{Higher Number of Branes}
Using the formalism of \cite{Butti:2007jv} explained in the previous subsection, we can compute the $N=2$ generating function in terms of characters of the global symmetry. The computation is somewhat lengthy but the result is relatively simple:
\beq\label{g2dP0}
\ba{rcl}
g_2 \left (t, f_1, f_2, a_1, a_2;~dP_0 \right ) &=& \sum\limits_{n=0}^\infty \sum\limits_{k=0}^{\lfloor \frac {n} {2} \rfloor }\left [2 n - 4 k, 2k, 0, n \right ] t^{ 2 n } +\\
&& + \sum\limits_{n_2=0}^\infty \sum\limits_{n_3=1}^\infty \sum\limits_{k=0}^{n_2} \left [2 n_2 + 3n_3 - 2 k, k, 0, n_2 \right ] t^{ 2 n_2 + 3 n_3 } ,
\ea\eeq
with coefficient 1 for each representation which appears in the expansion. It is important to identify the generators of this expression and a quick computation reveals the order $t^2$ and order $t^3$ generators to be the representations, $[2, 0, 0, 1]$, and $[3, 0, 0, 0]$, respectively. We can therefore sum the series and obtain
\begin{equation}
g_2 \left (t, f_1, f_2, a_1, a_2;~dP_0 \right ) =
A_2 PE \left [ \left [2, 0, 0 , 1 \right ] t^{ 2 } + \left [ 3, 0, 0, 0 \right ] t^{ 3} \right ] ,
\end{equation}
where $A_2$ is a complicated polynomial of order 58 in $t$ which has the first 10 terms:
\beq\ba{lll}
A_2 \left (t, f_1, f_2, a_1, a_2\right )
&=& 1- [2, 1, 1, 0] t^4 - [1, 2, 0, 1] t^5 + \\
&+& ( [4, 1, 0, 0] + [1, 1, 0, 0] + [3, 0, 1, 1] + [0, 3, 1, 1] - [0, 0, 0, 3] ) t^6  \\
&+& ([3, 2, 1, 0] + [2, 1, 1, 0] + [1, 3, 1, 0] + [1, 0, 1, 0] + [0, 2, 0, 2] + [0, 2, 1, 0]) t^7  \\
&-& ([5, 0, 0, 1] - [2, 0, 1, 2] - [2, 0, 0, 1] + [1, 2, 1, 2]) t^8  \\
&-& \left ([5, 2, 0, 0] + [3, 3, 0, 0] + [2, 2, 0, 0] + [1, 1, 0, 0] + [0, 3, 0, 0] + 2 [3, 0, 0, 0] \right .  \\
&& - \left . \,\, [3, 0, 0, 3] + [0, 0, 0, 3] + [2, 2, 1, 1] + [1, 1, 1, 1] + [0, 0, 1, 1] \right ) t^9 + \ldots
\ea\eeq

Similarly, we can obtain generic expressions for any $N$.
Recall from \cite{Butti:2007jv} that the generating function for a fixed integral K\"ahler modulus $\beta$ is equal to
\begin{equation}
g_{1, \beta} (t) = \sum_{n=0}^\infty {3 n + \beta + 2 \choose 2} t^{3 n + \beta} ;
\end{equation}
this can be easily written in terms of representations of the global symmetry as\begin{equation}
g_{1, \beta} \left (t, f_1, f_2, a_1, a_2; dP_0 \right ) = \sum_{n=0}^\infty [3 n + \beta, 0, 0, 0] t^{3 n + \beta} \ .
\end{equation}

The auxiliary partition function also admits an expression in terms of representations of the global symmetry:
\begin{equation}
Z_{\rm aux}(t, a_1, a_2; dP_0 ) = (1 - t^3) PE \left [ [0, 0, 0, 1] t \right ] ,\end{equation}
which has an expansion as
\begin{equation}\label{ZauxdP0}
Z_{\rm aux}(t, a_1, a_2; dP_0 ) = \sum_{\beta = 0}^\infty [0, 0, 0, \beta ] t^\beta - \sum_{ \beta = 3 }^\infty  [0, 0, 0, \beta - 3 ] t^\beta \ .
\end{equation}
As with \sref{s:F0rev}, we can use \eref{g1betadP0} and \eref{ZauxdP0} to compute the generating function for one D3-brane:
\begin{equation}\ba{rcl}
& &g_{1} \left (t, f_1, f_2, a_1, a_2; dP_0 \right ) = \sum\limits_{\beta = 0}^\infty \sum\limits_{n=0}^\infty [3 n + \beta, 0, 0, \beta ] t^{3 n + \beta} - \sum\limits_{\beta = 3}^\infty \sum\limits_{n=0}^\infty [3 n + \beta, 0, 0, \beta -3 ] t^{3 n + \beta}   \\
&=& \sum\limits_{\beta = 0}^\infty \sum\limits_{n=0}^\infty [3 n + \beta, 0, 0, \beta ] t^{3 n + \beta} - \sum\limits_{\beta = 0}^\infty \sum\limits_{n=1}^\infty [3 n + \beta, 0, 0, \beta ] t^{3 n + \beta} =  \sum\limits_{\beta = 0}^\infty [\beta, 0, 0, \beta ] t^{\beta} \ .
\ea\eeq
In the second equality we shifted the index $\beta$ by 3 and an opposite shift of the index $n$ by 1 in the second term. This allows us, in the third equality, to cancel all terms for $n$ except for the term in $n=0$.
Thus we reproduce \eref{g1dP0-2} in a remarkable cancellation that leaves only positive coefficients.

Subsequently, we can obtain the expression $g_N$, by computing the $\nu$-inserted plethystic exponential and series expansion:
\begin{equation}\label{gfulldP0}\ba{rcl}
&& 
\sum\limits_{N=0}^\infty \nu^N g_N\left ( \nu; t, f_1, f_2, a_1, a_2; dP_0 \right ) = 
g \left ( \nu; t, f_1, f_2, a_1, a_2; dP_0 \right )
\\ 
&=& \sum\limits_{\beta = 0}^\infty [0, 0, 0, \beta ] \left ( PE \left [ \nu \sum\limits_{n=0}^\infty [3 n + \beta, 0, 0, 0 ] t^{3 n + \beta} \right ] - PE \left [ \nu \sum\limits_{n=1}^\infty [3 n + \beta, 0, 0, 0] t^{3 n + 2 \beta} \right ] \right ) .
\ea\end{equation}
Note that the first $PE$ contains all the terms in the second $PE$ and hence all the coefficients in the expansion are positive.
As a check, we can expand \eref{gfulldP0} to second order in $\nu$:
\beq
g_2 \left ( t, f_1, f_2, a_1, a_2; dP_0 \right ) = \sum_{\beta = 0}^\infty \sum_{k = 0}^{\lfloor \frac{\beta}{2} \rfloor} \left [ 2 \beta - 4k, 2k, 0, \beta \right ] t^{ 2 \beta} \\ 
+ \sum_{\beta = 0}^\infty [\beta, 0, 0, \beta ] \sum_{n=1}^\infty [3 n + \beta, 0, 0, 0] t^{3 n + 2 \beta}
\eeq
which indeed agrees with \eref{g2dP0}.

\subsection{$dP_1$ Revisited}\label{s:dp1rev}
Let us now re-examine the $dP_1$ theory, studied in \sref{s:dP1}. The
hidden symmetry expected for the del Pezzo surfaces as an enhancement
of the non-anomalous baryonic symmetry \cite{Franco:2004rt} here is
still trivial $E_1=U(1)$. On the other hand, from the matrix of charges
$Q^t$ for the symplectic action, we realize that the symmetry of 
$\firr{dP_1}$ is $SU(2)\times SU(2)\times U(1)^4$. One $U(1)$ is the R-symmetry and the first $SU(2)$ is the natural one acting on the mesonic moduli space. The second $SU(2)$ is a ``hidden" symmetry coming from one of the two anomalous baryonic $U(1)$ symmetries. 

The four fields $U$ transform in the $({\bf 2},{\bf 2})$ representation of $SU(2)\times SU(2)$, the fields $V$ in the $({\bf 2},0)$ representation, $(Y_1,Y_3)$ in the $(0,{\bf 2})$ representation while $Y_2$ and
$Z$ are $SU(2)\times SU(2)$ singlets. We can use the full symmetry to compute the refined Hilbert series for this space. 
This can be done with any of the methods discussed in \sref{s:master} and the result is
\begin{equation}
g_1(t; dP_1) = P(t)  PE \left [ ( [1,1] + [0,1] + 2 [0,0]) t + [1,0] t^2
\right ]
\end{equation}
with
\begin{eqnarray}
P(t) &=& 1 - ([0,0] + [0,1]) t^2 - [1,1] t^3 + ([2,0] + [0,2] + 2 [0,1] +2 [0,0]) t^4 \nonumber\\
&-& [1,1]  t^5 - ([0,0] +[0,1]) t^6 + t^8  \ ,
\end{eqnarray}
where $[n,m]$ denotes the representation of dimension $(n+1,m+1)$ of $SU(2)\times SU(2)$. The variable $t$ is as in \sref{s:dP1} and, for simplicity, we have
suppressed the weights under the remaining $U(1)$ symmetries. Note that this Hilbert series is palindromic, as expected.
Although there are minus signs in the numerator of the refined Hilbert series,
it is easy to see that $g_1(t; dP_1)$ has
an expansion in terms of irreducible representations of $SU(2)\times U(1)$ that have
non-negative coefficients.

\subsection{$dP_2$ Revisited}
Let us now re-examine the $dP_2$ theory, studied in \sref{s:dP2}. This is
the first example where we expect to see a non trivial hidden 
symmetry $E_n$, extending the non-anomalous baryonic symmetries of
the quiver theory for $dP_n$ \cite{Franco:2004rt}. The expected symmetry
for $dP_2$ is $E_2=SU(2)\times U(1)$. We choose the following assignment of 
charges and weights,
\begin{equation}
\{ X_{14}, X_{23}, X_{31} , X_{45} , X_{52} , X_{42} , Y_{23} , X_{34} , X_{53} , Y_{31} , X_{15}\} \rightarrow \{ 3,-1,-1,-4,3,-1,-1,2,2,-1,-1\}
\end{equation}
for the $U(1)$ action with weight $q$, and
\begin{equation}
\{ X_{14}, X_{23}, X_{31} , X_{45} , X_{52} , X_{42} , Y_{23} , X_{34} , X_{53} , Y_{31} , X_{15}\} \rightarrow \{ 1,1,-1,0,-1,-1,1,0,0,-1,1\}
\end{equation}
for the action of the Cartan generator of $SU(2)$. 

We can again compute the refined Hilbert series with any of the methods discussed in \sref{s:master}. The result is
\begin{equation}
g_1(t, q; dP_2) = P(t,q,x)  PE \left [ 3 [1] q^{-1} + [1] q^3 + 2 [0] q^2 + [0] q^{-4} \right ] \ ,
\end{equation}
with
\begin{equation}\ba{l}
P(t,q,x) = 1 - t^3 ( [2] q^{-2} + [1] q  + [0] q^{-2} ) - t^4 ([1] (q-q^{-3}) + [0] q^4) + \\
+t^5 ( 2 [2] + [1] (q^3+q^{-3}) + 2 [0])
-t^6 ([1] (q^{-1} -q^{3}) + [0] q^{-4}) - t^7 ( [2] q^2 + [1] q^{-1} +  [0] q^2) + t^{10} \ ,
\ea\end{equation}
where $[n]$ denotes the representation of dimension $n+1$ of $SU(2)$. 

Although there are minus signs in the numerator of the refined Hilbert series,
one can check by explicit computation that $g_1(t, q, x; dP_2)$ has
an expansion in terms of irreducible representations of $SU(2)\times U(1)$ with
non-negative coefficients.

\subsection{$dP_3$ Revisited}

Let us now show that the $dP_3$ theory, studied in \sref{s:dP3}, has 
a hidden symmetry $E_3\equiv SU(2)\times SU(3)$, which should be related 
to the $E_3$ symmetry discussed in \cite{Franco:2004rt}. 
As in \cite{Franco:2004rt}, the first six fields in \eref{weightsdP3} with 
weight 
$t$ transform in the $({\bf 2},{\bf 3})$ representation of $SU(2)\times 
SU(3)$ while the other six, with weight $t^2$, transform as two copies of 
the representation $({\bf 1},\bar {\bf 3})$.
The Hilbert series, which we recall from \eref{dP3CY} can now be refined 
using weights of $SU(2)\times SU(3)$ representations and thus represent 
the explicit transformation rules under this group.


To achieve this we will introduce some notation.
We set $\chi_{({\bf 2},{\bf 3})}$ as the character of the fundamental representation of $SU(2)\times SU(3)$. This character can depend on the three different chemical potentials associated with the Cartan sub-algebra of $SU(2)\times SU(3)$. The choice of the weights is not important and is subject to personal preference. To be explicit we can denote the weights of the $SU(2)$ representation with spin $j$ and dimension $2j+1$ by a chemical potential $x=e^{i\theta}$ as
\begin{equation}
\chi_{\bf j}(x) = U_{2j}(\cos \theta),
\end{equation}
where $U_n(x)$ is the Chebyshev polynomial of the second kind; for example, $\chi_{\bf\frac{1}{2}}(x) = x+\frac{\bf 1}{x}$, $\chi_{1}(x) = x^2+1+\frac{1}{x^2}$, etc. The reader is referred to the Chebyshev gymnastics of \cite{Forcella:2007wk}.
Similarly, the $SU(3)$ characters can be chosen by selecting weights $\lambda_1, \lambda_2$ for the fundamental representation which can be conveniently taken as
\begin{equation}
\chi_{\bf 3}(\lambda_1, \lambda_2) = \lambda_1 + \frac{\lambda_2}{\lambda_1} + \frac{1}{\lambda_2} \ .
\end{equation}
Other characters can be computed by using this basic character. For completeness we write the few of direct use below:
\begin{equation}
\chi_{\bar {\bf 3}}(\lambda_1, \lambda_2) = \frac {1}{\lambda_1}+ \frac{\lambda_1}{\lambda_2} +\lambda_2 \ .
\end{equation}

The character for the fundamental representation of $SU(2)\times SU(3)$ then takes the form
\begin{equation}
\chi_{({\bf 2},{\bf 3})}(x, \lambda_1,\lambda_2) = \chi_{\bf\frac{1}{2}}(x) \chi_{\bf 3}(\lambda_1, \lambda_2) \ ,
\end{equation}
and more generally a representation of $SU(2)\times SU(3)$ with dimensions $({\bf 2j+1},{\bf \dim R})$ has a character
\begin{equation}
\chi_{({\bf j},{\bf R})}(x, \lambda_1,\lambda_2) = \chi_{\bf j}(x) \chi_{\bf R}(\lambda_1, \lambda_2) \ .
\end{equation}

We proceed by recasting the Hilbert series \eref{dP3CY} in a way more suitable to reflect the representation structure, by multiplying the numerator and denominator of $H(t;~\firr{dP_3})$ by $(1-t^2)^4$:
\begin{equation}\label{dP3CYdim}
g_1(t;~dP_3) = \frac{1 - 9 t^4 + 16 t^6 - 9 t^8 + t^{12}}{(1- t)^6(1-t^2)^6}  \ .
\end{equation}
Now the coefficients of the numerator are dimensions of representations while the exponents of the denominator are also dimensions of representations. The denominator takes a form of a plethystic exponential for the function $6t+6t^2$, which after refinement by $SU(2)\times SU(3)$ weights takes the form
$\chi_{({\bf 2},{\bf 3})}(x, \lambda_1,\lambda_2) t + 2 \chi_{({\bf 1},\bar {\bf 3})} (\lambda_1,\lambda_2) t^2$. This agrees with Table (3.8) of \cite{Franco:2004rt} and is indeed consistent with the general expectation that each field in the quiver is a generator of the Hilbert series for the master space.
The refinement in terms of weights also reveals the different characters for the numerator which takes the form
$1 - 3 \chi_{({\bf 1}, {\bf 3})} (\lambda_1,\lambda_2) t^4 + (8 + \chi_{({\bf 1}, {\bf 8})} (\lambda_1,\lambda_2) )t^6 - 3 \chi_{({\bf 1},\bar {\bf 3})} (\lambda_1,\lambda_2) t^8 + t^{12}$.
The representations again are chosen such that they agree with the expected generators and relations, as well as the palindromic property.

Collecting all of these together we find
\begin{eqnarray}\label{PEdP3}
g_1(t;~dP_3)={(1 - 3 \chi_{({\bf 1}, {\bf 3})} t^4 + (8 + \chi_{({\bf 1}, {\bf 8})} )t^6 - 3 \chi_{({\bf 1},\bar {\bf 3})} t^8 + t^{12})}
{{\rm PE} [ \chi_{({\bf 2},{\bf 3})} t + 2 \chi_{({\bf 1},\bar {\bf 3})} t^2 ]}  \ .
\end{eqnarray}
This has the nice feature that the terms in $t^a$ and in $t^{12-a}$ are symmetric with respect to conjugation of the representation, the term in $t^6$ being self conjugate.
Equation (\ref{PEdP3}) constitutes the explicit demonstration that the Hilbert series for the non-trivial Calabi-Yau component of the master space of $dP_3$ decomposes into representations of $SU(2)\times SU(3)$ and the basic building blocks are given by the simplest representations of this group.

Expanding \eref{PEdP3} in powers of $t$ we find
\beq\ba{rcl}
g_1(t;~dP_3)&=&1 + \chi_{({\bf 2},{\bf 3})} t + ( \chi_{({\bf 3},{\bf 6})} + 3 \chi_{({\bf 1},\bar {\bf 3})} ) t^2 + ( \chi_{({\bf 4},{\bf 10})} + 3 \chi_{({\bf 2}, {\bf 8})} + 2 \chi_{({\bf 2}, {\bf 1})} ) t^3 \\
&+&( \chi_{({\bf 5},{\bf 15})} + 3 \chi_{({\bf 3}, {\bf 15'})} + 6 \chi_{({\bf 1}, \bar {\bf 6})} + 2 \chi_{({\bf 3}, {\bf 3})} ) t^4 + \ldots
\ea\eeq

We will now re-write the Hilbert series in a suggestive form which uses a highest weight representation as was done in the preceding subsections. A representation of $SU(2)\times SU(3)$ will now be denoted by 3 non-negative integers $[n_1, n_2, n_3]$ such that $n_1=2j$ denotes an $SU(2)$ representation with spin $j$ and $n_2, n_3$ are the highest weights of an $SU(3)$ representation given by a Young diagram such that $n_2$ is the difference between the first row and the second row, while $n_3$ is the difference between the second row and the third row. Using this notation \eref{PEdP3} becomes
\begin{equation}
g_1(t;~dP_3)={(1 - 3 [0,1,0] t^4 + ( 8 + [0,1,1] ) t^6 - 3 [0,0,1] t^8 + t^{12} )}
{PE \left [ [1,1,0] t + 2 [0,0,1] t^2 \right ]} \ ,
\end{equation}
which admits an explicit expansion as
\beq\ba{rcl}
g_1(t;~dP_3)
&=&1 + [1,1,0] t + ( [2,2,0] + 3 [0,0,1] ) t^2
+ ( [3,3,0] + 3 [1,1,1] + 2 [1,0,0] ) t^3 \\ 
&+& ( [4,4,0] + 3 [2,2,1] + 6 [0,0,2] + 2 [2,1,0] ) t^4 \\ 
&+& ( [5,5,0] + 3 [3,3,1] + 6 [1,1,2] + 2 [3,2,0] + 3 [1,0,1] ) t^5 + \ldots
\ea\eeq

In fact, one can collect all the terms in the expansion and get an expression to all orders in powers of $t$:
\beq\label{HWdP3}
\ba{rcl}
g_1(t;~dP_3)
&=& \sum\limits_{n=0}^\infty \left ( \sum\limits_{k=0}^{\lfloor \frac{n}{2} \rfloor} \frac{ (k+1) (k+2) } {2} [n - 2k, n - 2k, k] \right ) t^{n} \\ 
&+& \sum\limits_{n=0}^\infty \left ( \sum\limits_{j=1}^{\lfloor \frac{n}{3} \rfloor} \sum\limits_{k=0}^{\lfloor \frac{n - 3j}{2} \rfloor} (k+j+1)[n-2k-2j, n-2k-3j, k] \right ) t^{n} ,
\ea\eeq
where $\lfloor x \rfloor$ represents the integer part of $x$. The restrictions in the summation over $j$ and $k$ are chosen such that every integer in the highest weight representations above is a non-negative integer.
Using the formula for the dimension of a representation of $SU(3)$ of highest weight $[n_2, n_3]$, namely,
\begin{equation}
{\rm dim}\, [n_2, n_3] = \frac{(n_2+1)(n_3+1)(n_2+n_3+2)}{2} ,
\end{equation}
and recalling the dimension of the $SU(2)$ representations, we have
\begin{equation}
{\rm dim}\, [n_1, n_2, n_3] = \frac{(n_1+1)(n_2+1)(n_3+1)(n_2+n_3+2)}{2} ,
\end{equation}
which can be substituted into \eref{HWdP3} by setting all weights to 1. The result is remarkably simple and reproduces \eref{dP3CYdim} as expected.

\subsection{A Prediction for True $dP_4$}
We have addressed the Pseudo-del Pezzo theories above in \sref{s:case} because they are still toric; the true del Pezzo theories above $n=3$ are non-toric and still a {\it terra incognita} as gauge theories \footnote{See \cite{Butti:2006nk} for a discussion of how to obtain non toric theories and their Hilbert series via deformation of toric ones. See also \cite{Wijnholt:2002qz} for some of the non-toric theories derived from exceptional collections.}. Nevertheless, armed with our hidden symmetry techniques and using the exact expression for the Hilbert series of $dP_3$ we can come up with arguments using the Higgs mechanism \cite{Feng:2002fv}
and simple group decomposition in order to propose an exact expression for the Hilbert series of true $dP_4$. 

The matter fields will be taken to be 10 fields of weight $t$ transforming in the second rank antisymmetric tensor ${\bf 10}=[0,1,0,0]$ and 5 fields of weight $t^2$ transforming in the anti-fundamental representation ${\bar {\bf 5}}=[0,0,0,1]$. The hidden symmetry is now expected to be $E_4=SU(5)$ and the under the decomposition $SU(5)\supset SU(2)\times SU(3)$ we have
\beq\ba{rcl}
{\bf 10}&\rightarrow& ({\bf 1},{\bf 1})+ ({\bf 1},{\bar {\bf 3}}) + ({\bf 2},{\bf 3}) \\
{\bar {\bf 5}}&\rightarrow& ({\bf 2},{\bf 1})+({\bf 1},{\bar {\bf 3}}) \ .
\ea\eeq
As above, we can denote a character of a representation by its highest weight, $[n_1, n_2, n_3, n_4]$, where $n_i$ are non-negative integers, where $n_1$ is the difference between the first row of the Young diagram to the second row, $n_2$ is the difference between second and third rows, etc., we propose:
\begin{equation}\ba{rcl}
H(t;~\firr{dP_4})
&=&\left( 1- [1,0,0,0] t^3 + 2 t^5 + [0,1,0,0] t^6 - [0,0,0,1] t^7 \right. \\
&& \left. -[1,0,0,0] t^8 + 2 t^{10} \right)
{PE [ \chi_{\bf 10} t + \chi_{\bar {\bf 5}} t^2 ]} \ .
\ea\end{equation}
This Hilbert series is conjectured to have a nice expansion in terms of the symmetric representations of $SU(5)$:
\begin{equation}\ba{rcl}
&&H(t;~\firr{dP_4})
=1 + [0,1,0,0] t + ( [0,2,0,0] + 2 [0,0,0,1] ) t^2
+ ( [0,3,0,0] + 2 [0,1,0,1] )t^3 \\ 
&&~~+( [0,4,0,0] + 2 [0,2,0,1] + 3 [0,0,0,2]) t^4
+ ( [0,5,0,0] + 2 [0,3,0,1] + 3 [0,1,0,2]) t^5 + \ldots
\ea\end{equation}
This can be used to evaluate the expression to all orders in $t$,
\begin{equation}\ba{rcl}
&&H(t;~\firr{dP_4}) = 
\sum\limits_{n=0}^\infty \left ( \sum\limits_{k=0}^{\lfloor \frac {n}{2} \rfloor } (k+1) [0,n - 2k,0,k] \right ) t^{n} \\ 
&=& \sum\limits_{n=0}^\infty \left ( \sum\limits_{k=0}^n (k+1) [0,2n - 2k,0,k] \right ) t^{2n} + \sum\limits_{n=0}^\infty \left ( \sum\limits_{k=0}^n (k+1) [0,2n - 2k+1,0,k] \right ) t^{2n+1} \ .
\ea\end{equation}

The above expression is explicitly checked up to order 15 in $t$. It is not hard to extend this to higher order. In fact, it is easy to compute the dimension of the representation $[0,m,0,l]$, which is
\begin{equation}
\frac{ (m+l+4)(m+l+3)(l+2)(l+1)(m+3)(m+2)^2(m+1)}{288} \ .
\end{equation}
We can now replace the characters of each representation by its dimension and obtain an expression for the Hilbert series of the Calabi-Yau component of the master space of $dP_4$:
\begin{equation}
H(t;~\firr{dP_4})
=\frac{1- 5 t^3 + 2 t^5 + 10 t^6 - 5 t^7 - 5 t^8 + 2 t^{10}}
{(1-t)^{10} (1 - t^2 )^5 } .
\end{equation}
An interesting aspect of this series is that it gives a dimension 11 for the master space of $dP_4$ and not a dimension 9 as what we expect from all the toric cases. This deserves a further inspection.

Another proposal for the Hilbert series can be
\begin{equation}
H(t;~\firr{dP_4}) = 
\sum\limits_{n=0}^\infty \left ( \sum\limits_{k=0}^{\lfloor \frac {n}{2} \rfloor } [0,n - 2k,0,k] \right ) t^{n} \ ,
\end{equation}
which sums to 
\begin{equation}
( 1-[0,0,0,1] t^2+[0,0,1,0] t^4- [0, 1, 0, 0] t^6 + [1, 0,0,0] t^8 - t^{10} ) PE [ [0,1,0,0] t + [0,0,0,1] t^2 ] .
\end{equation}
The current form is not satisfactory and we need more data in order to get the right answer. This is left for future work.

\section{Conclusions and Prospectus}\label{s:conc}\setall
We have enjoyed a long theme and variations in $\f$, touching upon diverse motifs. Let us now part with a recapitulatory cadence. We have seen that for a single brane, the master space is the algebraic variety formed by the space of F-terms. In the case of the singularity $\cX$ which the D3-brane probes being a toric Calabi-Yau threefold, we have a wealth of techniques to study $\f$: direct computation, toric cones via the $K$ and $T$ matrices, symplectic quotients in the Cox coordinates as well as dimer models and perfect matchings. Using these methods we have learned that
\begin{itemize}
\item $\cX$ is the mesonic branch of the full moduli space $\CM$ of a single  D3-brane gauge theory and is the symplectic quotient of $\f$ by the Abelian D-terms;
\item For a $U(1)^g$ toric quiver theory, $\f$ is a variety of dimension $g+2$;
\item The master space $\f$ is generically reducible, its top dimensional component, called the coherent component $\firr{}$, is a Calabi-Yau variety, of the same dimension and degree as $\f$. The lower-dimensional components are linear pieces $L_i$, composed of coordinate hyperplanes;
\item $\firr{}$ is generated by the perfect matchings in the dimer model (brane tiling) corresponding to the quiver theory. This should follow from the Birkhoff-von Neumann theorem;
\item In the field theory, $\firr{}$ often realizes as the Higgs branch, and the hyper-planes $L_i$, the Coulomb branch of the moduli space $\CM$. The acquisition of VEVs by the fields parametrising $L_i$ can cause one theory to flow to another via the Higgs mechanism, an archetypal example is the chain of $dP_n$ theories;
\item Under Seiberg/toric duality, we conjecture that $\firr{}$ remains invariant; 
\item According to the plethystic programme, the Hilbert series of $\cX$ is the generating function for the BPS mesonic operators. In order to count the full chiral BPS operators, including mesons and baryons, we need to find the refined (graded by various chemical potentials) Hilbert series of $\f$;
\item The Hilbert series of the various irreducible pieces of $\f$, obtained by primary decomposition, obey surgery relations;
\item The numerator of the Hilbert series of $\firr{}$, in second form, is palindromic. This follows from the Stanley theorem;
\item The gauge theory possesses hidden global symmetries corresponding to the symmetry of $\f$ which, though not manifest in the Lagrangian, is surprisingly encoded in the algebraic geometry of $\f$. In particular, we can re-write the terms of the single brane generating function, i.e., the refined Hilbert series, of $\firr{}$ in the weights of the representations of the Lie algebra of the hidden symmetry in a selected set of examples. Via the plethystic exponential, this extends to an arbitrary number $N$ of branes.
\end{itemize}

\begin{table}[t]
$
\begin{array}{|c|c|c|c|c|} \hline
\cX & \f & \firr{} & H(t;~\firr{}) & \mbox{Global Symmetry} \\ \hline \hline
\IC^3 & \IC^3 & \IC^3 & (1-t)^{-3} & U(3) \\ \hline
\cC & \IC^4 & \IC^4 & (1-t)^{-4} & U(1)_R \times SU(4)_H \\ \hline
(\IC^2 / \IZ_2) \times \IC & (4,2) & \cC \times \IC & \frac{1 + t}{(1-t)^4} &
  U(1)_R \times SU(2)_R \times U(1)_B \times SU(2)_H \\ \hline
\IC^3 / \IZ_2\times \IZ_2 & (6,14) & - & \frac{1+6 t+6 t^2+t^3}{(1-t)^6} &
  U(1)_R \times U(1)^2 \times SU(2)^3_H \\ \hline
SPP & (5,2) & \cC \times \IC^2 & \frac{1 + t}{(1-t)^5} &
  U(1)_R \times U(1)_M \times SU(2)_H^3 \\ \hline
dP_0 & (5,6) & \simeq \f & \frac {1+4t+t^2}{(1-t)^5} &
  U(1)_R \times SU(3)_M \times SU(3)_H \\ \hline
F_0 & (6,4) & \cC \times \cC & \frac{(1 + t)^2}{(1-t)^6} &
  U(1)_R \times U(1)_B \times SU(2)^2_M \times SU(2)_H^2 \\ \hline
dP_1 & (6,17) & -& \frac{1 + 4 t + 7 t^2 + 4 t^3 + t^4}{(1 - t)^6(1+t)^2} &
  U(1)_R \times SU(2)_M \times U(1)^3 \times SU(2)_H \\ \hline
dP_2 & (7,44) & -& \frac{1 + 2t + 5 t^2 + 2t^3 + t^4}{(1-t)^7(1+t)^2} &
  U(1)_R \times SU(2)_H  \times U(1)^5 \\ \hline
dP_3 & (8,96) & -& \frac{1 + 4t^2 + t^4}{(1 - t)^8(1+t)^2} &
  (SU(2) \times SU(3))_H \times U(1)^5 \\ \hline
\end{array}
$
\caption{{\sf    
The master space, its coherent component and Hilbert space as well
as the global symmetry of the gauge theory. The notation $(n,d)$ denotes the
dimension and degree respectively of $\f$. For the symmetries, the subscript $R$ denotes R-symmetry, $M$ denotes the symmetry of the mesonic branch, $B$ denotes baryon charge, while $H$ denotes the hidden global symmetry. Note that the rank of the global symmetry group is equal to the dimension of $\f$.}}
\label{t:sum}
\end{table}

In Table \ref{t:sum}, we illustrate some of the above points by suumarizing various results for our host of examples encountered throughout the paper. We present the single-brane master space, its coherent component (by name if familiar), the assciated Hilbert series as well as the global symmetry, standard as well as hidden. In passing, noticing the last few rows of the table, we see that for general toric $dP_{n=0,1,2,3}$, the coherent component of the master space has Hilbert series
\beq
H(t~\firr{dP_n}) = 
\frac{1 + (6-2n) t + (10 - \frac72 n + \frac12n^2) t^2 + (6-2n) t^3 + t^4}
{(1-t)^{5+n} (1+t)^2} \ .
\eeq
Indeed, the numerator is explicitly palindromic and there is no need to fret over the appearance of the $\frac12$ therein since $n^2-7n$ always divides 2 for integer $n$.

For a general number $N$ of D3-branes, the situation is more subtle. The moduli space is now the variety of F-flatness quotiented by the special unitary factors of the gauge group, so that when quotiented by the $U(1)$ factors as a symplectic quotient we once more arrive at the mesonic branch, which here is the $N$-th symmeterized product of the Calabi-Yau space $\cX$. However, the plethystic programme persists through and we can still readily extract the generating functions for any $N$. Furthermore, we still see the representation of the hidden symmetries in the plethystic exponential.

We are at the portal to a vast subject. The algebraic geometry of the master space at $N > 1$ number of branes deserves as detailed a study as we have done for the single-brane example; we have given its form on physical grounds and mathematically the structure is expected to be complicated, doubtlessly Hilbert schemes will arise since we are dealing with symmetrised tensor products.  Plethystics are expected to elucidate the situation.

The top-dimensional irreducible component of the master space is seen to be an important object and we have shown a few of its properties for $N=1$. We have conjectured that Seiberg duality preserves this with our example, it would be important to prove this in general, for higher number of branes, and indeed for generic $\CN=1$ gauge theories as well. Using the encoding of hidden symmetries by the refined Hilbert series, we have also been able to make predictions about gauge theories, such as the true non-toric del Pezzo theories, whose details have yet to be completely settled. We need to check these predictions with more data. Indeed, as emphasized above, our systematic analysis should apply to not only D-brane theories but supersymmetric gauge theories in general; the master space and its associated physical insight need to be thus investigated panoramically. The full symphony based on our motif in $\f$ awaits to be composed.

\begin{center}
${~}^{\underline{~~~~~~~~~~~~~~}}$~~
$\Diamond$
${~}^{\underline{~~~~~~~~~~~~~~}}$
\end{center}

\section*{Acknowledgements}

We are indebted to Nathan Broomhead, Vadim Kaplunovsky, Alastair King, Balazs Szendr\"oi and David Tong for many enlightening discussions.
A.~H., Y.-H.~H., and A.~Z.~would like to thank the Newton
Institute in Cambridge for hospitality and support during part of this
work.
A.~H.~is grateful to the Perimeter Institute, the physics and mathematics departments at the University of Texas at Austin, and Stanford University where parts of this project were completed.
D.~F.~is supported in part by INFN and the Marie Curie
fellowship under the programme EUROTHEPHY-2007-1.
Y.-H.~H.~bows to the gracious patronage of Merton College, Oxford through the FitzJames Fellowship.
A.~Z.~ is supported
in part by INFN and MIUR under contract 2005-024045-004 and
2005-023102 and by the European Community's Human Potential Program
MRTN-CT-2004-005104.

\newpage 

\noindent {\LARGE {\bf Appendices}}
\appendix
\section{Hilbert Series of Second Kind and the Reeb Vector}\label{app:reeb}
In this appendix, let us study, in a further detail, the properties of the Hilbert series of a dimension $n$ variety $M$, in light of its pole structure and the subsequent relation to the geometry of $\CM$.
A Laurent expansion for the Hilbert series of second kind in \eref{hs12}
can be developed, as a partial fraction expansion:
\beq\label{frac-f}
H(t,M)= \frac{V_n}{(1-t)^{n}} + \ldots
\frac{V_3}{(1-t)^{3}} + \frac{V_2}{(1-t)^{2}} + \frac{V_1}{1-t}
+ V_0 + \cO(1-t) \ ,
\eeq
where we see explicitly that the Hilbert series is a rational
function and the degree of its most singular pole is the dimension of
$M$. In the case of $M$ being a toric variety of dimension 3,
the coefficients $V_{0,1,2,3}$ are related directly to the Reeb
vector of $M$ and in particular, $V_3$ is the volume of the
spherical Sasaki-Einstein horizon. The relation to the Reeb vector, at
least for toric $M$, is
as follows. Refine the generating function into tri-variate (this can
always be done for toric $M$), in terms of $t_{i=1,2,3}$ and set
\beq
t_i := \exp( - b_i q) \ , \qquad
\vec{b} = (b_1, b_2, b_3) \mbox{ is the Reeb vector} 
\eeq
and then Laurent expand $f(t_1,t_2,t_3)$ near $q \to 0$ to compare
with \eref{frac-f}. A full discussion on the relation of the volume in
terms of the Reeb vector is nicely presented in \cite{Martelli:2006yb}.
We will call this multi-variate Laurent expansion around $q \to 0$ the
expansion of a {\bf refined Hilbert series of Second kind}.

Let us first illustrate with the simplest case of $\IC^3$. We recall
from \cite{pleth} that the refined fundamental generating function, is
simply
\beq
H(t_1,t_2,t_3;~\IC^3) = \left( (1-t_1)(1-t_2)(1-t_3) \right)^{-1} \ .
\eeq
Hence, the Laurent expansion gives
\bean
&& \qquad H(\exp( - b_1 q), \exp( - b_2 q), \exp( - b_3 q); \IC^3) = 
\frac{1}{b_1\,b_2\,b_3\,q^3} + 
  \frac{b_1 + b_2 + b_3}{2\,b_1\,b_2\,b_3\,q^2} + \\
&&  \frac{{b_1}^2 + {b_2}^2 + {b_3}^2 + 3 \left( b_1 b_2 + b_2
    b_3 + b_1 b_3 \right) }
     {12\,b_1\,b_2\,b_3\,q} + 
  \frac{\left( b_1 + b_2 + b_3 \right) \,
     \left( b_1 b_2 + b_2
    b_3 + b_1 b_3 \right) }{24\,
     b_1\,b_2\,b_3} + \cO(q)
\eean
Therefore, we can read off the volume as $V_3 = (b_1 b_2
b_3)^{-1}$. In general, the values of $V_{i}$ are read off as the
coefficients of $q^{-i}$ from above.

Let us study another example, viz., the conifold.
We recall from \cite{pleth} that the refined Hilbert series for the
conifold is
\beq
H(t_1,t_2,t_3;~\cC) =
\frac{t_1\,t_2\,\left( 1 - t_3 \right) }
    {\left( 1 - t_1 \right) \,\left( 1 - t_2 \right) \,
      \left( t_1 - t_3 \right) \,\left( t_2 - t_3 \right) } \ .
\eeq
Therefore, Laurent expansion gives us
\bean
&& \qquad H(\exp( - b_1 q), \exp( - b_2 q), \exp( - b_3 q); \cC) = 
\frac{b_3}{b_1\,\left( b_1 - b_3 \right) \,
     \left( {b_2}^2 - b_2\,b_3 \right) \,q^3} + \\
&&  + \frac{{b_3}^2}
   {2\,b_1\,\left( b_1 - b_3 \right) \,
     \left( {b_2}^2 - b_2\,b_3 \right) \,q^2} + 
  \frac{b_3\,\left( -{b_1}^2 - {b_2}^2 + 
       b_1\,b_3 + b_2\,b_3 + {b_3}^2 \right) }
     {12\,b_1\,b_2\,\left( b_1 - b_3 \right) \,
     \left( b_2 - b_3 \right) \,q} - \\
&&  - \frac{{b_3}^2\,\left( {b_1}^2 + {b_2}^2 - 
       b_1\,b_3 - b_2\,b_3 \right) }{24\,b_1\,
     \left( b_1 - b_3 \right) \,
     \left( {b_2}^2 - b_2\,b_3 \right) } + \cO(q)
\eean
Therefore, here the volume is $V_3 = \frac{b_3}{b_1\,\left( b_1 - b_3
  \right) \, \left( {b_2}^2 - b_2\,b_3 \right)}$, which up to
permutation of the definition of $b_{1,2,3}$, agrees with Eq 7.29 of
\cite{Martelli:2006yb}. The remaining $V_{2,1,0}$ can be similarly
obtained.

\section{Refined Hilbert Series: {\it Macaulay2} Implementation}\label{ap:M2}
In this appendix, we present the Macaulay2 routine (cf.~\cite{m2,M2book}) which computes the refined Hilbert series of a toric variety, given its $K$-matrix of charges.
\begin{verbatim}
toBinomial = (b,R) -> ( 
 top := 1_R; bottom := 1_R; 
 scan(#b, i->if b_i > 0 then top =top*R_i^(b_i) 
      else if b_i < 0 then bottom =bottom*R_i^(-b_i));
top - bottom);

toricIdeal = (A) -> (
 n :=#(A_0);
 R= QQ[vars(0..n-1),Degrees=>transpose A,MonomialSize=>16];
 B:= transpose LLL syz matrix A; 
 J:= ideal apply(entries B, b -> toBinomial(b,R)); 
 scan(gens ring J, f -> J=saturate(J,f)); 
J);
\end{verbatim}

The input is the $g+2$ by $E$ matrix $K$ and the output of the command
\verb|toricIdeal(K)| 
is the refined Hilbert series for the coherent component $\firr{~}$ weighted
by all the $g+2$ charges (chemical potentials). 
The algorithm can be easily generalized
to compute the Hilbert series depending on only one, or fewer, charges.

\section{Refined Hilbert Series using Molien Formula}\label{ap:ref}
In this Appendix we give an explicit example of computation of the refined 
Hilbert series using the Molien formula. 
The method works well for quivers
with relatively small number of fields.

Consider the example of  $\IC^3 / \IZ_2\times \IZ_2$. The Hilbert series 
depending on one parameter $t$ was computed in \sref{s:Molien} from the
symplectic quotient description with charges \eref{chargesZ2}.
To highlight the result for the refined Hilbert series we will exploit the full symmetry of the moduli space.
The symmetry of the master space can be readily determined to be $SU(2)^3\times U(1)^3$, the three $SU(2)$'s coming from repetition of columns in the charge matrix \eref{chargesZ2}. This is another example of the 
general phenomenon related to the
existence of hidden symmetries which is discussed in detail in \sref{s:hidden}.

For now, we want to learn how to compute the refined Hilbert series using 
\eref{Molien}. To this purpose we introduce nine homogeneous variables $y_\alpha, \alpha=1,...,9$ acted on by $(\mathbb{C}^{*})^3$ with charges $Q_i$ given
by the rows of the matrix \eref{chargesZ2}. The Molien formula \eref{Molien} then reads
\begin{equation}\ba{l}
   H(\underline{y},\firr{\IC^3 / \IZ_2\times \IZ_2})
=\int \frac{dr dw ds}{r w s}\frac{1}{(1-\frac{y_1}{r} w)(1- \frac{y_2}{r} s)(1- \frac{y_3}{w} s)(1-y_4 r)(1-y_5 r)(1-y_6 w)(1-y_7 w)(1- y_8 s )(1-y_9 s)} \ .
\ea\end{equation}
The result of the three integrations is a lengthy rational expression that 
we do not report here; it depends on 6 independent quantities that can
be matched with the  six chemical potentials $t_i$ by solving \eref{toricsympaction}. In this case we can use the $SU(2)^3\times U(1)^3$ symmetry
and introduce a set of adapted chemical potentials 
respecting the symmetry of the matrix $Q$:
\begin{equation}
\{y_1,y_2,y_3,y_4,y_5,y_6,y_7,y_8,y_9\} = \{t_1,t_2,t_3,  z, 1 /z, y , 1/y  , x,1/x \} \ .
\end{equation} 
The three charges $t_i$ parameterize the R-charge and the two non-anomalous flavor $U(1)$ symmetries of the theory; these are obtained by assigning a different chemical potential to the three external perfect matchings in the dimer description as discussed in \sref{s:dimer}. The variables $x,y,z$ are the
chemical potentials for the Cartan subgroup of $SU(2)^3$ and the refined 
Hilbert series can be then computed as
\begin{equation}
H(t_1,t_2,t_3,x,y,z,\firr{\IC^3 / \IZ_2\times \IZ_2}) = P(t_1,t_2,t_3,x,y,z) PE \left [t_1 [0,1,1] + t_2 [1,0,1] + t_3 [1,1,0] \right ]  \ ,
\end{equation}
with
\begin{equation}
\ba{rcl}
 P(t_1,t_2,t_3,x,y,z) &=& 1 - t_1^2-t_2^2 -t_3^2 + 2 t_1 t_2 t_3 +2 t_1^2 t_2^2 t_3^2 -  t_1 t_2^3 t_3^3 
-t_1^3 t_2 t_3^3 -t_1^3 t_2^3 t_3  + t_1^3 t_2^3 t_3^3\\ 
&+& [0,1,1] \left(- t_2 t_3 + t_1 t_2^2 + t_1 t_3^2- t_1^2 t_2 t_3- t_1 t_2^2 t_3^2
+t_1^2 t_2^3 t_3+ t_1^2 t_2 t_3^3- t_1^3 t_2^2 t_3^2 \right)  \\
&+& [1,0,1] \left(- t_1 t_3 + t_1^2 t_2 +t_2 t_3^2 - t_1 t_2^2 t_3- t_1^2 t_2 t_3^2 + t_1^3 t_2^2 t_3+t_1 t_2^2 t_3^3 - t_1^2 t_2^3 t_3^2 \right)\\
&+& [1,1,0] \left(- t_1 t_2+ t_1^2 t_3 + t_2^2t_3 - t_1 t_2 t_3^2- t_1^2 t_2^2 t_3
+ t_1^3 t_2 t_3^2 +t_1 t_2^3 t_3^2 - t_1^2 t_2^2 t_3^3 \right)\\
&+& [2,0,0] \left(t_1 t_2 t_3 - t_2^2 t_3^2-   t_1^3 t_2 t_3+t_1^2 t_2^2 t_3^2 \right)\\
&+& [0,2,0] \left(t_1 t_2 t_3 - t_1^2 t_3^2  - t_1 t_2^3 t_3 +t_1^2 t_2^2 t_3^2 \right)\\
&+& [0,0,2] \left(t_1 t_2 t_3 - t_1^2 t_2^2 - t_1 t_2 t_3^3  +t_1^2 t_2^2 t_3^2 \right) \ ,
\end{array}
\label{rHS}
\end{equation}
where $[n,m,l]$ is the character of the representation of dimension 
$(n+1,m+1,l+1)$ of $SU(2)^3$ expressed in the variables $x,y,z$ with the specified order. For example, $[1,0,0]=x+1/x$. 

The plethystic 
exponential $PE$ in the above is defined in \sref{s:plet} which we recall is such that for polynomials or power series $f(t)$ with $f(0)=0$, it is given by 
\begin{equation}
PE \left [ f(t)\right ] = \mbox{exp}\left (\sum_{k=1}^\infty \frac{ f(t^k)}{k} \right ) .
\end{equation}
In our case, the plethystic exponential counts  all possible symmetric products of the twelve elementary
fields organized as the $[0,1,1],[1,0,1]$ and $[1,1,0]$ representations
of $SU(2)^3$. The symmetry of the master space is manifest in the expression
of the refined Hilbert series; this is the subject of \sref{s:hidden}. 
Notice also the remarkable palindromic 
symmetry of the numerator of the refined Hilbert series as a
polynomial in three variables $t_1,t_2,t_3$ under the exchange
$t_1^i t_2^j t_3^k \leftrightarrow t_1^{3-i} t_2^{3-j} t_3^{3-k}$. This point is discussed in detail in \sref{recaptoric}.

On the practical side, it is sometimes convenient to perform the integration in the Molien formula
before substituting the expression of  the dummy variables $y_\alpha$ in terms 
of $t_i$ since in this way there is no ambiguity in the integration: 
the contour integral is performed on unit circles and take contributions
from the poles inside the unit circles, where we consider all $|y_\alpha|<1$.

\newpage

\end{document}